\documentclass[twoside,11pt]{article}
\usepackage{amsthm}          
\usepackage[preprint]{jmlr2e}  
\usepackage[T1]{fontenc}         
\usepackage{lmodern}            
\hypersetup{hidelinks}          
\usepackage{lastpage}
\usepackage{bm}
\usepackage{booktabs}
\makeatletter\renewenvironment{proof}[1][Proof]{\par\noindent{\bf #1\ }}%
  {\unskip\nobreak\hfil\penalty50\hskip2em\hbox{}\nobreak\hfill\BlackBox\par\addvspace{2mm}}\makeatother
\usepackage{amsmath}
\usepackage{graphicx}
\usepackage{caption}
\usepackage{enumerate}
\usepackage{url} 
\usepackage{amsfonts}
\usepackage{float} 
\usepackage{amssymb}
\usepackage[utf8]{inputenc}
\usepackage{multirow}
\usepackage{xcolor}

\usepackage{adjustbox} 
\usepackage{subcaption}
\usepackage{placeins}
\usepackage{tikz}
\usetikzlibrary{arrows.meta, positioning, calc}
\usetikzlibrary{arrows.meta,positioning,shadows.blur}
\newtheorem{assumption}{Assumption}
\newtheorem{lem}{Lemma}
\newtheorem{prop}{Proposition}
\newtheorem{rem}{Remark}

\providecommand{\symbf}[1]{\boldsymbol{#1}}
\newcommand{\bY}{\mathbf{Y}}\newcommand{\bX}{\mathbf{X}}\newcommand{\bU}{\mathbf{U}}\newcommand{\bW}{\mathbf{W}}
\newcommand{\bE}{\mathbf{E}}\newcommand{\bA}{\mathbf{A}}\newcommand{\bB}{\mathbf{B}}\newcommand{\bC}{\mathbf{C}}
\newcommand{\bI}{\mathbf{I}}\newcommand{\bS}{\mathbf{S}}
\newcommand{\bzero}{\mathbf{0}}
\newcommand{\bSig}{\boldsymbol{\Sigma}}\newcommand{\bSigs}{\boldsymbol{\Sigma}^{*}}\newcommand{\bPi}{\boldsymbol{\Pi}}
\newcommand{\bpi}{\boldsymbol{\pi}}

\newcommand{\E}{\mathbb{E}}\newcommand{\Prob}{\mathbb{P}}\newcommand{\R}{\mathbb{R}}
\newcommand{\T}{\top}
\newcommand{\Normal}{\mathrm{N}}\newcommand{\Bern}{\mathrm{Bernoulli}}\providecommand{\Beta}{\mathrm{Beta}}\DeclareMathOperator*{\argmax}{arg\,max}
\newcommand{\MRRGM}{\texttt{IV-RGM}}\newcommand{\MRRGMp}{\texttt{IV-RGM+}}\newcommand{\MRRGMnc}{\texttt{IV-RGM\_NoConf}}

\providecommand{\supp}{\operatorname{supp}}
\DeclareMathOperator{\tr}{tr}\DeclareMathOperator{\diag}{diag}\DeclareMathOperator{\Cov}{Cov}\DeclareMathOperator{\Var}{Var}
\DeclareMathOperator{\rank}{rank}

\jmlrheading{}{2026}{1-\pageref{LastPage}}{}{}{}{Bitan Sarkar, Yuchao Jiang, Tian Ge and Yang Ni}
\ShortHeadings{Cyclic Graphical Models with Instruments}{Sarkar, Jiang, Ge and Ni}
\firstpageno{1}

\usepackage{pdfpages}
\begin{document}

\title{Bayesian Instrumented Directed Graphical Models for Cyclic Causal Discovery with Unmeasured Confounding}

\author{\name Bitan Sarkar \email bitansarkar010899@gmail.com \\
       \addr Department of Statistics, Texas A\&M University, College Station, TX, USA
       \AND
       \name Yuchao Jiang \email yuchaojiang@tamu.edu \\
       \addr Department of Statistics, Texas A\&M University, College Station, TX, USA
       \AND
       \name Tian Ge \email tge1@mgb.org \\
       \addr Psychiatric and Neurodevelopmental Genetics Unit, Center for Genomic Medicine,\\
       Massachusetts General Hospital, Boston, MA, USA
       \AND
       \name Yang Ni\thanks{Corresponding author.} \email yangni87@gmail.com \\
       \addr Department of Statistics and Data Science, The University of Texas at Austin,\\
       105 E 24th St, D9800, Austin, TX 78705, USA}

\editor{}

\maketitle

\begin{abstract}
In networks with feedback, a pairwise instrumental-variable estimate can combine direct and indirect effects. We propose a Bayesian cyclic structural equation model with spike-and-slab priors on direct effects and on error covariances representing unmeasured confounding. Its joint posterior quantifies support for edges, confounding links, and feedback loops. The model has two versions. \texttt{IV-RGM} takes as known the map that lists which responses each instrument affects directly. \texttt{IV-RGM+} learns this map. A rank condition identifies the model with a known map. For an unknown map, we give conditions on instrument effects and cycle gains that identify the representation with the fewest direct instrument effects. We establish posterior consistency and directed graph recovery under additional conditions. In smaller simulated samples with strong instruments, \texttt{IV-RGM} ranked direct edges more accurately than competing methods when the map was correct. When the supplied map omitted shared targets, \texttt{IV-RGM+} ranked edges most accurately and reduced false discoveries relative to fixing the incorrect map. Weak instruments limited recovery across methods. In B cells, the analysis with a known map gave limited support to feedback among inhibitory receptor genes. Both versions are available in the R package \texttt{MR.RGM}.
\end{abstract}


\section{Introduction}\label{sec:intro}

Many systems of scientific interest are networks with feedback. Genes regulate each other in loops, price and demand influence each other in markets, and physiological variables control one another through feedback circuits. An unmeasured common cause can induce an association between two responses even when neither affects the other. When one response does affect another, the effect may pass through intermediate responses or return to its starting point along a feedback loop.

In a linear model, instrumental variables (IVs) can identify the total effect of an exposure on an outcome in the presence of unmeasured confounding \citep{angrist1996identification}. A valid instrument affects the exposure (relevance), is independent of the unmeasured confounders (independence), and affects the outcome only through the exposure (exclusion restriction). An instrument that violates the exclusion restriction is called invalid. Consider the chain $Y_1\to Y_2\to Y_3$. An instrument acting only on $Y_1$ can reveal a causal effect of $Y_1$ on $Y_3$. That effect passes entirely through $Y_2$, so drawing a direct edge from $Y_1$ to $Y_3$ would misrepresent the network. Our goal is to use instruments to learn which effects are direct in a network with feedback and unmeasured confounding. Randomized perturbations in single-cell screens provide one source of instruments. In Mendelian randomization (MR), genetic variants serve as instruments \citep{davey2003}.

\paragraph{Related work.}

In pairwise analyses, two-stage least squares \citep{theil1953repeated} and inverse-variance weighting (IVW, \citealp{burgess2013mendelian}) estimate the effect of one exposure on one outcome. Extensions allow nonlinear structural functions \citep{hartford2017deep,singh2019kernel} or allow violations of the exclusion restriction through sparsity assumptions \citep{kang2016instrumental,guo2018confidence}, MR-Egger regression \citep{bowden2015mendelian}, or median-based estimation \citep{bowden2016}. Multivariable MR (\texttt{MVMR}) can estimate the direct effects of several prespecified exposures on one outcome \citep{burgess2015multivariable}. For a pair of traits, LHC-MR jointly estimates reciprocal effects and the contribution of a latent heritable confounder from genome-wide association study (GWAS) summary statistics \citep{darrous2021simultaneous}.

Several methods use instruments to estimate a network. \texttt{MR2} jointly estimates the effects of prespecified exposures on multiple outcomes with correlated residuals \citep{zuber2023multi}. \texttt{MrDAG} learns a directed acyclic graph (DAG) within and between prespecified exposure and outcome groups \citep{zuber2025bayesian}. \citet{li2023unspecified} test directed relations in a DAG when intervention targets are unknown. \texttt{GrIVET} learns a Gaussian DAG in the presence of unmeasured confounding and identifies candidate instruments for each response \citep{chen2023discovery}. \citet{zou2026semiparametric} allow nonlinear instrument effects and potentially invalid instruments in an acyclic network. Graph-MRcML combines bidirectional pairwise MR with network deconvolution \citep{lin2023combining}. Its successor, \texttt{MR2G}, reconstructs the direct effects in a potentially cyclic network from pairwise MR estimates and GWAS summaries \citep{lin2026mr2g}. It allows correlated errors but does not estimate which residual covariances are nonzero.

Systems of simultaneous equations can be fitted jointly by three-stage least squares \citep{zellner1962three}. Penalized estimators use prespecified instruments to select direct effects in sparse systems \citep{cai2013inference,chen2018two}. In genetics, \citet{logsdon2010gene} use gene-specific perturbations to identify cyclic regulatory networks, and \citet{ni2018reciprocal} use external molecular measurements in a Bayesian reciprocal graphical model. Other methods learn acyclic phenotype networks jointly with the genetic architecture of each phenotype \citep{chaibubneto2010causal}. MRPC uses genetic variants to orient the edges of an estimated network skeleton \citep{badsha2019learning}. Findr scores pairs of genes using a genetic variant associated with one gene of the pair \citep{wang2017efficient}. For targeted perturbation screens, \citet{brown2025inspre} treat guide assignments as instruments and estimate a sparse graph by inverting the matrix of intervention effects.

Work on cyclic causal discovery includes graphical characterizations and discovery algorithms \citep{spirtes1995directed,richardson1996discovery}, foundations for models with latent variables \citep{bongers2021foundations}, and identification under structural restrictions with correlated errors \citep{drton2011global,foygel2012half}. Experimental interventions can identify linear cyclic models with latent variables \citep{hyttinen2012learning}. When shift targets are unknown, \texttt{backShift} uses changes in covariance across environments \citep{rothenhausler2015backshift}. Joint Causal Inference also treats context and intervention targets as part of the discovery problem \citep{mooij2020joint}. Bayesian approaches use longitudinal instruments to distinguish lagged effects from instantaneous feedback \citep{jin2025directed}, or non-Gaussian errors and a latent-factor model to identify disjoint cycles under additional restrictions \citep{jin2025baycausal}. Non-Gaussian errors also underlie cyclic discovery by independent component analysis \citep{lacerda2008discovering}. Here the identifying information comes from the association of measured instruments with the response means in a single sample. Once the direct effects are identified, the reduced-form covariance determines the structural error covariance.

\paragraph{Contributions.}
We propose a Bayesian model for a system of linear structural equations in which each response can affect any other. A nonzero coefficient of one response in the equation for another represents a directed edge. A nonzero covariance between two structural errors indicates unmeasured confounding between the corresponding responses. We call such a nonzero covariance a confounding link. Spike-and-slab priors on the coefficients and error covariances give posterior probabilities for individual edges, confounding links, and sets of edges such as feedback loops.

The model has two versions, \MRRGM{} and \MRRGMp{}, which differ in what is known about the instruments. The instrument--response map lists, for each instrument, the responses on which it may act directly. In \MRRGM{}, this map is known. A known map may allow an instrument to act on more than one response. \MRRGMp{} learns the map by placing a spike-and-slab prior on the direct instrument effects. This allows an instrument that also acts directly on a response other than its assigned one. Both versions assume that the instruments are independent of the unmeasured causes and structural errors. Both can be fitted using second-moment matrices computed from a single sample.%

With a known map, the conditional distribution of the responses given the instruments uniquely determines the structural parameters under the classical rank condition. In a network of $p$ responses, the instruments excluded from each equation must shift the other $p-1$ responses in $p-1$ linearly independent directions. With an unknown map, we compare representations of the same conditional distribution in which the product of the effects around each directed cycle is smaller than one in absolute value. If one representation satisfies instrument diversity, it is the unique one with the fewest nonzero direct instrument effects. Instrument diversity means that any linear combination involving two or more response equations has more nonzero instrument effects than any single equation.

Under additional conditions on the model and prior, we establish posterior consistency for the structural parameters in both versions and for the instrument--response map in \MRRGMp{}. In both versions, the posterior also recovers the directed graph when the ratio of the spike variance to the slab variance decreases at an appropriate rate as the sample size grows.

We compare both versions with ten competing methods in simulations that vary the network structure, instrument strength, and sample size, including settings in which instruments act on multiple responses. We then use genetic variants as instruments to estimate a regulatory network among ten genes in B cells from the OneK$1$K cohort \citep{onek1k2022}. Both versions are available in the R package \texttt{MR.RGM} on CRAN. The package extends the earlier implementation described by \citet{sarkar2025mrrgm} by estimating a sparse structural error covariance and allowing the full instrument--response map to be learned without prespecified exclusions.%

Section~\ref{sec:method} presents the model, the identification results, the priors, and the consistency results. Section~\ref{sec:sim} reports the simulations, and Section~\ref{sec:realdata} the B-cell analysis. Section~\ref{sec:discussion} discusses limitations. Appendices~\ref{sec:ch3:app-ident} and~\ref{sec:ch3:app-consistency} contain the proofs. The Online Appendix, whose sections are numbered OA.1--OA.6, gives the sampler, the settings used for each method in the simulations, the complete simulation results, and further details of the B-cell analysis.

\section{Method}\label{sec:method}

\subsection{Background: From Pairwise IV Estimation to a Network Model}\label{sec:overview}

Consider two responses, $Y_1$ and $Y_2$, each with its own instrument. Bidirectional IV analysis uses $X_2$ to estimate the effect of $Y_2$ on $Y_1$, and $X_1$ to estimate the effect in the other direction. A structural equation model (SEM) for this setting is
\begin{align}\label{eq:mr5}
  \left\{ \begin{array}{l}
  Y_1 = a_{12}Y_2+b_1X_1+c_1W+E_1,\\
  Y_2 = a_{21}Y_1+b_2X_2+c_2W+E_2,
  \end{array}\right.
\end{align}
where $W$ is an unmeasured confounder and $X_1,X_2,W,E_1,E_2$ are mutually independent, with $\Var(X_1),\Var(X_2)>0$. Write $r_{jh}$ for the population regression coefficient of $Y_j$ on $X_h$. When $1-a_{12}a_{21}\ne0$, solving \eqref{eq:mr5} gives $r_{12}=a_{12}b_2/(1-a_{12}a_{21})$ and $r_{22}=b_2/(1-a_{12}a_{21})$. Provided $b_2\ne0$, the feedback factor cancels in the IV ratio $r_{12}/r_{22}=a_{12}$. Similarly, if $b_1\ne0$, then $r_{21}/r_{11}=a_{21}$. \citet{hwang2024note} derive these reciprocal-effect identities for bidirectional MR.

Now introduce a third response with its own instrument under the same independence assumptions. The resulting SEM is
\begin{align}\label{eq:mr6}
  \left\{ \begin{array}{l}
  Y_1 = a_{12}Y_2+a_{13}Y_3+b_1X_1+c_1W+E_1,\\
  Y_2 = a_{21}Y_1+a_{23}Y_3+b_2X_2+c_2W+E_2,\\
  Y_3 = a_{31}Y_1+a_{32}Y_2+b_3X_3+c_3W+E_3,
  \end{array}\right.
\end{align}
where $a_{jh}$ is the direct effect of $Y_h$ on $Y_j$, represented by the edge $Y_h\to Y_j$. Suppose that the system is invertible and $r_{22}\ne0$. Since $X_2$ enters only the equation for $Y_2$, the first and third equations imply $r_{12}=a_{12}r_{22}+a_{13}r_{32}$ and $r_{32}=a_{31}r_{12}+a_{32}r_{22}$. Thus the same IV ratio is now
\[
    \frac{r_{12}}{r_{22}}
    = \frac{a_{12}+a_{13}a_{32}}{1-a_{13}a_{31}}.
\]
The numerator includes both the direct effect $a_{12}$ and the effect along $Y_2\to Y_3\to Y_1$. The denominator accounts for feedback between $Y_1$ and $Y_3$. The ratio therefore identifies the total effect of $Y_2$ on $Y_1$ \citep{pearl1995causal}, which can be nonzero even when $a_{12}=0$. To learn the directed edges, we model the responses jointly.

\subsection{Proposed Model}\label{subsec:model}\label{sec:likelihood}

Let $\mathbf{Y} = (Y_{1}, \ldots, Y_{p})^{\T}$ denote the $p$ responses and $\mathbf{X} = (X_{1}, \ldots, X_{k})^{\T}$ the $k$ instrumental variables. In the application of Section~\ref{sec:realdata}, the responses are the expression levels of genes and the instruments are genetic variants (single-nucleotide polymorphisms, SNPs). Let $\mathbf{U} = (U_{1}, \ldots, U_{l})^{\T}$ denote the $l$ measured covariates. Let $\mathbf{W} = (W_{1}, \ldots, W_{t})^{\T}\sim\text{N}_t(\mathbf{0}, \mathbf{I}_t)$ denote $t$ unmeasured confounders. The data-generating process is the SEM
\begin{align} \label{Eq1}
    \mathbf{Y} = \mathbf{A}\mathbf{Y} + \mathbf{B}\mathbf{X} + \mathbf{C}\mathbf{U} + \boldsymbol{\Lambda}\mathbf{W} + \mathbf{E},
\end{align}
where $\mathbf{A} \in \mathbb{R}^{p \times p}$ contains the direct causal effects among the responses, $\mathbf{B} \in \mathbb{R}^{p \times k}$ the instrument effects, $\mathbf{C} \in \mathbb{R}^{p \times l}$ the covariate effects, and $\boldsymbol{\Lambda} \in \mathbb{R}^{p \times t}$ the confounder loadings. The exogenous error $\mathbf{E} \sim \text{N}_p(\mathbf{0}, \boldsymbol{\Sigma})$ has a diagonal covariance matrix $\boldsymbol{\Sigma}$. We assume that there are no self-loops, $\text{diag}(\mathbf{A}) = \mathbf{0}$, that $\mathbf{I}_p-\mathbf{A}$ is invertible, and that $\mathbf{W}$, $\mathbf{E}$ and the pair $(\mathbf{X},\mathbf{U})$ are mutually independent. The instruments and covariates are allowed to be correlated with each other. In \texttt{IV-RGM}, structural zeros are imposed on $\mathbf{B}$ through a known binary instrument--response map $\boldsymbol{\Delta}=(\delta_{jh})\in\{0,1\}^{p\times k}$. If $\delta_{jh}=0$, then $b_{jh}=0$. If $\delta_{jh}=1$, the instrument $X_h$ may act directly on $Y_j$, and its effect is estimated. This encodes the exclusion restriction. In \texttt{IV-RGM+}, the map is not assumed, and every entry of $\mathbf{B}$ is subject to selection, which allows invalid instruments (Section~\ref{subsec:priors}).

An instrument that acts directly on more than one response is called shared. Shared instruments are allowed in both versions. In \MRRGM{}, their allowed targets must be specified in advance. In \MRRGMp{}, a direct effect on another response can be learned. Section~\ref{subsec:identifiability} gives the population conditions under which these instruments identify the structural parameters.

Integrating out the unmeasured confounders gives the structural error
\[
\mathbf{E}^* := \boldsymbol{\Lambda}\mathbf{W} + \mathbf{E} \sim \text{N}_p(\mathbf{0}, \boldsymbol{\Lambda}\boldsymbol{\Lambda}^{\T} + \boldsymbol{\Sigma}) = \text{N}_p(\mathbf{0}, \boldsymbol{\Sigma}^*).
\]
A nonzero off-diagonal entry $\sigma^*_{jh}=\sum_{m=1}^t\lambda_{jm}\lambda_{hm}$ indicates that $Y_j$ and $Y_h$ share at least one unmeasured cause. Contributions from several shared causes can cancel. We model $\boldsymbol{\Sigma}^*$ directly, so the number of unmeasured confounders need not be specified. With $\mathbf{E}^*$, the SEM reads $(\mathbf{I}_p - \mathbf{A})\mathbf{Y} = \mathbf{B}\mathbf{X} + \mathbf{C}\mathbf{U} + \mathbf{E}^*$, so that
\begin{align} \label{ConditionalDistn}
    \mathbf{Y} \mid \mathbf{X}, \mathbf{U} \sim \text{N}_p \left\{ (\mathbf{I}_p - \mathbf{A})^{-1} (\mathbf{B}\mathbf{X} + \mathbf{C}\mathbf{U}),\, (\mathbf{I}_p - \mathbf{A})^{-1} \boldsymbol{\Sigma}^* (\mathbf{I}_p - \mathbf{A})^{-\T} \right\},
\end{align}
which allows feedback loops through $\mathbf{A}$ (for example, if $a_{jh}\neq0$ and $a_{hj}\neq0$, then $Y_j\leftrightarrows Y_h$) and unmeasured confounding through $\boldsymbol{\Sigma}^*$. 

Write $\mathbf{y}_i$, $\mathbf{x}_i$ and $\mathbf{u}_i$ for the observations of individual $i=1,\ldots,n$. The six second-moment matrices of the data are
\begin{align*}
\mathbf{S}_{yy}&=\frac1n\sum_{i=1}^n\mathbf{y}_i\mathbf{y}_i^{\T}, &
\mathbf{S}_{yx}&=\frac1n\sum_{i=1}^n\mathbf{y}_i\mathbf{x}_i^{\T}, &
\mathbf{S}_{yu}&=\frac1n\sum_{i=1}^n\mathbf{y}_i\mathbf{u}_i^{\T},\\
\mathbf{S}_{xx}&=\frac1n\sum_{i=1}^n\mathbf{x}_i\mathbf{x}_i^{\T}, &
\mathbf{S}_{xu}&=\frac1n\sum_{i=1}^n\mathbf{x}_i\mathbf{u}_i^{\T}, &
\mathbf{S}_{uu}&=\frac1n\sum_{i=1}^n\mathbf{u}_i\mathbf{u}_i^{\T}.
\end{align*}
These six matrices suffice to evaluate the likelihood. Let $\mathbf{v}_i=(\mathbf{y}_i^\T,\mathbf{x}_i^\T,\mathbf{u}_i^\T)^\T$ and let $\mathbf{S}=n^{-1}\sum_i\mathbf{v}_i\mathbf{v}_i^\T$, whose blocks are these six matrices and their transposes. With $\mathbf{M}=\mathbf{I}_p-\mathbf{A}$ and $\mathbf{H}=[\,\mathbf{M}\mid-\mathbf{B}\mid-\mathbf{C}\,]$, the structural error for observation $i$ is $\mathbf{H}\mathbf{v}_i$. Its sample second moment is $\mathbf{H}\mathbf{S}\mathbf{H}^\T$. The change of variables from $\mathbf{y}_i$ to the structural error has Jacobian $|\det\mathbf{M}|$, so the conditional log-likelihood, up to an additive constant, is
\[
    n\log|\det\mathbf{M}|
    -\frac n2\log\det\boldsymbol{\Sigma}^*
    -\frac n2\tr\!\left\{(\boldsymbol{\Sigma}^*)^{-1}
       \mathbf{H}\mathbf{S}\mathbf{H}^{\T}\right\}.
\]
Thus a sample of known size $n$ can be analyzed using second moments alone, provided all six matrices are computed from the same individuals. Once these matrices have been computed, working with them reduces the cost of one likelihood evaluation from $\mathcal{O}\!\big(n\{p^{2}+p(k+l)\}+p^{3}\big)$ to $\mathcal{O}\!\big(p^{3}+p^{2}(k+l)+p(k+l)^{2}\big)$, which does not depend on $n$. 

\subsection{Identifiability}
\label{subsec:identifiability}

We first remove the measured covariates $\bU$ by taking the residuals from the population linear projections of $\bY$ and $\bX$ on $\bU$. Because $\bU$ enters \eqref{Eq1} linearly and is independent of the structural errors, the residuals satisfy the same structural equations with the same $\bA$, $\bB$, and $\bSigs$. Writing $\bY$ and $\bX$ for these residuals gives
\begin{equation}\label{eq:ch3:semident}
\bY = \bA\bY + \bB\bX + \bE^*,
\qquad
\bE^* \sim \Normal_p(\bzero,\bSigs),
\end{equation}
where $\diag(\bA)=\bzero$ and $\mathbf M=\bI_p-\bA$ is invertible. Throughout, $p\ge2$. Both versions of the model use the following assumption.

\begin{assumption}[Independence]\label{ass:ch3:indep}
The residualized instruments $\bX$ are independent of the unmeasured confounders $\bW$ and exogenous errors $\bE$ jointly, and $\Var(\bX)$ is invertible.
\end{assumption}
These conditions hold when the original instruments and measured covariates are jointly independent of $(\bW,\bE)$ and have a nonsingular joint covariance matrix. Under Assumption~\ref{ass:ch3:indep}, $\bY\mid\bX\sim\Normal_p(\bPi\bX,\symbf{\Omega})$, where
\begin{equation}\label{eq:ch3:Omega}
\bPi=\mathbf M^{-1}\bB,
\qquad
\symbf{\Omega}=\mathbf M^{-1}\bSigs\mathbf M^{-\T}.
\end{equation}
The distribution determines both quantities. In particular, $\bPi=\Cov(\bY,\bX)\Var(\bX)^{-1}$ and $\symbf{\Omega}=\Var(\bY)-\bPi\Var(\bX)\bPi^\T$. Thus two structural parameters $\theta=(\bA,\bB,\bSigs)$ and $\widetilde\theta=(\widetilde\bA,\widetilde\bB,\widetilde\bSigs)$ induce the same conditional distribution exactly when they have the same reduced form $(\bPi,\symbf{\Omega})$. We ask when this pair uniquely determines the structural parameters.

\paragraph{\MRRGM{}: known instrument--response map.}

Let $H_j=\{h:\delta_{jh}=0\}$ denote the instruments excluded from the equation for response $j$. The identity $\bB=(\bI_p-\bA)\bPi$ gives
\begin{equation}\label{eq:ch3:bfrompi}
b_{jh}=\pi_{jh}-\sum_{i\ne j} a_{ji}\pi_{ih}.
\end{equation}
Every known zero $b_{jh}=0$ therefore supplies one linear equation in the $p-1$ off-diagonal entries of row $j$ of $\bA$. Write $\bPi_{-j,H_j}$ for the matrix obtained by removing row $j$ and keeping the columns indexed by $H_j$, and use the same convention for $\bB$.

\begin{theorem}[Identifiability of \MRRGM{}]
\label{thm:ch3:identrgm}
Suppose Assumption~\ref{ass:ch3:indep} holds. Let a conditional distribution of $\bY$ given $\bX$ with reduced form $(\bPi,\symbf{\Omega})$ admit a representation in model~\eqref{eq:ch3:semident} satisfying the exclusions $b_{jh}=0$ for every $j$ and $h\in H_j$. If $\rank(\bPi_{-j,H_j})=p-1$ for every $j$, then the representation $(\bA,\bB,\bSigs)$ satisfying the exclusions is unique. Equivalently, the rank condition can be written as $\rank(\bB_{-j,H_j})=p-1$ for every $j$.
\end{theorem}

The classical rank condition for simultaneous equations \citep{koopmans1950} requires the instruments excluded from each response to shift the other $p-1$ responses in $p-1$ linearly independent directions. This requires at least $p-1$ exclusions per response. Under the rank condition, the equations in \eqref{eq:ch3:bfrompi} for $h\in H_j$ have a unique solution for each row of $\bA$. Once $\bA$ is determined, $\bB=\mathbf M\bPi$ and $\bSigs=\mathbf M\symbf{\Omega}\mathbf M^\T$ follow. The rank condition in $\bPi$ depends only on the observable distribution. Its equivalent form in $\bB$ is easier to interpret. For instance, in the design of Section~\ref{sec:realdata}, each instrument is assigned to a single response. Under such a map, the rank condition holds exactly when each response has at least one instrument with a nonzero direct effect on it (Remark~\ref{rem:ch3:ciseqtl}). If an instrument acts directly on only one response, we call it an \emph{own instrument} of that response.

\paragraph{\MRRGMp{}: unknown instrument--response map.}

Without known exclusions, a conditional distribution can admit many structural representations. To see how they are related, consider two parameters $\theta$ and $\widetilde\theta$ with the same reduced form, and write $\widetilde{\mathbf M}=\bI_p-\widetilde{\bA}$. The invertible matrix $\symbf{\Xi}=\widetilde{\mathbf M}\mathbf M^{-1}$ satisfies
\begin{equation}\label{eq:ch3:xiform}
\widetilde{\mathbf M}=\symbf{\Xi}\mathbf M,
\qquad
\widetilde{\bB}=\symbf{\Xi}\bB,
\qquad
\widetilde{\bSigs}=\symbf{\Xi}\bSigs\symbf{\Xi}^{\T}.
\end{equation}
Each row of $\widetilde{\bB}$ is a linear combination of the rows of $\bB$. When combining two or more rows introduces additional direct instrument effects, the sparsity of $\bB$ can distinguish the representations. This use of sparse directions in the row space is related to identification of a dictionary from sparse coefficient vectors \citep{spielman2012exact}. Here we impose a deterministic condition on the instrument effects. Let $\|\cdot\|_0$ count nonzero entries and define
\[
s(\bB)=\max_j\|\bB_{j,:}\|_0,
\qquad
c(\bB)=\min_{\mathbf w\in\R^p:\,\|\mathbf w\|_0\ge2}\|\mathbf w^{\T}\bB\|_0 .
\]
Thus $s(\bB)$ counts the largest number of instruments acting directly on any one response. The quantity $c(\bB)$ counts the fewest that remain after taking a linear combination of two or more response equations, allowing for cancellation.

\begin{assumption}[Instrument diversity]
\label{ass:ch3:diversity}
The instrument-effect matrix $\bB$ satisfies $s(\bB)<c(\bB)$.
\end{assumption}
Under this condition, every row of $\bB$ is nonzero, the rows are linearly independent, and the support of $\bB$ gives an instrument--response map satisfying the rank condition of Theorem~\ref{thm:ch3:identrgm} (Lemma~\ref{lem:ch3:rank}). Own instruments illustrate how diversity can arise even when some instruments are shared. If response $i$ has $m_i\ge1$ own instruments with nonzero effects, their effects cannot cancel against effects in another row. Hence $c(\bB)$ is at least the sum of the two smallest $m_i$. For example, with three own instruments per response and at most two additional shared instruments per response, $s(\bB)\le5$ and $c(\bB)\ge6$, as in the main simulation study.

Even if the rows of $\bB$ are distinguished by their instrument effects, the same equations can sometimes be reassigned to different responses. We use the cycle-product condition of \citet{rothenhausler2015backshift} to distinguish these assignments. For a cycle $j_1\to j_2\to\cdots\to j_r\to j_1$ with $r\ge2$ distinct responses, its \emph{gain} is the product $a_{j_2j_1}a_{j_3j_2}\cdots a_{j_1j_r}$.

\begin{assumption}[Cycle gains]
\label{ass:ch3:cycle}
Every directed cycle in the graph of $\bA$ has gain smaller than one in absolute value.
\end{assumption}
This condition holds for acyclic graphs. It also holds if all direct effects have absolute value below one, or if the matrix $|\bA|$ of entrywise absolute values has spectral radius below one. The spectral radius of $\bA$ itself can be below one even when some cycle violates the bound, and it can exceed one when every cycle satisfies the bound (Remark~\ref{rem:ch3:tightness} in Appendix~\ref{sec:ch3:app-ident}).

\begin{theorem}[Identifiability of \MRRGMp{}]
\label{thm:ch3:ident}
Suppose Assumption~\ref{ass:ch3:indep} holds. Let $\theta=(\bA,\bB,\bSigs)$ and $\widetilde\theta=(\widetilde\bA,\widetilde\bB,\widetilde\bSigs)$ be two parameter values of model~\eqref{eq:ch3:semident} that induce the same conditional distribution of $\bY$ given $\bX$. Suppose that every directed cycle in either graph has gain smaller than one in absolute value (Assumption~\ref{ass:ch3:cycle}).
If $\bB$ satisfies instrument diversity (Assumption~\ref{ass:ch3:diversity}), then the following hold.
\begin{enumerate}
\item[(i)] If every row of $\widetilde\bB$ has fewer than $c(\bB)$ nonzero entries, then $\widetilde\theta=\theta$.
\item[(ii)] $\|\widetilde\bB\|_0\ge\|\bB\|_0$, with equality if and only if $\widetilde\theta=\theta$. Here $\|\bB\|_0$ denotes the total number of nonzero entries in $\bB$.
\end{enumerate}
\end{theorem}

Part~(i) gives uniqueness among representations with fewer than $c(\bB)$ direct instrument effects per response. Part~(ii) allows competing representations with any number of effects per response. Among representations of the same conditional distribution satisfying the cycle-gain bound, $\theta$ is the unique one with the fewest direct instrument effects in total.

The two conditions play different roles. By \eqref{eq:ch3:xiform}, any row of $\widetilde\bB$ that combines two or more rows of $\bB$ has at least $c(\bB)>s(\bB)$ nonzero entries. Such a row violates the bound in part~(i). It also increases the total number of effects in part~(ii), because the rows of $\widetilde\bB$ that are multiples of single rows of $\bB$ must correspond to distinct rows of $\bB$ by invertibility of $\symbf{\Xi}$. Once row combinations are ruled out, the equations agree up to rescaling and reassignment. For example, exchanging the two equations along a feedback loop $Y_i\leftrightarrows Y_j$ and rescaling each to give its response coefficient one preserves the reduced form and the number of instrument effects. The cycle gain changes from $g=a_{ij}a_{ji}$ to $1/g$. More generally, any reassignment other than the identity reverses a directed cycle in this way. If $|g|<1$, the reassigned equations violate the cycle-gain bound. With each equation assigned to its original response, the unit diagonal of $\mathbf M$ fixes its scale.

Diversity is sufficient for both conclusions. For part~(ii), it is enough that each linear combination involving two or more rows has more nonzero entries than each of its contributing rows. This weaker condition holds whenever every response has an instrument with a nonzero direct effect and no instrument is shared, even if the responses differ greatly in their numbers of instruments. Appendix~\ref{sec:ch3:app-ident} gives the proofs and further examples.

\begin{rem}[Weak instruments]\label{rem:ch3:weakident}
Theorem~\ref{thm:ch3:ident} relies on exact zeros in $\bB$. Adding a multiple of one structural equation to another can change $\bA$ and introduce new instrument effects in $\bB$ while preserving the observational distribution. When those effects are small, setting them to zero can yield a sparse model with different causal effects and nearly the same observable distribution. The data may then provide little information about $\bA$, and we call such a fit \emph{weakly identified}. In \MRRGM{}, the known instrument--response map rules out changes that violate its exclusions. Weak instruments can still limit precision.
\end{rem}

\subsection{Priors and Posterior Summaries}\label{subsec:priors}

The priors serve both parameter estimation and structure learning. Spike-and-slab priors \citep{george1993variable,ishwaran2005spike} on the causal graph ($\mathbf{A}$) and on the confounding structure ($\boldsymbol{\Sigma}^*$) provide inclusion indicators whose posterior probabilities quantify the support for each edge and link.

Each off-diagonal entry $a_{jh}$ of $\mathbf{A}$ receives a spike-and-slab prior. The binary indicator $\gamma_{jh}$ assigns the direct effect of $Y_h$ on $Y_j$ to a narrow spike ($\gamma_{jh}=0$) or a wider slab ($\gamma_{jh}=1$), with
\begin{align*}
    &a_{jh} \sim \gamma_{jh} \cdot \text{N}(0, \tau_{jh}) + (1 - \gamma_{jh}) \cdot \text{N}(0, \nu_1 \cdot \tau_{jh}),\\
   & \gamma_{jh} \sim \text{Bernoulli}(\rho_{jh}),~~
    \rho_{jh} \sim \text{Beta}(a_\rho, b_\rho),~~
    \sqrt{\tau_{jh}} \sim \mathcal{C}^+(0, 1),
\end{align*}
where $\mathcal{C}^+(0, 1)$ is the standard half-Cauchy distribution and $\nu_1 \ll 1$ is the ratio of the spike variance to the slab variance. We call $\nu_1$ the \emph{variance ratio}. The spike concentrates $a_{jh}$ near zero, and the slab allows larger effects. Both components are continuous. The half-Cauchy prior admits an inverse-gamma representation with an auxiliary variable $\kappa_{jh}$, $\tau_{jh} \mid \kappa_{jh} \sim \text{IG}(1/2, 1/\kappa_{jh})$, $\kappa_{jh} \sim \text{IG}(1/2, 1)$ \citep{makalic2016simple}, which gives closed-form Gibbs updates of $\tau_{jh}$.

For $\mathbf{B}$, \texttt{IV-RGM} fixes the structural zeros given by $\boldsymbol{\Delta}$ and places $b_{jh} \sim \text{N}(0, \eta_{jh})$, $\sqrt{\eta_{jh}} \sim \mathcal{C}^+(0, 1)$, on the free entries. \texttt{IV-RGM+} places a spike-and-slab prior on every entry,
\begin{align*}
    &b_{jh} \sim \phi_{jh} \cdot \text{N}(0, \eta_{jh}) + (1 - \phi_{jh}) \cdot \delta_0, \\
    &\phi_{jh} \sim \text{Bernoulli}(\psi_{jh}),~~
    \psi_{jh} \sim \text{Beta}(a_\psi, b_\psi),~~
    \sqrt{\eta_{jh}} \sim \mathcal{C}^+(0, 1),
\end{align*}
where $\delta_0$ is the point mass at zero and $\phi_{jh}$ indicates whether $X_h$ acts on $Y_j$. An instrument effect that is not selected is exactly zero. The identification result of Section~\ref{subsec:identifiability} and the consistency result of Section~\ref{subsec:asymptotics} for \texttt{IV-RGM+} both require such exact zeros. The prior enables the selection of the instrument--response map from the data, and an instrument may act on more than one response.

The covariate effects receive the conjugate matrix-normal prior $\mathbf{C} \sim \text{MN}_{p \times l}(\mathbf{0},\, \boldsymbol{\Sigma}^*,\, \tau_C \mathbf{I}_l)$, with row covariance $\boldsymbol{\Sigma}^*$ and column covariance $\tau_C \mathbf{I}_l$. We introduce a binary indicator matrix $\mathbf{Z} = (z_{jh}) \in \{0,1\}^{p \times p}$ whose off-diagonal indicator $z_{jh}=z_{hj}$ assigns $\sigma^*_{jh}$ to a narrow spike or a wider slab. We use the stochastic search structure learning (SSSL) prior of \citet{wang2015},
\begin{align*}
   & \sigma^*_{jh}  \sim z_{jh} \cdot \text{N}(0, \omega_1^2) + (1 - z_{jh}) \cdot \text{N}(0, \omega_2^2), \quad j < h, \\
   & z_{jh} \sim \text{Bernoulli}(\varpi),~~ \sigma^*_{jj}\sim \text{Exp}\left(\frac{\lambda}{2}\right).
\end{align*}
Here $\omega_1^2>\omega_2^2$, and $\text{Exp}(\lambda/2)$ is the exponential distribution with rate $\lambda/2$. The joint prior of $(\boldsymbol{\Sigma}^*,\mathbf{Z})$ is the product of these densities, restricted to positive definite matrices and normalized. 

\paragraph{Hyperparameters and computation.} Unless stated otherwise, we use $a_\rho=3$, $b_\rho=1$, $a_\psi=b_\psi=0.5$, $\nu_1=10^{-3}$, $\tau_C=2$, $\omega_1^2=5$, $\omega_2^2=0.005$, $\varpi=0.5$ and $\lambda=1$. The prior probability of assigning a causal effect to the slab is then $a_\rho/(a_\rho+b_\rho)=0.75$. For a confounding link the mixture weight is $\varpi=0.5$ before the restriction to positive definite matrices. The prior probability of an instrument effect in \texttt{IV-RGM+} is $a_\psi/(a_\psi+b_\psi)=0.5$. The SSSL variances are fixed, so every column of $\mathbf{Y}$, $\mathbf{X}$ and $\mathbf{U}$ is centered and scaled before fitting. The direct effect $a_{jh}$ is therefore expressed in standard deviations of $Y_j$ per standard deviation of $Y_h$.

Posterior draws are obtained by a Metropolis-within-Gibbs sampler with $50{,}000$ iterations, a burn-in of $10{,}000$ and thinning $10$. Only the entries of $\mathbf{A}$ require Metropolis--Hastings steps, with proposal variances tuned during the burn-in \citep{roberts2009examples}. The chain starts at $\mathbf{A}=\mathbf{0}$, and in \texttt{IV-RGM+} also at $\mathbf{B}=\mathbf{0}$ with no instrument selected. The remaining updates use full conditional distributions. Section~OA.1 gives the sampler and the details of the following diagnostics.

\paragraph{Diagnostics.} For \MRRGM{}, we check instrument strength by regressing each response on its allowed instruments without adjustment for the measured covariates. A first-stage $F<10$ flags weak instruments \citep{staiger1997instrumental}. This statistic describes how strongly the assigned instruments predict that response. The rank condition in Section~\ref{subsec:identifiability} concerns the joint effects of the instruments excluded from each equation. Conditional first-stage tests \citep{sanderson2016weak} check this joint condition in IV regressions with several endogenous regressors. They could complement the marginal $F$ check. \MRRGMp{} has no prespecified map for the first-stage check.

Two further warnings concern the fitted structural equations. The first, the error-variance warning, flags a fitted structural error variance greater than five times the observed variance of the corresponding response, an empirical sign of a potentially weakly identified fit. The second flags a spectral radius of at least one for the posterior mean $\widehat{\mathbf A}$. At such values the series $\sum_{m\ge0}\widehat{\mathbf A}^{\,m}$ diverges, so iterating the estimated feedback equations need not converge \citep{lacerda2008discovering}. A unique simultaneous solution requires only that $\mathbf{I}_p-\mathbf{A}$ be invertible.

We also check the mixing of the chain. This check needs at least $200$ retained draws. A warning is issued if the posterior inclusion probabilities for any directed edge in the two halves of the chain differ by more than $0.2$ and by more than four Monte Carlo standard errors of the difference. A separate warning is issued when the acceptance rate for the updates of $\mathbf A$, pooled over its off-diagonal entries after burn-in, is below $10\%$.

\paragraph{Posterior summaries.} We report the posterior means of $\mathbf{A}$, $\mathbf{B}$, $\mathbf{C}$ and $\boldsymbol{\Sigma}^*$ as point estimates. The posterior inclusion probability (PIP) of a directed edge or a confounding link is the posterior probability that its coefficient or covariance belongs to the slab. For a direct instrument effect in \MRRGMp{}, the PIP is the posterior probability that the coefficient is nonzero. An edge is selected if its PIP exceeds $0.5$ (the median probability model, \citealp{barbieri2004optimal}). 
A confounding link is selected by the same rule. Under the default SSSL prior, the conditional slab probability for a given $\sigma^*_{jh}$ exceeds $0.5$ when $|\sigma^*_{jh}|$ exceeds approximately $0.186$, where the two weighted normal densities cross. This corresponds to an absolute structural error correlation near $0.19$ when both error variances are near one. The reported PIP averages this conditional probability over the posterior of $\sigma^*_{jh}$.

We also report the posterior probability of network motifs, such as $Y_1\to Y_2\to Y_3$, which is the posterior probability that all edges of the motif are present.

\subsection{Posterior Consistency}\label{subsec:asymptotics}

We now consider repeated sampling from the model with a fixed structural parameter $\theta_0=(\bA_0,\bB_0,\bSig_0^*)$ and fixed dimensions $p$ and $k$. Write $\mathcal D_n=\{(\bY_i,\bX_i)\}_{i=1}^n$ for the sample from \eqref{eq:ch3:semident}, $\mathcal P_n(\cdot\mid\mathcal D_n)$ for the posterior, and $\Prob_{\theta_0}$ for the sampling distribution. We use $\|\theta-\theta_0\|_\infty$ for the largest absolute difference between corresponding entries. For a vector, $\|\cdot\|_2$ is the Euclidean norm. For a matrix, $\|\cdot\|_2$ is the spectral norm, $\|\cdot\|_F$ is the Frobenius norm and $\|\cdot\|_{\max}$ is the largest absolute entry. For symmetric matrices, $\mathbf M_1\preccurlyeq\mathbf M_2$ means that $\mathbf M_2-\mathbf M_1$ is positive semidefinite. Let $a^0_{jh}$ and $b^0_{jh}$ denote the entries of $\bA_0$ and $\bB_0$. The directed graph has indicators $\gamma^0_{jh}=\mathbb I(a^0_{jh}\ne0)$, and the direct instrument effects have indicators $\phi^0_{jh}=\mathbb I(b^0_{jh}\ne0)$. With known instrument exclusions, the rank condition identifies $\theta_0$ within the specified map. With an unknown map, several maps can give the same conditional distribution. We give conditions under which the posterior selects the map of $\bB_0$ and concentrates on $\theta_0$ within that map.

The results use the priors of Section~\ref{subsec:priors} with the following restrictions. We write $\nu_{1,n}$ for the ratio of the spike variance to the slab variance of a directed edge, allowing this ratio to depend on sample size.

\begin{assumption}[Instrument design]\label{ass:ch3:design}
The observations are independent and identically distributed. The instruments $\bX$ are centered, are independent of $\bE^*$, and have a positive definite covariance matrix $\bSig_{XX}$. For some $c_X>0$, $\E\{\exp(c_X\|\bX\|_2^2)\}<\infty$.
\end{assumption}

\begin{assumption}[Compact parameter set]\label{ass:ch3:compact}
The prior is restricted to the set $\Theta_M$ of structural parameters satisfying
\[
\|(\bI_p-\bA)^{-1}\|_2\le C_M,\quad
\|\bA\|_{\max}\le C_A,\quad \|\bB\|_{\max}\le C_B,\quad
\underline c_\Sigma\bI_p\preccurlyeq\bSigs\preccurlyeq\overline c_\Sigma\bI_p,
\]
where the constants are finite, $\underline c_\Sigma>0$, and $\|\cdot\|_{\max}$ denotes the largest absolute entry. The parameter $\theta_0$ lies in the interior of $\Theta_M$ relative to matrices with zero diagonal in $\bA$ and symmetric $\bSigs$.
\end{assumption}

\begin{assumption}[Error covariance prior]\label{ass:ch3:sigmaprior}
The prior on $\bSigs$ is supported on the positive definite cone and has a density, in the distinct covariance entries, bounded below by a positive constant in a neighborhood of $\bSig_0^*$.
\end{assumption}

\begin{assumption}[Coefficient priors]\label{ass:ch3:priorpos}
The beta shape parameters $a_\rho,b_\rho$ and, for \MRRGMp{}, $a_\psi,b_\psi$ are positive. The half-Cauchy priors on the slab scales $\sqrt{\tau_{jh}}$ and $\sqrt{\eta_{jh}}$ are truncated to a fixed interval $[\underline s,\overline s]\subset(0,\infty)$ on which their densities are positive and continuous. The variance ratio satisfies $0<\nu_{1,n}<1$.
\end{assumption}

\paragraph{\MRRGM{} with a known map.} The first result concerns estimation of the structural parameters. The posterior can estimate a zero edge coefficient with arbitrarily small error while still assigning positive probability to both components of its continuous spike-and-slab prior. Recovering the corresponding indicator requires a further condition on the spike variance.

\begin{theorem}[Posterior consistency of \MRRGM{}]\label{thm:ch3:thetacons}
Suppose that $\bB_0$ satisfies the exclusions of $\symbf{\Delta}$, that the rank condition of Theorem~\ref{thm:ch3:identrgm} holds for $(\bPi_0,\symbf{\Delta})$, and that Assumptions~\ref{ass:ch3:design}--\ref{ass:ch3:priorpos} hold. Then, for any $\varepsilon>0$,
\begin{equation*}
\mathcal{P}_n\!\left(\|\theta-\theta_0\|_\infty>\varepsilon\;\Big|\;\mathcal{D}_n\right)\to 0
\quad\text{in }\Prob_{\theta_0}\text{-probability.}
\end{equation*}
\end{theorem}

\begin{assumption}[Rate of the variance ratio]\label{ass:ch3:nurate}
As $n\to\infty$, $\nu_{1,n}\to0$ and $n\nu_{1,n}/\log n\to\infty$.
\end{assumption}

\begin{theorem}[Graph recovery of \MRRGM{}]\label{thm:ch3:indicators}
Under the conditions of Theorem~\ref{thm:ch3:thetacons} and Assumption~\ref{ass:ch3:nurate}, the posterior recovers the directed graph, in the sense that
\begin{equation*}
\mathcal{P}_n\!\left(\{\gamma_{jh}\}_{j\neq h}=\{\gamma^0_{jh}\}_{j\neq h}\;\Big|\;\mathcal{D}_n\right)\to 1
\quad\text{in }\Prob_{\theta_0}\text{-probability.}
\end{equation*}
\end{theorem}

Theorem~\ref{thm:ch3:indicators} extends selection consistency for spike-and-slab priors in regression \citep{ishwaran2005spike,narisetty2014bayesian,castillo2012needles} to this structural model. The structural parameter contracts at rate $n^{-1/2}\log n$ in the maximum-entry norm (Lemma~\ref{lem:ch3:localrate}). Under Assumption~\ref{ass:ch3:nurate}, the posterior inclusion probability of each directed edge tends to one if its coefficient is nonzero and to zero otherwise.

\paragraph{\MRRGMp{} with an unknown map.} We use a uniform bound on the cycle gains over the support of the prior and a condition on the sets of instruments that can enter each structural equation.

\begin{assumption}[Uniform bound on cycle gains]\label{ass:ch3:compactplus}
The prior is restricted to $\Theta_M^+\subseteq\Theta_M$, the set of parameters for which every directed cycle has absolute gain at most $c_g$, for some $0<c_g<1$. Every directed cycle of $\bA_0$ has absolute gain strictly smaller than $c_g$. The prior density of $\bSigs$ is bounded above over the covariance matrices in $\Theta_M$.
\end{assumption}

Let $S_{0i}=\{h:b^0_{ih}\ne0\}$ be the set of instruments that act directly on response $i$. For a vector $\mathbf v$, let $\supp(\mathbf v)$ denote the set of positions of its nonzero entries. For any set of instruments $T\subseteq\{1,\ldots,k\}$, define
\[
\mathcal V(T)=\{\mathbf w\in\R^p:\supp(\mathbf w^\T\bB_0)\subseteq T\},\qquad
d(T)=\dim\mathcal V(T).
\]
The vectors in $\mathcal V(T)$ are the weights of linear combinations of structural equations whose direct instrument effects are confined to $T$. We say that response $i$ enters $\mathcal V(T)$ if some combination in this space gives its equation a nonzero coefficient.

\begin{assumption}[Minimality of the instrument--response map]\label{ass:ch3:overid}
For every $T$ and every response $i$ that enters $\mathcal V(T)$, $|T|-d(T)\ge |S_{0i}|-1$, with equality only when $T=S_{0i}$.
\end{assumption}

Since $\mathbf e_i\in\mathcal V(S_{0i})$, response $i$ enters $\mathcal V(S_{0i})$, and the inequality with $T=S_{0i}$ forces $d(S_{0i})=1$. For any other $T$ such that response $i$ enters $\mathcal V(T)$, the inequality is strict. This accounts for the fact that several combinations of structural equations may give the same distribution. Among the maps that reproduce that distribution, the parameters of the map of $\bB_0$ then trace out a reduced-form set of the smallest dimension (Lemma~\ref{lem:ch3:plusrank}). One sufficient design gives each response at least two instruments of its own. If the smallest number of own instruments is $m$, each response may also share fewer than $m$ instruments with other responses. The instrument assignments in the simulations of Section~\ref{sec:sim} satisfy this sufficient condition (Example~\ref{ex:ch3:overid} in Appendix~\ref{sec:ch3:app-consistency}). At the other extreme, if any response has just one direct instrument, the minimality assumption fails.

\begin{theorem}[Posterior consistency of \MRRGMp{}]\label{thm:ch3:pluscons}
Consider \MRRGMp{} with the point-mass spike on the instrument effects and suppose that Assumptions~\ref{ass:ch3:design}--\ref{ass:ch3:priorpos}, \ref{ass:ch3:compactplus} and~\ref{ass:ch3:overid} hold. Suppose also that $\nu_{1,n}\ge n^{-\alpha}$ for all sufficiently large $n$ and some $0<\alpha<1/\{p(p-1)\}$. Then, in $\Prob_{\theta_0}$-probability,
\begin{enumerate}
\item[(i)] $\mathcal{P}_n\big(\{\phi_{jh}\}=\{\phi^0_{jh}\}\mid\mathcal{D}_n\big)\to1$, that is, the instrument--response map is recovered,
\item[(ii)] $\mathcal{P}_n\big(\|\theta-\theta_0\|_\infty>\varepsilon\mid\mathcal{D}_n\big)\to0$ for any $\varepsilon>0$, and
\item[(iii)] under Assumption~\ref{ass:ch3:nurate}, $\mathcal{P}_n\big(\{\gamma_{jh}\}_{j\neq h}=\{\gamma^0_{jh}\}_{j\neq h}\mid\mathcal{D}_n\big)\to1$.
\end{enumerate}
\end{theorem}

The point mass at zero allows the posterior to select the instrument--response map even with a fixed variance ratio for the directed effects. Under these restricted priors, the default $\nu_1=10^{-3}$ is compatible with parameter consistency in both versions and map recovery in \MRRGMp{}. A sequence such as $\nu_{1,n}=1/\log n$ also gives recovery of the directed graph. The parameter results cover the entries of $\bSigs$. Selection of its nonzero off-diagonal entries by the confounding indicators is a separate question.

The implementation uses untruncated slab scales and does not impose the compact parameter bounds or the cycle-gain bound. For \MRRGMp{}, the cycle-gain bound can distinguish two equally sparse representations of the same distribution. It therefore matters for interpreting the consistency result. Appendix~\ref{sec:ch3:app-consistency} gives the proofs.

\section{Simulation Studies}\label{sec:sim}

We compare methods for estimating direct effects in networks with feedback and unmeasured confounding. In \textbf{Case~1} (scale-free), each edge of a Barab\'asi--Albert tree is made reciprocal \citep{barabasi1999emergence}. In \textbf{Case~2} (small-world), the edges of a Watts--Strogatz network are oriented at random \citep{watts1998collective}. In \textbf{Case~3} (shared instruments), we add a shared instrument to each pair $(1,2),(3,4),\ldots$ of the Case~2 network, with one additional pair $(1,5)$ when $p=5$. Each shared instrument acts directly on both responses, with the same effect as the own instruments, but the supplied map assigns it only to the first. Under the supplied map, the shared instrument therefore violates the exclusion restriction, because it also acts directly on the second response. The figures and the Online Appendix label Case~3 ``invalid instruments'' and Cases~1 and~2 ``valid instruments''.

Each response has three instruments of its own, each with direct effect $1$. Nonzero effects between responses are $\pm0.1$. The instruments and five unmeasured confounders are independent standard normal variables. Half the confounder loadings are set to zero at random and the others are $\pm1$. The independent Gaussian errors have variance $9$. For each case and $p\in\{5,10\}$, we generate $100$ networks and retain each network and its loadings across sample sizes $n\in\{500,1{,}000,10{,}000,30{,}000\}$. Within each replicate, the methods use the same observations or summary statistics calculated from them. We evaluate the direct effects on the standardized scale. The main-study results reported here are for $p=10$. Section~OA.2 gives the results for $p=5$.

\paragraph{Methods compared.}
\MRRGM{} uses the supplied map, which is correct in Cases~1 and~2. \MRRGMp{} considers every instrument for every response. In all three cases, each response has three own instruments and at most two shared instruments, so instrument diversity holds. The direct effects of $\pm0.1$ keep every cycle gain below one in absolute value (Example~\ref{ex:ch3:satisfied} in Appendix~\ref{sec:ch3:app-ident}). We also fit \MRRGM{} with diagonal error covariance (\MRRGMnc{}) to assess the contribution of modeling confounding.

Of the ten competing methods, \texttt{MR2G} and \texttt{GrIVET} fit the whole network. \texttt{MR2G} receives the supplied map. \texttt{GrIVET} selects its instruments. \texttt{MVMR} and \texttt{MrDAG} take each response in turn as the outcome. \texttt{MR2} is included only in the separate exposure--outcome experiment. The five pairwise estimators are IVW, the simple and weighted medians \citep{R-MendelianRandomization}, two-stage residual inclusion in \texttt{OneSampleMR} \citep{terza2008two,R-OneSampleMR}, and Bayesian IVW in \texttt{mrbayes} \citep{R-MrBayes}. The figures label the first three \texttt{MR-IVW}, \texttt{MR-SimpleMedian}, and \texttt{MR-WeightedMedian}. We apply each to every ordered response pair using the instruments assigned to its exposure. With valid instruments, the pairwise estimators target total effects, so an indirect path can contribute to their error when evaluated against direct effects. A shared instrument that violates the pairwise exclusion restriction can also bias the estimated total effect.

\paragraph{Evaluation.}
The area under the receiver operating characteristic curve (AUC) measures how well edge scores rank the $p(p-1)$ ordered pairs against the directed graph. We rank by decreasing PIP, increasing $p$-value (two-sided posterior tail probability for \texttt{mrbayes}), or decreasing absolute effect for \texttt{GrIVET}. We select edges with PIP above $0.5$, an unadjusted $p$-value or tail probability below $0.05$, or a nonzero \texttt{GrIVET} estimate. The true positive rate (TPR) is the proportion of edges recovered. The false discovery rate (FDR) is the proportion of selected edges that are false, averaged over replicates and assigned zero for an empty selection. Mean absolute deviation (MAD) measures estimation error over all off-diagonal entries of $\mathbf A$ and separately on edges and non-edges. The MAD of the all-zero estimate provides a reference for sparse graphs.

For confounding, the AUC assesses whether the PIPs rank pairs with above-average absolute structural error covariance ahead of the remaining pairs. The average is over unordered response pairs on the generating scale within each replicate. For instrument selection, the AUC compares the PIPs from \MRRGMp{} with the nonzero entries of $\mathbf B$. We summarize each method over its successful fits in settings with at least ten successes out of $100$ replicates. A fit fails when the method returns no result, for example because it stops with an error or does not finish within $48$ hours (Section~OA.2). Sections~OA.2--OA.4 give implementation details, failure counts and full results.

\subsection{Main Study}\label{sec:sim_main}

\begin{figure}[!htb]
  \centering
  \includegraphics[width=\textwidth,height=0.61\textheight,keepaspectratio]{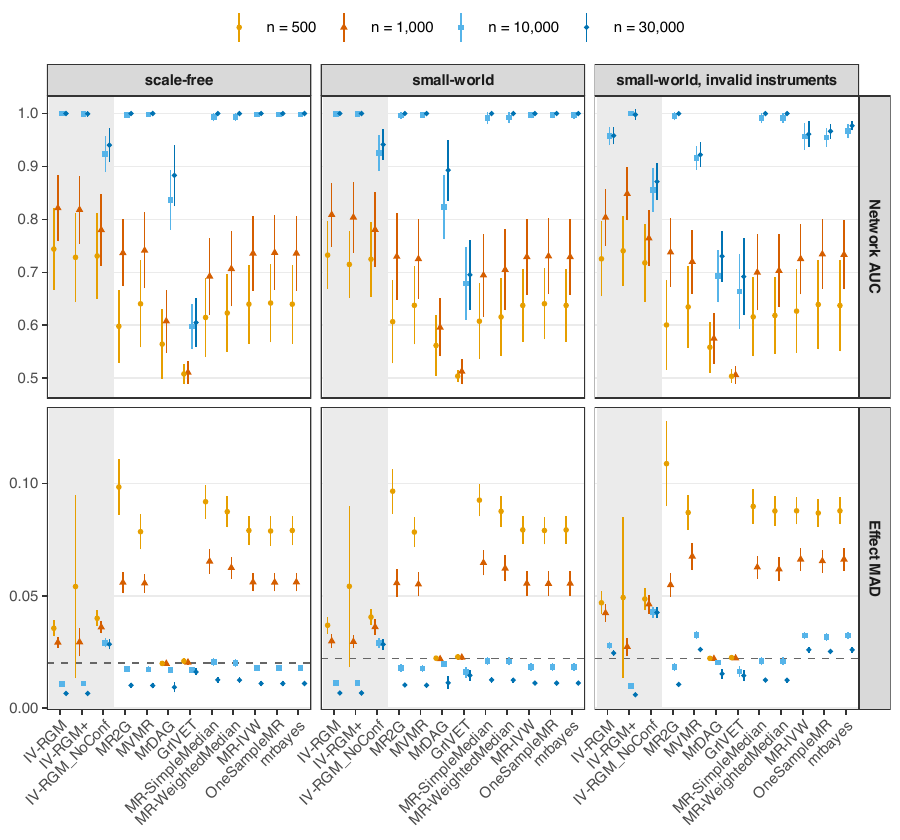}
  \caption{Network recovery and direct-effect estimation for $p=10$. Columns show Case~1 (scale-free), Case~2 (small-world), and Case~3 (shared instruments, labeled invalid instruments in the panel title). Within each method, points give means at the four sample sizes and vertical lines extend one standard deviation in each direction across successful fits. Higher network AUC (top) and lower effect MAD (bottom) indicate better performance. Shading identifies \texttt{IV-RGM}, \texttt{IV-RGM+} and \texttt{IV-RGM\_NoConf}, and the dashed line gives the MAD of the all-zero estimate.}
  \label{fig:main_summary}
\end{figure}

\paragraph{Cases 1 and 2.}
When the supplied instrument--response map is correct, \MRRGM{} ranks the edges more accurately than the competing methods at the two smaller sample sizes (Figure~\ref{fig:main_summary}). Its network AUC is $0.73$--$0.74$ at $n=500$ and $0.81$--$0.82$ at $n=1{,}000$, compared with at most $0.64$ and $0.74$, respectively, among the competing methods. The AUC of \MRRGMp{}, which learns the map, is within $0.02$ of that of \MRRGM{} in these settings. At $n=10{,}000$, both \MRRGM{} and \texttt{MR2G} recover about $98\%$ of the edges in each design. The corresponding false discovery rates are $0.017$ for \MRRGM{} and $0.10$--$0.11$ for \texttt{MR2G}.

With a sparse network, errors on the few direct effects can be hidden by an average over all ordered pairs. In Case~2 at $n=1{,}000$, for example, \texttt{MrDAG} has the same overall MAD as the all-zero estimate ($0.022$), compared with $0.030$ for \MRRGM{}. On the true edges, however, MAD is $0.098$ for \texttt{MrDAG} and $0.051$ for \MRRGM{}. On absent edges, their mean absolute estimates are below $0.001$ and $0.024$, respectively.

\paragraph{Case 3.}
With shared instruments, \MRRGMp{} has the highest mean network AUC at both smaller sample sizes (Figure~\ref{fig:main_summary}). At $n=500$, its AUC is $0.741$, compared with $0.639$ for the best-performing competing method, \texttt{OneSampleMR}. At $n=1{,}000$, its AUC is $0.849$, compared with $0.738$ for the leading competing method, \texttt{MR2G}. At this sample size, \MRRGM{} with the misspecified map has an AUC of $0.804$. Learning the map reduces the FDR from $0.43$ to $0.30$, with a small decrease in TPR from $0.60$ to $0.57$.

At the two larger sample sizes, \MRRGMp{}, \texttt{MR2G} and the simple and weighted median estimators all rank the edges nearly perfectly (AUC $\ge 0.99$). At $n=10{,}000$, \MRRGMp{} and \texttt{MR2G} recover $99\%$ and $97\%$ of the true edges. The FDR is $0.009$ for \MRRGMp{}, $0.117$ for \texttt{MR2G} and about $0.14$ for either median estimator. The FDR for \MRRGM{} with the misspecified map is $0.152$. Learning the map also reduces the overall effect MAD from $0.028$ to $0.010$. The lowest MAD among the competing methods is $0.016$ for \texttt{GrIVET}. Across all three main-study designs and all sample sizes, \MRRGMp{} ranks instrument--response pairs with AUC at least $0.997$.

For confounding, the AUC of both versions rises from $0.65$--$0.67$ at $n=500$ to above $0.99$ at $n=30{,}000$ in Cases~1 and~2. In Case~3 at $n\ge 10{,}000$ it is $0.96$--$0.99$ for \MRRGMp{} and $0.81$--$0.84$ for \MRRGM{}.

\subsection{Instrument Strength}\label{sec:sim_strength}

\begin{figure}[!htb]
  \centering
  \includegraphics[width=\textwidth,height=0.61\textheight,keepaspectratio]{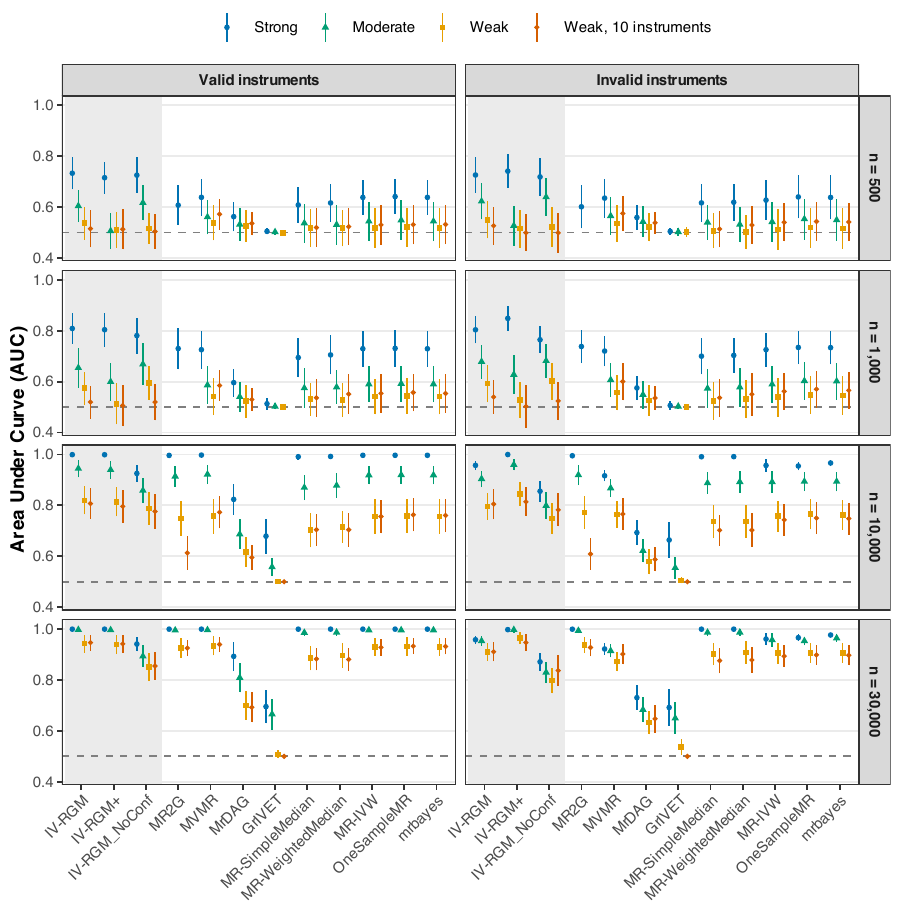}
  \caption{Network AUC by instrument strength for small-world networks with $p=10$. Columns distinguish designs without and with shared instruments (labeled valid and invalid instruments), and rows give the sample size. The legend gives the four instrument designs. Points show means across successful fits, and vertical lines extend one standard deviation in each direction. Shading identifies \texttt{IV-RGM}, \texttt{IV-RGM+}, and \texttt{IV-RGM\_NoConf}. The dashed line marks chance ranking ($0.5$). Estimates are shown for settings with at least ten successful fits.}
  \label{fig:strength_p10}
\end{figure}

To assess how instrument strength affects network recovery, we repeat Cases~2 and~3 with weaker instruments (Figure~\ref{fig:strength_p10}). With a correct map, the three instruments assigned to each response explain about $20\%$ of its variance in the main study. Reducing each direct instrument effect from $1$ (strong) to $0.5$ (moderate) or $0.3$ (weak) lowers this share to about $6\%$ or $2\%$, respectively. A fourth design (weak, 10 instruments) spreads the same weak total strength over ten instruments per response. At $n=500$, the implied first-stage $F$ statistic falls from $42$ with three strong instruments to $4$ with three weak instruments and about $1$ with ten weak instruments.

At $n\le 1{,}000$, the mean network AUCs in both weak-instrument designs range from about $0.50$ to $0.60$ across methods. With moderate instruments and a correct map at $n=1{,}000$, \MRRGM{} issues at least one diagnostic warning in only $5\%$ of fits. Its mean AUC is $0.65$ (Section~OA.4), so the diagnostics rarely flag this loss of accuracy. At $n=10{,}000$, \MRRGM{} and \MRRGMp{} have similar network AUCs ($0.82$ and $0.81$) when each response has three weak instruments of its own and none are shared. When instruments are shared and the supplied map omits their second target, the AUC is $0.80$ with the fixed map and $0.84$ when the map is learned.

\subsection{Further Experiments}\label{sec:sim_more}

\paragraph{Exposure--outcome design.}
We simulate four exposures and four outcomes at $n=1{,}000$ and $n=10{,}000$, allowing a directed acyclic graph among the outcomes. Only \texttt{MR2} and \texttt{MrDAG} receive the exposure--outcome split. On the 16 possible exposure--outcome effects, \MRRGM{} and \texttt{MR2} both rank the edges accurately (AUC at least $0.98$ at both sample sizes). Their effect MADs are $0.030$ and $0.043$, respectively, at $n=1{,}000$, and $0.011$ and $0.020$ at $n=10{,}000$ (Section~OA.3).

\paragraph{Error variance and confounding.}
In Case~2 with $p=10$ and $n=1{,}000$, increasing the error variance from $1$ to $9$ reduces the network AUC of \MRRGM{} from $0.97$ to $0.81$. A separate experiment varies the confounder loadings. At $n=1{,}000$ with no confounding, \MRRGMnc{} ranks edges more accurately than \MRRGM{} (AUC $0.95$ versus $0.88$). With the loadings at $1.5$ times their original values, \MRRGM{} ranks edges more accurately ($0.74$ versus $0.68$). At $n=10{,}000$ under this stronger confounding, the FDR is $0.04$ for \MRRGM{} and $0.67$ for \MRRGMnc{} (Section~OA.3).

\paragraph{Prior sensitivity.}
In Case~2 at $p=10$ and $n=1{,}000$, replacing the default $\Beta(3,1)$ prior on the causal edge inclusion probabilities with a uniform prior reduces the FDR from $0.37$ to $0.07$ for \MRRGM{} and from $0.36$ to $0.06$ for \MRRGMp{}. For both versions the network AUC changes by at most $0.003$. Varying the variance ratio $\nu_1$ from $10^{-3}$ to $10^{-2}$ or $10^{-4}$ also changes the AUC by less than $0.01$ for either method (Section~OA.3).

\subsection{Large Samples and Computation Time}\label{sec:sim_time}

Both versions recover every edge without false discoveries in the ten-response small-world design at $n=10^5$. At $n=10^6$, \MRRGM{} again recovers the graph. For \MRRGMp{}, edge probabilities differ enough between the two halves of the $50{,}000$-iteration chain to trigger a warning in $93$ of the $100$ fits. Extending the chains to $250{,}000$ iterations on the same data reduces the warning rate to $6\%$ and raises the TPR from $0.873$ to $0.996$, with no false discoveries at either chain length.

\texttt{MR2G} and \texttt{MVMR} also select every true edge at both sample sizes, with mean FDRs of about $0.13$ and $0.12$, respectively. At $n=10^6$, \texttt{MrDAG} recovers $97.7\%$ of the true edges with an FDR of $0.005$. All five pairwise methods recover every true edge at both sample sizes, but their FDRs rise from $0.22$--$0.30$ at $n=10^5$ to $0.50$--$0.53$ at $n=10^6$. Larger samples give these methods more power to detect total effects along indirect paths. Selecting such a pair counts as a false discovery when the direct edge is absent.

Once the second-moment matrices have been computed, the cost of a likelihood evaluation is independent of $n$. Across sample sizes from $500$ to $10^6$, a $50{,}000$-iteration fit of \MRRGM{} with $p=10$ takes $43$--$45$ seconds on average on one core. At $n=10^6$, \MRRGMp{} takes $56$ seconds on average with $50{,}000$ iterations and $241$ seconds with $250{,}000$. On the same data, \texttt{MVMR} averages less than a second, \texttt{MR2G} averages $29$ seconds and \texttt{OneSampleMR} averages $5.4$ hours. None of the $100$ \texttt{GrIVET} fits finishes within $48$ hours. With $p=20$ and $n=30{,}000$, median fitting times are $63$ seconds for \texttt{MR2G}, $121$ for \MRRGM{}, $180$ for \MRRGMp{} and $525$ for \texttt{OneSampleMR} (Section~OA.3).

\FloatBarrier

\section{Application: Gene Regulation in B Cells}\label{sec:realdata}

We use genotype and single-cell RNA sequencing data from the OneK$1$K cohort \citep{onek1k2022} to study how expression of one gene in the B-cell receptor (BCR) signaling pathway may affect expression of another. In the primary analysis, we fit \MRRGM{} under the assumption that each genetic instrument acts directly on the expression of its assigned gene and can affect the other analyzed genes only through that gene. We refer to these exclusions as the \emph{own-gene map}. We also examine how the estimated network changes when we relax the exclusions, and Section~OA.6 reports the effect of a uniform prior on the selected edges.

\subsection{Gene and Instrument Selection}

Of the $961$ donors with B-cell expression data considered for analysis, $891$ passed a filter on the total read count per donor (Section~OA.6). For each donor, we sum the read counts of each gene over its B cells, divide by the donor's total count and log-transform the result. The donor is the unit of analysis. We adjust for sex, age group and the first three genotype principal components. The IV analysis assumes that the genotypes and these covariates are independent of the unmeasured causes of expression (Assumption~\ref{ass:ch3:indep}).

We start with the $82$ genes in the BCR signaling pathway of the Kyoto Encyclopedia of Genes and Genomes (KEGG, \citealp{kanehisa2000kegg}). For each gene, we test SNPs within $1$ million base pairs of its transcription start site for association with expression, adjusting for the same covariates. We screen at a single-SNP first-stage $F>10$ \citep{staiger1997instrumental} and retain at most ten SNPs per gene. When two SNPs have squared genotype correlation $r^2>0.5$, or a SNP qualifies for two genes, we keep the SNP with the stronger association. Of the $35$ genes with a SNP above the screening threshold, $14$ have an association that remains significant after Bonferroni correction at level $0.05$ across the SNPs tested for that gene (Section~OA.6).

To choose how many of the $14$ genes to analyze, we rank them by the concentration parameter $\mu^2=nR^2/(1-R^2)$, where $R^2$ is the partial $R^2$ of a gene's instruments given the covariates \citep{stock2002survey}. Before fitting the expression network, we simulate networks of the top $5$, $8$, $10$, $12$ and $14$ genes, matching the donor count, the number of instruments per gene and their observed strength. For each size we use $100$ replicates with direct effects of $\pm0.2$ on the generating scale. We retain the largest size for which that set and all smaller tested sets meet three criteria. The mean network AUC must reach $0.80$, the mean correlation of edge PIPs between the two halves of the chain must reach $0.90$, and at most $10\%$ of fits may trigger the error-variance warning. With ten genes these values are $0.86$, $0.94$ and $6\%$, respectively. At twelve genes the warning rate rises to $58\%$ (Section~OA.5).

The resulting ten genes have $48$ instruments. Their partial $R^2$ values range from $0.026$ for \textit{VAV3} to $0.193$ for \textit{BLK}. Selecting the SNPs and estimating their strength in the same donors can inflate these values. No retained SNP reaches $F>10$ for association with another analyzed gene. This screen can miss weaker associations, and a cross-gene association could also arise through an indirect path. The own-gene map therefore remains an assumption. Under that map, one instrument with a nonzero direct effect per gene suffices for the population rank condition of Theorem~\ref{thm:ch3:identrgm}. Example~\ref{ex:ch3:satisfied} explains which identification conditions apply when the map is unknown and the genes have different numbers of instruments.

\subsection{The Estimated Expression Network}

We fit four \MRRGM{} chains with different random seeds, each for $100{,}000$ iterations. After discarding the first $20{,}000$ iterations and retaining every tenth draw, we pool $32{,}000$ draws. Across the directed-edge indicators, the largest rank-normalized split-$\widehat R$ is $1.002$ and the smallest bulk effective sample size is $3{,}581$ \citep{vehtari2021rank}. The package flags the unadjusted first stage for \textit{CD81} ($F=9.5$). Each of its seven SNPs passes the individual screen. In a separate regression adjusting for the covariates, their joint first-stage $F$ is $9.4$ (Section~OA.6).

\begin{figure}[htbp]
    \centering
    \includegraphics[width=.78\textwidth]{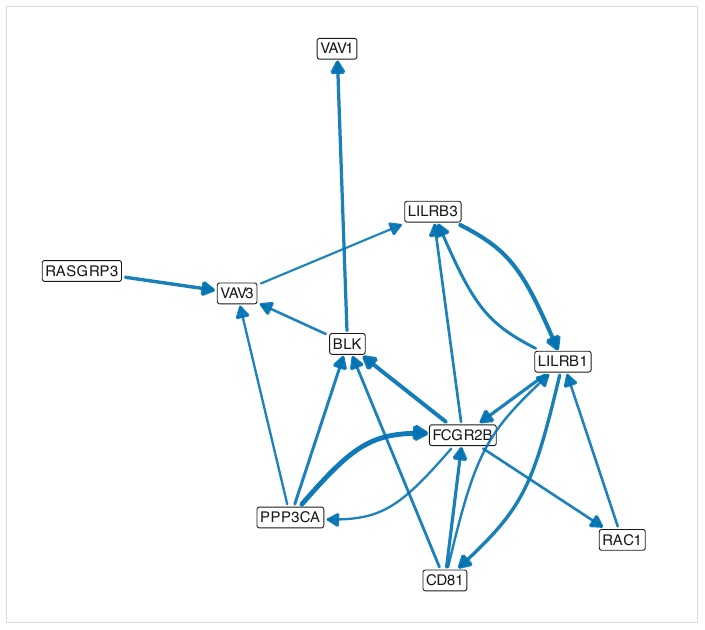}
    \caption{Estimated network of B-cell expression in $891$ OneK$1$K donors under the own-gene map. Each arrow represents a direct effect with PIP above $0.5$ under \MRRGM{}. Arrow width increases with PIP on a scale from $0.5$ to $1$. Section~OA.6 gives the PIP and posterior mean effect for each arrow.}
    \label{fig:onek_network}
\end{figure}

Of the $90$ possible directed edges, $19$ have PIP above $0.5$ (Figure~\ref{fig:onek_network}). The highest-ranked edge runs from \textit{PPP3CA}, which encodes a catalytic subunit of calcineurin, to the inhibitory receptor gene \textit{FCGR2B}. Its posterior mean direct effect is $-0.21$ in standardized units. Its PIP is $0.776$, and the other $18$ selected edges have PIPs between $0.506$ and $0.695$.

The selected graph connects \textit{FCGR2B} to \textit{BLK}, a Src-family kinase, and \textit{BLK} to both \textit{VAV1} and \textit{VAV3}. The positive \textit{BLK}~$\to$~\textit{VAV1} estimate (PIP $0.61$) has the widest agreement across methods (Section~\ref{sec:realdata_comparison}). In primary mouse B cells, antigen-receptor stimulation induces phosphorylation of Vav proteins \citep{fujikawa2003vav}. This provides pathway context for the association between the two human transcripts.

Three gene pairs have selected edges in both directions, with posterior mean effects of opposite signs. They are \textit{LILRB1} and \textit{LILRB3}, \textit{LILRB1} and \textit{CD81}, and \textit{PPP3CA} and \textit{FCGR2B}. These same pairs have the three largest confounding PIPs. The highest is only $0.19$, for \textit{LILRB1} and \textit{CD81}. A parametric bootstrap illustrates the difficulty of recovering reciprocal effects in this data set \citep{franklin2014plasmode}. Taking the fitted $19$-edge network as the generating model, we simulate expression for the same donors and refit \MRRGM{} to $100$ data sets. The \textit{PPP3CA}~$\to$~\textit{FCGR2B} edge is selected in $97$ refits, but the reverse effect is selected with its fitted sign in only $40$. Across refits, an average of $55\%$ of the selected edges are absent from the generating network (Section~OA.6).

We also report motif probabilities (Section~\ref{subsec:priors}, \citealp{milo2002motifs}). A cascade is a chain $a\to b\to c$, a feed-forward loop adds the edge $a\to c$, and a feedback loop is $a\to b\to c\to a$. Among motifs formed from the selected edges, the highest-probability cascade is \textit{LILRB3}~$\to$~\textit{LILRB1}~$\to$~\textit{CD81} ($0.44$). The highest-probability feed-forward loop links \textit{PPP3CA} to \textit{BLK} both directly and through \textit{FCGR2B} ($0.31$). The leading three-gene feedback loop connects the inhibitory receptors \textit{LILRB1}~$\to$~\textit{FCGR2B}~$\to$~\textit{LILRB3}~$\to$~\textit{LILRB1} ($0.23$). Their role in regulating immune activation makes this loop a candidate for follow-up \citep{ravetch2000inhibitory}.

\subsection{Comparison with Other Methods}\label{sec:realdata_comparison}

On the same donors, genes and instruments, \texttt{MVMR} shares $7$ of its $15$ selected edges with \MRRGM{}, and \texttt{mrbayes} shares $8$ of $16$. \texttt{MrDAG} selects one edge, which falls outside the \MRRGM{} network, and \texttt{GrIVET} selects none. \texttt{MR2G} requires for every gene at least one instrument with association $p<5\times10^{-8}$ (genome-wide significance), a requirement met by only four of the ten genes. All competing methods that report an edge shared with \MRRGM{} agree on the sign of its effect. In particular, \texttt{MVMR}, IVW, both median estimators and \texttt{mrbayes} all give a positive estimate for \textit{BLK}~$\to$~\textit{VAV1}. Four of these five estimate total effects, so their agreement also allows for an effect mediated by other genes. Section~OA.6 gives the full comparison, including the pairs evaluated for each method.

We examine the influence of the model assumptions in two further fits. With a diagonal error covariance, \MRRGMnc{} shares $17$ of its $18$ selected edges with \MRRGM{}. Allowing \MRRGMp{} to learn the instrument--response map leads to a different problem. Even after $500{,}000$ iterations per chain, the four chains disagree about individual edges. Their PIPs differ by as much as $0.45$, the largest split-$\widehat R$ for the edge indicators is $1.17$, and fitted error variances are about $14$ times the response variances. As in the instrument-strength simulations (Section~\ref{sec:sim_strength} and Section~OA.4), the available instruments provide little information about the network once the exclusions are relaxed. We base the reported edge and motif probabilities on the fit with the own-gene map.

\FloatBarrier

\section{Discussion}\label{sec:discussion}

Instrumental variables provide a way to estimate direct effects among responses that can influence one another through feedback and share unmeasured causes. When the instrument--response map is known, the exclusions in each equation identify its coefficients under the classical rank condition. When the map is unknown, \MRRGMp{} allows an instrument to act directly on several responses. Under instrument independence and diversity, the observed distribution determines a unique representation with the fewest direct instrument effects among those satisfying the cycle-gain bound. For either version, the joint posterior gives the probability that all the edges of a proposed feedback loop are present.

The posterior consistency results use the restricted priors of Section~\ref{subsec:asymptotics} and, for \MRRGMp{}, a minimality condition on the instrument--response map. Consistent selection of the directed graph also requires the variance ratio $\nu_{1,n}$ to decrease at a suitable rate.

The simulations illustrate what is gained or lost by specifying the instrument targets. With strong instruments and a correct map, \MRRGM{} ranked the edges more accurately than the competing methods at $n=500$ and $1{,}000$. When some instruments acted on a second response omitted from the supplied map, learning the map reduced false discoveries. At $n=10{,}000$ in that design, \MRRGMp{} and \texttt{MR2G} both recovered more than $97\%$ of the true edges, with mean FDRs of $0.009$ and $0.117$, respectively (Section~OA.2). Fixing the incorrect map gave an FDR of $0.152$. Modeling the error covariance also reduced false discoveries when confounding was strong. Estimating the additional covariances reduced ranking accuracy in the smaller sample when confounding was absent.

Instrument strength limits both versions in finite samples. In the weak-instrument designs at $n\le1{,}000$, mean network AUCs ranged from about $0.50$ to $0.60$ across methods. The OneK$1$K analysis illustrates the practical consequence. With the own-gene map, the four chains agreed on the edge probabilities, and the highest-ranked edge was \textit{PPP3CA}~$\to$~\textit{FCGR2B}. With the map estimated from the data, the chains gave different probabilities for individual edges. Even under the fixed map, the leading three-gene feedback loop had posterior probability $0.23$. In bootstrap samples generated from the fitted network, an average of $55\%$ of the selected edges were absent from the generating graph. The fitted network should therefore be used only to prioritize individual regulatory hypotheses for follow-up.

The present likelihood uses second moments computed from the same individuals. Extending it to GWAS summary statistics from separate samples would require additional information or assumptions about the covariances among responses. Another open question is whether shrinking the spike variance $\omega_2^2$ in the prior on the structural error covariances can yield consistent selection of confounding links. For larger networks, joint updates of blocks of coefficients or gradient-based updates could improve computation. A latent Gaussian formulation could handle binary responses by sampling the latent variables along with the network. Beyond genetics, randomized perturbations with known targets offer another setting in which to study feedback with a prespecified instrument--response map.

\section*{Data and Code Availability}

The R package \texttt{MR.RGM} (version \texttt{0.1.4}) implements \texttt{IV-RGM} and \texttt{IV-RGM+}, which it calls \texttt{MR.RGM} and \texttt{MR.RGM+}, together with the diagnostics and the motif probabilities. The package is distributed on CRAN (\url{https://cran.r-project.org/package=MR.RGM}) and GitHub (\url{https://github.com/bitansa/MR.RGM}). Version \texttt{0.1.4} is available from R-universe (\url{https://bitansa.r-universe.dev/MR.RGM}) and is archived on Zenodo (\url{https://doi.org/10.5281/zenodo.24044840}). All simulation and analysis scripts are in the \texttt{reproducibility} folder of the GitHub repository. The result files from which they regenerate every table and figure are archived on Zenodo (\url{https://doi.org/10.5281/zenodo.24043120}), as are the preprocessed OneK$1$K B-cell data (\url{https://doi.org/10.5281/zenodo.18489687}). The original OneK$1$K expression and array genotype data are available through the Gene Expression Omnibus (GEO) under accessions GSE196735 and GSE196829, respectively (series GSE196830, \citealp{onek1k2022}).

\acks{We thank the OneK$1$K consortium for making the single-cell expression and genotype data available. Portions of this research were conducted with the advanced computing resources provided by the Texas A\&M Department of Statistics Arseven Computing Cluster. Competing interests: the authors declare no competing interests. Use of artificial intelligence: large language models were used to help edit the text and to check the code and the proofs. The authors verified every statement, result, and reference and take full responsibility for the content.}

\FloatBarrier

\appendix
\section{Identifiability of \texorpdfstring{\MRRGM{}}{IV-RGM} and \texorpdfstring{\MRRGMp{}}{IV-RGM+}}
\label{sec:ch3:app-ident}

\subsection{Setup}
\label{sec:ch3:app-ident-setup}

Under Assumption~\ref{ass:ch3:indep}, two parameter values $\theta=(\bA,\bB,\bSigs)$ and $\widetilde\theta=(\widetilde\bA,\widetilde\bB,\widetilde\bSigs)$ induce the same conditional distribution of $\bY$ given $\bX$ precisely when they give the same reduced form $(\bPi,\symbf{\Omega})$ in \eqref{eq:ch3:Omega}. Write $\mathbf{M}=\bI_p-\bA$ and $\widetilde{\mathbf{M}}=\bI_p-\widetilde\bA$. For any representation of this conditional distribution, $\bB=\mathbf{M}\bPi$ and $\bSigs=\mathbf{M}\symbf{\Omega}\mathbf{M}^{\T}$. It therefore suffices to identify the rows of $\bA$.

For response $j$, let $\mathbf{a}_{j,-j}\in\R^{p-1}$ be the column vector of off-diagonal entries in row $j$ of $\bA$, and let $\mathbf{b}_j,\bpi_j\in\R^k$ be the transposes of row $j$ of $\bB$ and $\bPi$, respectively. With $\symbf{\Phi}_j=\bPi_{-j,:}^{\T}$, the $j$th row of $\bB=\mathbf{M}\bPi$ gives
\begin{equation*}
\mathbf{b}_j=\bpi_j-\symbf{\Phi}_j\mathbf{a}_{j,-j}.
\end{equation*}

\subsection{Proof of the Identifiability Theorem for \texorpdfstring{\MRRGM{}}{IV-RGM}}
\label{sec:ch3:app-ident-rgm}

\begin{proof}[Proof of Theorem~\ref{thm:ch3:identrgm}]
Let $\theta$ and $\widetilde\theta$ satisfy the exclusions of the known map and have the same conditional distribution. For response $j$, the excluded instruments $H_j=\{h:\delta_{jh}=0\}$ give the system
\begin{equation*}
[\bpi_j]_{H_j}=(\bPi_{-j,H_j})^{\T}\mathbf{a}_{j,-j}.
\end{equation*}
Both $\mathbf{a}_{j,-j}$ and $\widetilde{\mathbf{a}}_{j,-j}$ solve this system. Its coefficient matrix has full column rank by hypothesis, so the solutions agree. Applying this to each $j$ gives $\widetilde\bA=\bA$. The identities $\bB=\mathbf M\bPi$ and $\bSigs=\mathbf M\symbf{\Omega}\mathbf M^{\T}$ then give $\widetilde\theta=\theta$.

To see the equivalence of the two rank conditions, use $\bPi=\mathbf{M}^{-1}\bB$ and $\bB_{j,H_j}=\bzero$ to obtain
\begin{equation}\label{eq:ch3:pirank}
\bPi_{-j,H_j}=[\mathbf{M}^{-1}]_{-j,-j}\,\bB_{-j,H_j}.
\end{equation}
By Jacobi's identity for complementary minors, $\det\{[\mathbf{M}^{-1}]_{-j,-j}\}=M_{jj}/\det(\mathbf{M})=1/\det(\mathbf{M})\ne0$. The two matrices $\bPi_{-j,H_j}$ and $\bB_{-j,H_j}$ therefore have the same rank.
\end{proof}

\begin{rem}[Own instruments]\label{rem:ch3:ciseqtl}
Suppose the map assigns every instrument to exactly one response, as in Section~\ref{sec:realdata}, and write $G_i$ for the set assigned to response $i$. For every $\bB$ satisfying the exclusions, $H_j=\bigcup_{i\ne j}G_i$ and $\bB_{-j,H_j}$ consists of disjoint row blocks $\bB_{i,G_i}$, $i\ne j$. Its rank is $p-1$ if and only if all these blocks are nonzero. Thus \eqref{eq:ch3:pirank} gives full rank for every $j$ if and only if each response has at least one own instrument with a nonzero direct effect. With $\bA$ fixed, if all own effects for response $i$ approach zero, then for each $j\ne i$ the smallest singular value of $\bPi_{-j,H_j}$ approaches zero. In that case estimates of the linear system can be sensitive to sampling error.
\end{rem}

\subsection{Proof of the Identifiability Theorem for \texorpdfstring{\MRRGMp{}}{IV-RGM+}}
\label{sec:ch3:app-ident-plus}

Let $S_j=\supp(\bB_{j,:})$ denote the set of instruments acting directly on response $j$. We first establish the rank consequence of instrument diversity for a given $\bB$.

\begin{lem}[Instrument diversity implies the rank condition]
\label{lem:ch3:rank}
If $\bB$ satisfies Assumption~\ref{ass:ch3:diversity}, its rows are nonzero and linearly independent. For $H_j=\{h:b_{jh}=0\}$, $\rank(\bB_{-j,H_j})=\rank(\bPi_{-j,H_j})=p-1$ for every $j$.
\end{lem}

\begin{proof}
If row $i$ were zero, for any $j\ne i$ the combination $(\mathbf{e}_i+\mathbf{e}_j)^{\T}\bB$ would have $|S_j|\le s(\bB)$ nonzero entries, contradicting $c(\bB)>s(\bB)$. With all rows nonzero, a linear dependence would give a combination of at least two rows with empty support, also a contradiction. Now suppose the rows of $\bB_{-j,H_j}$ are linearly dependent. There is then a nonzero $\mathbf{w}$ with $w_j=0$ and $\mathbf{w}^{\T}\bB_{:,H_j}=\bzero$. Both $\mathbf{w}^{\T}\bB$ and $\mathbf{e}_j^{\T}\bB$ are supported on $S_j$, so their sum has at most $|S_j|$ nonzero entries. It is a combination of at least two rows, again contradicting instrument diversity. The rank of $\bPi_{-j,H_j}$ follows from \eqref{eq:ch3:pirank}.
\end{proof}

\begin{proof}[Proof of Theorem~\ref{thm:ch3:ident}]
Let $\theta$ and $\widetilde\theta$ satisfy the hypotheses of the theorem, with $\bB$ satisfying instrument diversity. Because the reduced form is common to both representations, $\symbf{\Xi}=\widetilde{\mathbf{M}}\mathbf{M}^{-1}$ is invertible and $\widetilde\bB=\symbf{\Xi}\bB$ by \eqref{eq:ch3:xiform}. Write $\symbf{\xi}_j^{\T}$ for row $j$ of $\symbf{\Xi}$. Call this row \emph{pure} if it has a single nonzero entry and \emph{mixed} if it has two or more.

\textit{Step 1. Counting instrument effects.} For a pure row $j$, let $r(j)$ be the position of its nonzero entry. Row $j$ of $\widetilde\bB$ is then a nonzero multiple of row $r(j)$ of $\bB$ and has $|S_{r(j)}|$ nonzero entries. The indices $r(j)$ are distinct by invertibility of $\symbf{\Xi}$. Write $R$ for their set. There are $p-|R|$ mixed rows, and each contributes at least $c(\bB)>s(\bB)$ nonzero entries to $\widetilde\bB$. Hence
\begin{equation*}
\|\widetilde\bB\|_0
=\sum_{i\in R}|S_i|+\sum_{j\text{ mixed}}\|\symbf{\xi}_j^{\T}\bB\|_0
\ \ge\ \sum_{i\in R}|S_i|+\sum_{i\notin R}|S_i|=\|\bB\|_0.
\end{equation*}
The inequality is strict if there is a mixed row. Part~(i) excludes mixed rows by its rowwise bound. In part~(ii), equality of the two totals also excludes them.

\textit{Step 2. Matching equations to responses.} We give the permutation argument of \citet[proof of Theorem~1]{rothenhausler2015backshift} in our notation. If every row is pure, then $\symbf{\xi}_j=d_j\mathbf{e}_{\sigma(j)}$ for a permutation $\sigma$ and nonzero scalars $d_j$. The diagonal of $\widetilde{\mathbf{M}}=\symbf{\Xi}\mathbf{M}$ gives $1=d_j M_{\sigma(j),j}$. Thus $d_j=1$ if $\sigma(j)=j$. If $\sigma(j)\ne j$, then $a_{\sigma(j),j}\ne0$ and $d_j=-1/a_{\sigma(j),j}$. Since $\widetilde M_{j,\sigma(j)}=d_j M_{\sigma(j),\sigma(j)}=d_j$, we have
\begin{equation}\label{eq:ch3:flip}
\widetilde a_{j,\sigma(j)}=-d_j=1/a_{\sigma(j),j}.
\end{equation}
If $\sigma$ has a cycle of length at least two, the corresponding edges $j\to\sigma(j)$ form a directed cycle in $\bA$. Let its gain be $g$. Equation~\eqref{eq:ch3:flip} gives a reversed cycle in $\widetilde\bA$ with gain $1/g$. The two absolute gains cannot both be less than one. Since the theorem requires the cycle bound in both representations, $\sigma$ is the identity. It follows that $d_j=1$ for every $j$, $\widetilde{\mathbf{M}}=\mathbf{M}$ and $\widetilde\bB=\bB$. Finally $\widetilde\bSigs=\widetilde{\mathbf{M}}\symbf{\Omega}\widetilde{\mathbf{M}}^{\T}=\bSigs$. This proves both conclusions.
\end{proof}

\begin{rem}[A weaker diversity condition]\label{rem:ch3:rowdiversity}
Lemma~\ref{lem:ch3:rank} and part~(ii) of Theorem~\ref{thm:ch3:ident} also hold under the following weaker form of instrument diversity. For every $\mathbf{w}$ with at least two nonzero entries, $\|\mathbf{w}^{\T}\bB\|_0>|S_i|$ for every $i$ with $w_i\ne0$. In the proof of the lemma, the combinations with weights $\mathbf{e}_i+\mathbf{e}_j$ and $\mathbf{w}+\mathbf{e}_j$ both involve row $j$ and have at most $|S_j|$ nonzero entries. A linear dependence among nonzero rows gives a combination with empty support. These contradictions still apply. In Step~1 of the theorem, the determinant expansion for $\symbf{\Xi}$ gives a permutation $\sigma$ with $\xi_{j,\sigma(j)}\ne0$ for all $j$. Each pure row contributes exactly $|S_{\sigma(j)}|$ nonzero entries to $\widetilde\bB$, and each mixed row contributes strictly more. Summing gives $\|\widetilde\bB\|_0\ge\|\bB\|_0$, with strict inequality whenever a row is mixed. Step~2 then applies. The weaker condition holds whenever the supports $S_i$ are nonempty and pairwise disjoint, since then $\|\mathbf{w}^{\T}\bB\|_0=\sum_{i:w_i\ne0}|S_i|$.
\end{rem}

\begin{rem}[The role of the identification conditions]\label{rem:ch3:tightness}
The following examples show what each condition of Theorem~\ref{thm:ch3:ident} rules out.
\begin{enumerate}
\item \textit{Cycle gains.} Suppose the graph of $\bA$ has a directed cycle $j_1\to\cdots\to j_r\to j_1$ of gain $g$. Let $\sigma$ send each vertex on the cycle to its successor and fix every other vertex. Form $\symbf{\Xi}$ with row $\symbf{\xi}_j^{\T}=-\mathbf{e}_{\sigma(j)}^{\T}/a_{\sigma(j),j}$ for $j$ on the cycle and $\mathbf{e}_j^{\T}$ elsewhere. Set $\widetilde{\mathbf M}=\symbf{\Xi}\mathbf M$, $\widetilde\bA=\bI_p-\widetilde{\mathbf M}$, $\widetilde\bB=\symbf{\Xi}\bB$ and $\widetilde\bSigs=\symbf{\Xi}\bSigs\symbf{\Xi}^{\T}$. Then $\widetilde{\mathbf M}$ is invertible with diagonal one, and $\widetilde\theta$ and $\theta$ induce the same conditional distribution. Since $\widetilde\bB$ consists of rescaled and permuted rows of $\bB$, $\|\widetilde\bB\|_0=\|\bB\|_0$. By \eqref{eq:ch3:flip}, reassigning the equations reverses the cycle and gives it gain $1/g$. When $|g|<1$ this excludes the reassignment from the class in Theorem~\ref{thm:ch3:ident}. When $|g|=1$ both representations violate the strict bound.
\item \textit{Sufficient conditions for the cycle bound.} If $|a_{jh}|<1$ for all $j\ne h$, every absolute cycle gain is smaller than one. If the spectral radius $\varrho$ of $|\bA|$ is smaller than one, a cycle of length $r$ with initial vertex $j_1$ and gain $g$ satisfies $|g|\le(|\bA|^r)_{j_1j_1}\le\varrho^r<1$. Here a diagonal entry of a nonnegative matrix is at most its spectral radius. The spectral radius of the signed matrix $\bA$ alone does not characterize the cycle bound. For example, suppose the only nonzero entries of $\bA$ are $a_{12}=a_{21}=a_{13}=a_{31}=0.95$. Both cycles have gain $0.9025$. The spectral radius of $\bA$ is about $1.34$. Conversely, take $p=4$ and nonzero entries $a_{12}=1$, $a_{14}=-1/2$, $a_{23}=-1/2$, $a_{31}=-3/2$, and $a_{43}=-1$. The two cycle gains are $3/4$ and $-3/4$, and $\bA$ has spectral radius zero. Reassigning equations along $1\to3\to4\to1$ as in item (i) gives nonzero entries $\widetilde a_{13}=-2/3$, $\widetilde a_{23}=-1/2$, $\widetilde a_{34}=-1$, $\widetilde a_{41}=-2$, and $\widetilde a_{42}=2$. The spectral radius of $\widetilde\bA$ is about $0.69$, yet its cycle $1\to4\to3\to1$ has gain $-4/3$.
\item \textit{Instrument diversity.} Suppose $\bA$ satisfies the cycle bound and $S_i\subseteq S_j$ for some $i\ne j$. A combination of rows $i$ and $j$ is supported on $S_j$, so $c(\bB)\le|S_j|\le s(\bB)$. For small $\epsilon\ne0$, form $\symbf{\Xi}_\epsilon$ by replacing row $j$ of the identity matrix with $(\mathbf{e}_j+\epsilon\mathbf{e}_i)^{\T}/(1-\epsilon a_{ij})$. Define $\widetilde{\mathbf M}$, $\widetilde\bA$, $\widetilde\bB$ and $\widetilde\bSigs$ as in the first part of this remark, using $\symbf{\Xi}_\epsilon$. The denominator makes $\widetilde M_{jj}=1$. This gives a distinct representation with the same reduced form, and row $j$ of $\widetilde\bB$ is $(\bB_{j,:}+\epsilon\bB_{i,:})/(1-\epsilon a_{ij})$. For all sufficiently small $\epsilon$, this row has support $S_j$, as none of its finitely many nonzero entries can vanish in a sufficiently small neighborhood of $\epsilon=0$. Every other row of $\widetilde\bB$ is unchanged. Invertibility and the strict cycle bound persist by continuity, and any new cycle created in row $j$ has gain of order $\epsilon$. The conditional distribution therefore admits arbitrarily close, distinct representations with the same pattern of direct instrument effects. This row transformation also underlies Remark~\ref{rem:ch3:weakident}.
\end{enumerate}
\end{rem}

\subsection{Examples of Instrument Diversity}
\label{sec:ch3:app-diversity-examples}

\begin{example}[Own instruments]
\label{ex:ch3:blockdiag}
Suppose response $i$ has $m_i\ge1$ own instruments, with indices $G_i$, that have nonzero effects on response $i$ alone. Other instruments may act on several responses. For $h\in G_i$, $(\mathbf{w}^{\T}\bB)_h=w_i b_{ih}$, so each participating row contributes all of its own instruments to the support of the combination. Thus $\|\mathbf{w}^{\T}\bB\|_0\ge\sum_{i:w_i\ne0}m_i$ and $c(\bB)\ge m_{(1)}+m_{(2)}$, where $m_{(1)}$ and $m_{(2)}$ are the two smallest counts. If all nonzero instrument effects are own effects, $s(\bB)=\max_i m_i$ and $c(\bB)=m_{(1)}+m_{(2)}$. Instrument diversity then holds precisely when $\max_i m_i<m_{(1)}+m_{(2)}$.
\end{example}

\begin{example}[Simulation designs and application]
\label{ex:ch3:satisfied}
In Cases 1 and 2 of Section~\ref{sec:sim}, every response has three own instruments and there are no shared instruments, so $s(\bB)=3$ and $c(\bB)=6$. In Case 3, each shared instrument acts on two responses. Then $s(\bB)=4$ for $p=10$ and $s(\bB)=5$ for $p=5$, where response 1 belongs to two pairs, and $c(\bB)\ge6$ by Example~\ref{ex:ch3:blockdiag}. In the design with ten own instruments per response of Section~\ref{sec:sim_strength}, $s(\bB)\le12$ and $c(\bB)\ge20$. Instrument diversity therefore holds in these designs. The nonzero effects among responses are $\pm0.1$ on the generating scale and $\pm0.3$ in the acyclic exposure--outcome design. Every cycle gain is at most $0.01$ in absolute value. Rescaling the responses and instruments preserves the cycle gains and the support sizes entering $s(\bB)$ and $c(\bB)$.

In the application of Section~\ref{sec:realdata}, the own-gene map assigns two to ten instruments to each gene, with two instruments assigned to each of three genes. If each assigned instrument has a nonzero direct effect on its gene, $s(\bB)=10$ and $c(\bB)=4$. Instrument diversity then fails, but the disjoint supports satisfy the weaker condition of Remark~\ref{rem:ch3:rowdiversity}. The precision of the estimated effects depends in addition on instrument strength and sample size (Remark~\ref{rem:ch3:weakident}).
\end{example}

\section{Asymptotic Posterior Consistency}
\label{sec:ch3:app-consistency}

This appendix proves Theorems~\ref{thm:ch3:thetacons}, \ref{thm:ch3:indicators} and~\ref{thm:ch3:pluscons}. Sections~\ref{sec:ch3:app-asymp-priors}--\ref{sec:ch3:app-gammacons} treat a prior with a fixed instrument--response map, under two abstract identification conditions (Assumptions~\ref{ass:ch3:asympident} and~\ref{ass:ch3:localbilip}), and prove Propositions~\ref{prop:ch3:thetagen} and~\ref{prop:ch3:indgen}. Section~\ref{sec:ch3:app-rgmcons} verifies these conditions for \MRRGM{}, and Section~\ref{sec:ch3:app-pluscons} treats \MRRGMp{}. The argument for a fixed map is the prior-mass and testing approach of \citet{Schwartz1965}, \citet{Ghosal2000} and \citet[Chapters~6--8]{ghosal2017fundamentals}. The steps are given in full because the prior may change with $n$ and the Bayes-factor argument for the indicators needs the contraction rate of Lemma~\ref{lem:ch3:localrate}. The covariates are dropped by projection as in Section~\ref{subsec:identifiability}, so $\bC=\bzero$. The estimation of $\bC$ with its conjugate prior is not analyzed. Throughout, $p$ and $k$ are fixed as $n\to\infty$.

\subsection{Setting, Likelihood and Prior}
\label{sec:ch3:app-asymp-priors}

The data are $\mathcal{D}_n=\{(\bY_i,\bX_i)\}_{i=1}^n$ from the SEM
\begin{equation}\label{eq:ch3:semasym}
\bY = \bA\bY + \bB\bX + \bE^*,\qquad \bE^* \sim \Normal_p(\bzero,\bSigs),
\end{equation}
with $\diag(\bA)=\bzero$ and $\bI_p-\bA$ invertible, under the true parameter $\theta_0=(\bA_0,\bB_0,\bSig_0^*)$ with indicators $\gamma^0_{jh}:=\mathbb{I}(a^0_{jh}\neq0)$ and $\phi^0_{jh}:=\mathbb{I}(b^0_{jh}\neq0)$. Let $\mathbf{M}:=\bI_p-\bA$, $\mathbf{M}_0:=\bI_p-\bA_0$ and $\mathbf{r}_i:=\mathbf{M}\bY_i-\bB\bX_i$. Since $\bY_i\mid\bX_i\sim\Normal_p(\mathbf{M}^{-1}\bB\bX_i,\mathbf{M}^{-1}\bSigs\mathbf{M}^{-\T})$, the log-likelihood is, up to an additive constant,
\begin{equation}\label{eq:ch3:loglik}
\begin{split}
\ell_n(\theta)&=n\log|\det\mathbf{M}|-\frac n2\log|\bSigs|-\frac12\sum_{i=1}^n \mathbf{r}_i^{\T}(\bSigs)^{-1}\mathbf{r}_i\\
&=n\log|\det\mathbf{M}|-\frac n2\log|\bSigs|-\frac n2\tr\{(\bSigs)^{-1}\mathbf{Q}_n(\bA,\bB)\},
\end{split}
\end{equation}
with $\mathbf{Q}_n(\bA,\bB)=\mathbf{M}\bS_{yy}\mathbf{M}^{\T}-\mathbf{M}\bS_{yx}\bB^{\T}-\bB\bS_{yx}^{\T}\mathbf{M}^{\T}+\bB\bS_{xx}\bB^{\T}$ and the sample second moments of Section~\ref{subsec:model}. The reduced form is $\bPi:=\mathbf{M}^{-1}\bB$, $\symbf{\Omega}:=\mathbf{M}^{-1}\bSigs\mathbf{M}^{-\T}$, so that $\bY\mid\bX\sim\Normal_p(\bPi\bX,\symbf{\Omega})$. The pair $(\bPi_0,\symbf{\Omega}_0)$ is the reduced form of $\theta_0$ and $\bSig_{XX}:=\E[\bX\bX^{\T}]$.

\paragraph{Prior.} The prior is the one of Section~\ref{subsec:priors} with two modifications. First, the variance ratio of the prior of $\bA$ may depend on $n$: for $j\neq h$,
\begin{equation}\label{eq:ch3:priorA}
\begin{gathered}
a_{jh}\mid \gamma_{jh},\tau_{jh}\sim\Normal(0,\tau_{jh})\ \text{if }\gamma_{jh}=1,\qquad a_{jh}\mid \gamma_{jh},\tau_{jh}\sim\Normal(0,\nu_{1,n}\tau_{jh})\ \text{if }\gamma_{jh}=0,\\
\gamma_{jh}\mid\rho_{jh}\sim \Bern(\rho_{jh}),\qquad \rho_{jh}\sim \Beta(a_\rho,b_\rho).
\end{gathered}
\end{equation}
The hyperparameters $\rho_{jh}$ are coordinate-specific, so after integrating them out the prior odds of each indicator are $a_\rho/b_\rho$, independently across coordinates. Second, the half-Cauchy priors on the slab scales $\sqrt{\tau_{jh}}$ and $\sqrt{\eta_{jh}}$ are truncated to a fixed interval $[\underline s,\overline s]$, $0<\underline s<\overline s<\infty$, so that $\tau_{jh},\eta_{jh}\in[\underline t,\overline t]:=[\underline s^2,\overline s^2]$. This gives uniform control in the prior-mass bounds. The prior on $\bSigs$ enters only through Assumption~\ref{ass:ch3:sigmaprior}, and its indicators $z_{jh}$ are not analyzed.

Let $\symbf{\Delta}\in\{0,1\}^{p\times k}$ be a binary map. The parameter set is
\begin{equation*}
\Theta':=\Theta_M^{\symbf{\Delta}}:=\{\theta\in\Theta_M: b_{jh}=0 \text{ whenever } \delta_{jh}=0\},
\end{equation*}
the set of parameters in the compact set $\Theta_M$ of Assumption~\ref{ass:ch3:compact} that satisfy the exclusions of $\symbf{\Delta}$. The entries $b_{jh}$ with $\delta_{jh}=1$ are called free and have the slab prior $b_{jh}\mid\eta_{jh}\sim\Normal(0,\eta_{jh})$. The prior $\mathcal{P}_n$ is the product of the priors of the entries of $\bA$, of the free entries of $\bB$ and of $\bSigs$, restricted to $\Theta'$ and normalized. Balls, densities, covering numbers and $\|\theta-\theta_0\|_\infty$ refer to the free coordinates, and $\theta_0\in\Theta'$ requires $\bB_0$ to satisfy the exclusions of $\symbf{\Delta}$. This is the prior of \MRRGM{}. The prior of \MRRGMp{} is a mixture of such priors over the maps (Section~\ref{sec:ch3:app-pluscons}).

\subsection{Consequences of the Regularity Conditions}
\label{sec:ch3:app-asymp-assumptions}

On the set $\Theta_M$ of Assumption~\ref{ass:ch3:compact} the reduced-form map $\theta\mapsto(\bPi,\symbf{\Omega})$ is smooth with bounded derivatives, the eigenvalues of $\symbf{\Omega}$ lie in a fixed interval $[\underline{c}_\Omega,\overline{c}_\Omega]\subset(0,\infty)$, and, with Assumption~\ref{ass:ch3:design}, $\sup_{\theta\in\Theta_M}\E_\theta(\|\bX\|_2^4+\|\bY\|_2^4)<\infty$ and $\bS_{xx}\to\bSig_{XX}$ in probability. These facts are used without further comment. If an implementation uses the unconstrained prior and its posterior assigns $o_{\Prob}(1)$ mass to $\Theta_M^c$, the two posteriors are asymptotically equivalent on events inside $\Theta_M$ (Remark~\ref{rem:ch3:theoryvspractice}).

The lower bound on the covariance density in Assumption~\ref{ass:ch3:sigmaprior} is used in Lemmas~\ref{lem:ch3:klmass} and~\ref{lem:ch3:localrate} and in the proof of Theorem~\ref{thm:ch3:pluscons}. It holds for the SSSL prior of Section~\ref{subsec:priors}, whose density is continuous and positive on the open cone, when $\bSig_0^*$ is in the interior of the cone. The SSSL prior density is also bounded above, as Assumption~\ref{ass:ch3:compactplus} requires.

\begin{assumption}[Identifiability]\label{ass:ch3:asympident}
Any two points of $\Theta'$ with reduced form $(\bPi_0,\symbf{\Omega}_0)$ coincide.
\end{assumption}

\begin{assumption}[Local bi-Lipschitz identification]\label{ass:ch3:localbilip}
There are a neighborhood $\mathcal U_0\subset\Theta'$ of $\theta_0$ and constants $0<c_L\le C_L<\infty$ such that, for all $\theta\in\mathcal U_0$,
$c_L\|\theta-\theta_0\|_\infty\le\|\bPi(\theta)-\bPi_0\|_F+\|\symbf{\Omega}(\theta)-\symbf{\Omega}_0\|_F\le C_L\|\theta-\theta_0\|_\infty$.
\end{assumption}

For use in the proofs, Assumption~\ref{ass:ch3:nurate} implies $\log(1/\nu_{1,n})=O(\log n)$ and $\log^2 n/(n\nu_{1,n})=o\{\log(1/\nu_{1,n})\}$. The first relation follows from $n\nu_{1,n}/\log n\to\infty$. For the second, consider separately the $n$ with $\nu_{1,n}\ge n^{-1/2}$, for which $\log^2n/(n\nu_{1,n})\to0$, and the other $n$, for which $\log(1/\nu_{1,n})\ge\tfrac12\log n$. In both cases the ratio of $\log^2 n/(n\nu_{1,n})$ to $\log(1/\nu_{1,n})$ tends to zero. The variance ratio thus shrinks more slowly than the $1/n$ scale of the likelihood curvature, which is used in Lemmas~\ref{lem:ch3:marginals} and~\ref{lem:ch3:bfA}. Examples are $\nu_{1,n}=n^{-1/2}$ and $\nu_{1,n}=1/\log n$. Proposition~\ref{prop:ch3:thetagen} and Lemmas~\ref{lem:ch3:klmass} and~\ref{lem:ch3:localrate} hold for any variance ratio in $(0,1)$, fixed or not. Assumption~\ref{ass:ch3:nurate} is used only for the edge indicators (Lemmas~\ref{lem:ch3:marginals} and~\ref{lem:ch3:bfA}). By Lemma~\ref{lem:ch3:rfkl}, Assumption~\ref{ass:ch3:asympident} is equivalent to a positive Kullback--Leibler divergence $\mathrm{KL}(\theta_0\|\theta)>0$ for every $\theta\in\Theta'\setminus\{\theta_0\}$. Assumption~\ref{ass:ch3:localbilip} is used only for the contraction rate of Lemma~\ref{lem:ch3:localrate} and the indicator results that build on it, not for Proposition~\ref{prop:ch3:thetagen}.

\paragraph{Why the exclusions are needed.} On the set $\Theta_M$, which imposes no exact zeros on $\bB$, Assumptions~\ref{ass:ch3:asympident} and~\ref{ass:ch3:localbilip} fail. For every $\bA'$ with zero diagonal close to $\bA_0$, the point $(\bA',(\bI_p-\bA')\bPi_0,(\bI_p-\bA')\symbf{\Omega}_0(\bI_p-\bA')^{\T})$ lies in $\Theta_M$ and has the reduced form of $\theta_0$. The two assumptions are therefore imposed on the set $\Theta'$ of parameters that satisfy the exclusions of a map, where the rank condition gives them.

\begin{lem}[Identification conditions under a known map]\label{lem:ch3:bilip}
Suppose that Assumptions~\ref{ass:ch3:design} and~\ref{ass:ch3:compact} hold, that $\bB_0$ satisfies the exclusions of $\symbf{\Delta}$ and that $\rank(\bPi_{0,-j,H_j})=p-1$ for every $j$. Then Assumptions~\ref{ass:ch3:asympident} and~\ref{ass:ch3:localbilip} hold on $\Theta'=\Theta_M^{\symbf{\Delta}}$.
\end{lem}

\begin{proof}
Uniqueness is Theorem~\ref{thm:ch3:identrgm}. The upper bound of Assumption~\ref{ass:ch3:localbilip} is the mean value inequality for the smooth map $F(\theta)=(\mathbf{M}^{-1}\bB,\mathbf{M}^{-1}\bSigs\mathbf{M}^{-\T})$ on a closed ball around $\theta_0$ (all norms on the finite-dimensional parameter space are equivalent). For the lower bound, $\Theta_M^{\symbf{\Delta}}$ is the intersection of $\Theta_M$ with the linear subspace $L$ of parameters with zero diagonal in $\bA$, $b_{jh}=0$ for $\delta_{jh}=0$, and symmetric $\bSigs$. For a direction $v=(\mathrm{d}\bA,\mathrm{d}\bB,\mathrm{d}\bSigs)\in L$, since $\mathrm{d}\mathbf{M}^{-1}=\mathbf{M}^{-1}(\mathrm{d}\bA)\mathbf{M}^{-1}$,
\begin{equation*}
\mathrm{d}\bPi=\mathbf{M}_0^{-1}\{(\mathrm{d}\bA)\bPi_0+\mathrm{d}\bB\},\qquad
\mathrm{d}\symbf{\Omega}=\mathbf{M}_0^{-1}(\mathrm{d}\bA)\symbf{\Omega}_0+\symbf{\Omega}_0(\mathrm{d}\bA)^{\T}\mathbf{M}_0^{-\T}+\mathbf{M}_0^{-1}(\mathrm{d}\bSigs)\mathbf{M}_0^{-\T}.
\end{equation*}
If $\mathrm{d}F(\theta_0)v=\bzero$, then $(\mathrm{d}\bA)\bPi_0=-\mathrm{d}\bB$. Row $j$ of $\mathrm{d}\bB$ vanishes on $H_j$ and $\mathrm{d}\bA$ has zero diagonal, so $(\mathrm{d}\mathbf{a}_{j,-j})^{\T}\bPi_{0,-j,H_j}=\bzero$ and the rank condition gives $\mathrm{d}\mathbf{a}_{j,-j}=\bzero$ for every $j$. Hence $\mathrm{d}\bA=\bzero$, $\mathrm{d}\bB=\bzero$ and $\mathrm{d}\bSigs=\mathbf{M}_0(\mathrm{d}\symbf{\Omega})\mathbf{M}_0^{\T}=\bzero$. The differential of $F$ at $\theta_0$ restricted to $L$ is thus injective, so $c_F:=\min\{\|\mathrm{d}F(\theta_0)v\|:v\in L,\|v\|_\infty=1\}>0$. Taylor's theorem gives $\|F(\theta)-F(\theta_0)\|\ge(c_F/2)\|\theta-\theta_0\|_\infty$ for $\theta\in L$ in a small ball around $\theta_0$, which lies in $\Theta_M^{\symbf{\Delta}}$ because $\theta_0$ is interior to $\Theta_M$.
\end{proof}

The lemmas of Sections~\ref{sec:ch3:app-asymp-likelihood}--\ref{sec:ch3:app-gammacons} are stated on $\Theta'$. Bounds that hold uniformly on the compact set $\Theta_M\supseteq\Theta'$ are used on $\Theta'$ without change. The proofs further use that $\theta_0$ is an interior point of $\Theta'$ in the free coordinates, and the product structure of the prior on the free coordinates. The two results for a fixed map are the following.

\begin{prop}[Parameter consistency for a fixed map]\label{prop:ch3:thetagen}
Suppose that Assumptions~\ref{ass:ch3:design}--\ref{ass:ch3:priorpos} and~\ref{ass:ch3:asympident} hold, with a variance ratio in $(0,1)$ that may be fixed or depend on $n$. Then $\mathcal{P}_n(\|\theta-\theta_0\|_\infty>\varepsilon\mid\mathcal{D}_n)\to 0$ in $\Prob_{\theta_0}$-probability, for any $\varepsilon>0$.
\end{prop}

\begin{prop}[Graph recovery for a fixed map]\label{prop:ch3:indgen}
Under the conditions of Proposition~\ref{prop:ch3:thetagen} and Assumptions~\ref{ass:ch3:nurate} and~\ref{ass:ch3:localbilip}, $\mathcal{P}_n(\{\gamma_{jh}\}_{j\neq h}=\{\gamma^0_{jh}\}_{j\neq h}\mid\mathcal{D}_n)\to 1$ in $\Prob_{\theta_0}$-probability.
\end{prop}

\begin{rem}[Theory versus implementation]\label{rem:ch3:theoryvspractice}
The results are proved for a version of the prior that differs from the implemented one in the following respects. (i) \emph{Compactness.} The analysis restricts the prior to the compact set $\Theta_M$ and truncates the half-Cauchy priors on the slab scales. The package uses the untruncated half-Cauchy prior and does not constrain $\bA$ or $\bSigs$. If the unconstrained posterior assigns $o_{\Prob}(1)$ mass outside a fixed compact set that contains $\theta_0$ in its interior, the two posteriors are asymptotically equivalent on events contained in that set. This is a hypothesis, not a consequence of the theorems, and it does not cover the truncation of the slab scales. (ii) \emph{Variance ratio.} Theorems~\ref{thm:ch3:indicators} and~\ref{thm:ch3:pluscons}(iii) use a variance ratio that tends to zero at the rate of Assumption~\ref{ass:ch3:nurate}. For the fixed ratio $\nu_1=10^{-3}$ used by the package, if the coefficient posterior concentrates at its true value, Section~OA.1 gives a limiting PIP strictly between zero and one. It exceeds $0.5$ for standardized effects of magnitude greater than about $0.044$. (iii) \emph{Covariates.} The theory drops the measured covariates by projection, whereas the package estimates $\bC$ jointly with a conjugate prior. (iv) \emph{Error covariance.} The SSSL prior enters the theory only through its support and its density near $\bSig_0^*$. The map-selection proof also uses an upper bound on this density on the compact set. The consistency statement covers each entry $\sigma^*_{jh}$. With fixed spike and slab variances, the conditional inclusion probability for $z_{jh}$ is a smooth function of $|\sigma^*_{jh}|$ (Section~\ref{subsec:priors}). It can remain strictly between zero and one as the posterior for the covariance concentrates. (v) \emph{Standardization.} The theory treats the variables on a fixed scale, whereas the analyses center and scale them by sample moments. (vi) \emph{Cycle gains.} The theory of \MRRGMp{} restricts the prior to parameters whose cycle gains are at most $c_g<1$ (Assumption~\ref{ass:ch3:compactplus}). The package does not impose this restriction. When the true graph has a directed cycle, the reassigned parameter of Remark~\ref{rem:ch3:tightness} has the same likelihood and the same number of instrument effects as $\theta_0$, so the unrestricted posterior need not concentrate at $\theta_0$. The chain of the package starts at $\bA=\bzero$, and the package warns when the spectral radius of the estimate is at least one (Section~OA.1). The statements concern the posterior, not the algorithm that samples from it.
\end{rem}

\subsection{Preliminary Lemmas}
\label{sec:ch3:app-asymp-likelihood}

\begin{lem}[Reduced-form KL divergence]\label{lem:ch3:rfkl}
Under Assumption~\ref{ass:ch3:design}, the Kullback--Leibler divergence per observation, $\mathrm{KL}(\theta_0\|\theta):=\E_{\bX}[\mathrm{KL}\{p_{\theta_0}(\cdot\mid\bX)\|p_\theta(\cdot\mid\bX)\}]$, equals
\begin{equation}\label{eq:ch3:rfkl}
\mathrm{KL}(\theta_0\|\theta)
=\frac12\Big\{\tr(\symbf{\Omega}^{-1}\symbf{\Omega}_0)-p+\log\frac{|\symbf{\Omega}|}{|\symbf{\Omega}_0|}\Big\}
+\frac12\tr\!\big\{\symbf{\Omega}^{-1}(\bPi-\bPi_0)\bSig_{XX}(\bPi-\bPi_0)^{\T}\big\}\ \ge0,
\end{equation}
with equality if and only if $(\bPi,\symbf{\Omega})=(\bPi_0,\symbf{\Omega}_0)$.
\end{lem}

\begin{proof}
Conditional on $\bX$ the two laws are Gaussian with means $\bPi_0\bX$, $\bPi\bX$ and covariances $\symbf{\Omega}_0$, $\symbf{\Omega}$, and the Gaussian KL formula and $\E[\bX\bX^{\T}]=\bSig_{XX}$ give \eqref{eq:ch3:rfkl}. The first term is zero only if $\symbf{\Omega}=\symbf{\Omega}_0$ and the second only if $\bPi=\bPi_0$, because $\symbf{\Omega}\succ0$ and $\bSig_{XX}\succ0$.
\end{proof}

\begin{lem}[One-coordinate score and curvature]\label{lem:ch3:onecoord}
Fix $j\neq h$ and treat the other coordinates as fixed. Then
\begin{equation}\label{eq:ch3:scoreA}
\begin{gathered}
\frac{\partial \ell_n}{\partial a_{jh}}=-n(\mathbf{M}^{-1})_{hj}+\sum_{i=1}^n Y_{ih}\,\mathbf{e}_j^{\T}(\bSigs)^{-1}\mathbf{r}_i,
\\
\frac{\partial^2 \ell_n}{\partial a_{jh}^2}=-n\{(\mathbf{M}^{-1})_{hj}\}^2-\{(\bSigs)^{-1}\}_{jj}\sum_{i=1}^nY_{ih}^2.
\end{gathered}
\end{equation}
\end{lem}

\begin{proof}
With $\mathbf{E}_{jh}=\mathbf{e}_j\mathbf{e}_h^{\T}$, $\partial\mathbf{M}/\partial a_{jh}=-\mathbf{E}_{jh}$, so $\partial\mathbf{r}_i/\partial a_{jh}=-Y_{ih}\mathbf{e}_j$, $\partial\log|\det\mathbf{M}|/\partial a_{jh}=-\tr(\mathbf{M}^{-1}\mathbf{E}_{jh})=-(\mathbf{M}^{-1})_{hj}$, and $\partial\mathbf{M}^{-1}/\partial a_{jh}=\mathbf{M}^{-1}\mathbf{E}_{jh}\mathbf{M}^{-1}$, whose $(h,j)$ entry is $\{(\mathbf{M}^{-1})_{hj}\}^2$. Differentiating \eqref{eq:ch3:loglik} gives \eqref{eq:ch3:scoreA}.
\end{proof}

\begin{lem}[Score and curvature at $\theta_0$]\label{lem:ch3:lan}
Under Assumptions~\ref{ass:ch3:design} and~\ref{ass:ch3:compact}, at $\theta_0$,
$\partial\ell_n/\partial a_{jh}=O_{\Prob}(\sqrt n)$ and $-\partial^2\ell_n/\partial a_{jh}^2=nI^{(A)}_{jh}+o_{\Prob}(n)$ with $I^{(A)}_{jh}:=\{(\mathbf{M}_0^{-1})_{hj}\}^2+\{(\bSig_0^*)^{-1}\}_{jj}\E_{\theta_0}(Y_h^2)>0$.
\end{lem}

\begin{proof}
At $\theta_0$, $\mathbf{r}_i=\bE_i^*$ and $\bY_i=\mathbf{M}_0^{-1}(\bB_0\bX_i+\bE_i^*)$, so $\E\{Y_{ih}\mathbf{e}_j^{\T}(\bSig_0^*)^{-1}\bE_i^*\}=\mathbf{e}_j^{\T}\mathbf{M}_0^{-\T}\mathbf{e}_h=(\mathbf{M}_0^{-1})_{hj}$ because $\E[\bE_i^*\bY_i^{\T}]=\bSig_0^*\mathbf{M}_0^{-\T}$. The score is therefore a sum of i.i.d.\ mean-zero terms with finite variance, hence $O_{\Prob}(\sqrt n)$. The curvature limit is the law of large numbers applied to \eqref{eq:ch3:scoreA}, with positivity from $(\bSig_0^*)^{-1}\succ0$ and $\E_{\theta_0}(Y_h^2)>0$.
\end{proof}

\subsection{Posterior Consistency for the Continuous Parameter}
\label{sec:ch3:app-thetacons}

\begin{lem}[Separation in reduced-form space]\label{lem:ch3:sep}
Under Assumptions~\ref{ass:ch3:compact} and~\ref{ass:ch3:asympident}, for every $\varepsilon>0$ there is $\delta(\varepsilon)>0$ such that $\|\theta-\theta_0\|_\infty>\varepsilon$ implies $\|\bPi(\theta)-\bPi_0\|_F+\|\symbf{\Omega}(\theta)-\symbf{\Omega}_0\|_F>\delta(\varepsilon)$ on $\Theta'$.
\end{lem}

\begin{proof}
Otherwise there is a sequence $\theta_m\in\Theta'$ with $\|\theta_m-\theta_0\|_\infty>\varepsilon$ and reduced forms converging to $(\bPi_0,\symbf{\Omega}_0)$. A convergent subsequence has a limit $\theta^\star\neq\theta_0$ with the reduced form of $\theta_0$, by continuity, contradicting Assumption~\ref{ass:ch3:asympident}.
\end{proof}

\begin{lem}[Prior mass of KL neighborhoods]\label{lem:ch3:klmass}
Under Assumptions~\ref{ass:ch3:design}--\ref{ass:ch3:priorpos}, for every $\kappa>0$ there are $c_\kappa>0$ and $N_\kappa$ such that $\mathcal{P}_n(\mathcal K_\kappa)\ge c_\kappa$ for all $n\ge N_\kappa$, where $\mathcal K_\kappa=\{\theta\in\Theta':\mathrm{KL}(\theta_0\|\theta)\le\kappa\}$.
\end{lem}

\begin{proof}
By continuity of the KL divergence there is $\delta_\ast>0$ such that the sup-norm ball $B(\theta_0,\delta_\ast)$ lies in $\mathcal K_\kappa$ ($\theta_0$ is interior to $\Theta'$ in the free coordinates). Let $\mathcal{P}_n^{\mathrm{prod}}$ denote the unrestricted product prior, of which $\mathcal{P}_n$ is the restriction to $\Theta'$. Since $\mathcal{P}_n(B)\ge\mathcal{P}_n^{\mathrm{prod}}(B)$ for $B\subseteq\Theta'$, it suffices to bound the product prior, under which the indicator pattern of $\bA$, the coefficients given the pattern and the scales, and $\bSigs$ are independent. The event that the indicators equal the true pattern has probability at least $c_\gamma>0$, free of $n$. Given this event, each truly nonzero entry of $\bA$ has a slab density that is continuous and positive at its true value, uniformly over the scale in $[\underline t,\overline t]$, so $|a_{jh}-a^0_{jh}|<\delta_\ast/2$ has probability at least a constant $c^{(1)}>0$. Each truly zero entry of $\bA$ has the spike law $\Normal(0,\nu_{1,n}t)$, where $t\in[\underline t,\overline t]$ is its slab variance, with $\nu_{1,n}t<\overline t$, so $|a_{jh}|<\delta_\ast/2$ has probability at least $c^{(0)}:=\Prob\{|\Normal(0,\overline t)|<\delta_\ast/2\}>0$. Each free entry of $\bB$ has the slab prior only. Its density is continuous and positive at $b^0_{jh}$, whether or not $b^0_{jh}=0$, so $|b_{jh}-b^0_{jh}|<\delta_\ast/2$ also has probability at least a positive constant. By Assumption~\ref{ass:ch3:sigmaprior}, $\|\bSigs-\bSig_0^*\|_F<\delta_\ast/2$ has probability at least $c_2>0$. The intersection of these finitely many events lies in $B(\theta_0,\delta_\ast)$ and has probability at least the product of the bounds, which is the constant $c_\kappa$.
\end{proof}

Under Assumptions~\ref{ass:ch3:design} and~\ref{ass:ch3:compact}, the single-observation log-likelihood ratio $\log\{p_{\theta_0}(\bY\mid\bX)/p_\theta(\bY\mid\bX)\}$ is a quadratic polynomial in $(\bY,\bX)$. Its coefficients are bounded uniformly over $\Theta_M$, because they are continuous functions of $\bPi$, $\symbf{\Omega}^{-1}$ and $\log|\symbf{\Omega}|$. By the fourth-moment bound,
\begin{equation}\label{eq:ch3:Vbound}
V:=\sup_{\theta\in\Theta_M}\Var_{\theta_0}\Big(\log\frac{p_{\theta_0}(\bY\mid\bX)}{p_\theta(\bY\mid\bX)}\Big)<\infty.
\end{equation}

\begin{lem}[Exponentially consistent tests]\label{lem:ch3:tests}
Under Assumptions~\ref{ass:ch3:design}, \ref{ass:ch3:compact} and~\ref{ass:ch3:asympident}, for every $\varepsilon>0$ there are tests $\varphi_n=\varphi_n(\mathcal D_n)$ and constants $c_\varepsilon,C_\varepsilon>0$ with
$\E_{\theta_0}[\varphi_n]\le C_\varepsilon e^{-c_\varepsilon n}$ and $\sup\{\Prob_\theta(1-\varphi_n):\theta\in\Theta',\|\theta-\theta_0\|_\infty>\varepsilon\}\le C_\varepsilon e^{-c_\varepsilon n}$.
\end{lem}

\begin{proof}
Let $\bar A_\varepsilon:=\{\theta\in\Theta':\|\theta-\theta_0\|_\infty\ge\varepsilon\}$, a compact set, and $\mathcal R_\varepsilon$ the compact image of $\bar A_\varepsilon$ under the reduced-form map, which by Lemma~\ref{lem:ch3:sep} is at distance at least $\delta(\varepsilon/2)>0$ from $\vartheta_0:=(\bPi_0,\symbf{\Omega}_0)$. Write $\vartheta=(\bPi,\symbf{\Omega})$, with norm $\|\vartheta\|=(\|\bPi\|_F^2+\|\symbf{\Omega}\|_F^2)^{1/2}$, $p_\vartheta(\mathbf{y}\mid\mathbf{x})$ for the $\Normal_p(\bPi\mathbf{x},\symbf{\Omega})$ density, and note that the likelihood ratio of $\mathcal D_n$ is the product of the conditional ratios because the law of $\bX$ does not depend on $\theta$. For $\vartheta_1=(\bPi_1,\symbf{\Omega}_1)\in\mathcal R_\varepsilon$ let $\varphi_{n,\vartheta_1}:=\mathbb{I}\{\sum_{i=1}^n\log[p_{\vartheta_1}(\bY_i\mid\bX_i)/p_{\vartheta_0}(\bY_i\mid\bX_i)]>0\}$. By Markov's inequality applied to the exponential of half the sum, $\E_{\theta_0}[\varphi_{n,\vartheta_1}]\le\mathcal{H}(\vartheta_1)^n$ with $\mathcal{H}(\vartheta_1):=\E\int\sqrt{p_{\vartheta_0}p_{\vartheta_1}}(\mathbf{y}\mid\bX)\,d\mathbf{y}$, the Hellinger affinity, which is $<1$ because $\vartheta_1\neq\vartheta_0$ and $\bSig_{XX}\succ0$. Likewise, for $\theta$ with reduced form $\vartheta$, $\Prob_\theta(1-\varphi_{n,\vartheta_1})\le g(\vartheta;\vartheta_1)^n$ with $g(\vartheta;\vartheta_1):=\E\int p_{\vartheta_0}^{1/2}p_{\vartheta_1}^{-1/2}p_\vartheta(\mathbf{y}\mid\bX)\,d\mathbf{y}$ and $g(\vartheta_1;\vartheta_1)=\mathcal{H}(\vartheta_1)$. The inner integral is a Gaussian integral whose quadratic coefficient is $\tfrac12(\symbf{\Omega}_0^{-1}+\symbf{\Omega}_1^{-1})\succ0$ at $\vartheta=\vartheta_1$, so for $\vartheta$ near $\vartheta_1$ it equals $\exp\{q_\vartheta(\bX)\}$ with $q_\vartheta$ quadratic in $\bX$, continuous in $\vartheta$, and $q_{\vartheta_1}\le0$. Dominated convergence with the exponential moment of Assumption~\ref{ass:ch3:design} shows that $g(\cdot;\vartheta_1)$ is continuous at $\vartheta_1$. Hence there is an open $U_{\vartheta_1}\ni\vartheta_1$ on which $g\le\{1+\mathcal{H}(\vartheta_1)\}/2<1$, and $\varphi_{n,\vartheta_1}$ has both error probabilities at most $e^{-c_{\vartheta_1}n}$ on $U_{\vartheta_1}$, $c_{\vartheta_1}:=-\log[\{1+\mathcal{H}(\vartheta_1)\}/2]$. Extract a finite subcover $U_1,\ldots,U_m$ of $\mathcal R_\varepsilon$ and let $\varphi_n:=\max_r\varphi_{n,r}$. Then $\E_{\theta_0}\varphi_n\le\sum_r\E_{\theta_0}\varphi_{n,r}\le me^{-(\min_rc_r)n}$, and for $\theta\in\bar A_\varepsilon$, whose reduced form lies in some $U_r$, $\Prob_\theta(1-\varphi_n)\le\Prob_\theta(1-\varphi_{n,r})\le e^{-c_rn}$. Take $C_\varepsilon=m$ and $c_\varepsilon=\min_rc_r$.
\end{proof}

\begin{lem}[Posterior consistency under changing priors]\label{lem:ch3:changingprior}
Suppose that Assumptions~\ref{ass:ch3:design} and~\ref{ass:ch3:compact} hold, let $\{\mathcal{P}_n\}$ be priors on $\Theta'$, fix $\varepsilon>0$ and let $A_\varepsilon:=\{\theta\in\Theta':\|\theta-\theta_0\|_\infty>\varepsilon\}$. If (i) there are tests $\varphi_n$ with $\E_{\theta_0}[\varphi_n]\le C_\varepsilon e^{-c_\varepsilon n}$ and $\sup_{\theta\in A_\varepsilon}\Prob_\theta(1-\varphi_n)\le C_\varepsilon e^{-c_\varepsilon n}$, and (ii) for every $\kappa>0$, $\mathcal{P}_n(\mathcal K_\kappa)\ge c_\kappa>0$ for all large $n$, then $\mathcal{P}_n(A_\varepsilon\mid \mathcal D_n)\to 0$ in $\Prob_{\theta_0}$-probability.
\end{lem}

\begin{proof}
Let $L_n(\theta):=p_\theta(\mathcal D_n)/p_{\theta_0}(\mathcal D_n)$, $N_n:=\int_{A_\varepsilon}L_n\,d\mathcal{P}_n$ and $Z_n:=\int_{\Theta'}L_n\,d\mathcal{P}_n$, so that $\mathcal{P}_n(A_\varepsilon\mid\mathcal D_n)=N_n/Z_n\le\varphi_n+(1-\varphi_n)N_n/Z_n$. By (i), $\varphi_n\to0$ in probability, and by the change of measure $\E_{\theta_0}[(1-\varphi_n)L_n(\theta)]=\E_\theta[1-\varphi_n]$, $\E_{\theta_0}[(1-\varphi_n)N_n]\le C_\varepsilon e^{-c_\varepsilon n}$, so $(1-\varphi_n)N_n=O_{\Prob}(e^{-c_\varepsilon n/2})$ by Markov's inequality.

For the denominator, fix $\kappa:=c_\varepsilon/8$ and write $\Lambda_n(\theta):=n^{-1}\log L_n(\theta)$. For $\theta\in\mathcal K_\kappa$, $\E_{\theta_0}\Lambda_n(\theta)=-\mathrm{KL}(\theta_0\|\theta)\ge-\kappa$ and $\Var_{\theta_0}\Lambda_n(\theta)\le V/n$ by \eqref{eq:ch3:Vbound}, so by Chebyshev's inequality $\Prob_{\theta_0}\{L_n(\theta)<e^{-2\kappa n}\}\le V/(n\kappa^2)$. Let $B_n:=\{\theta\in\mathcal K_\kappa:L_n(\theta)\ge e^{-2\kappa n}\}$. By Fubini's theorem $\E_{\theta_0}[\mathcal{P}_n(\mathcal K_\kappa\setminus B_n)]\le V/(n\kappa^2)\to0$, so $\mathcal{P}_n(B_n)\ge c_\kappa-o_{\Prob}(1)$ by (ii), and $Z_n\ge e^{-2\kappa n}\mathcal{P}_n(B_n)\ge\tfrac12c_\kappa e^{-c_\varepsilon n/4}$ with probability tending to one. On this event $(1-\varphi_n)N_n/Z_n\le2c_\kappa^{-1}e^{c_\varepsilon n/4}O_{\Prob}(e^{-c_\varepsilon n/2})=o_{\Prob}(1)$.
\end{proof}

\begin{proof}[Proof of Proposition~\ref{prop:ch3:thetagen}]
Lemma~\ref{lem:ch3:tests} gives the tests of condition (i) and Lemma~\ref{lem:ch3:klmass} the prior mass of condition (ii) in Lemma~\ref{lem:ch3:changingprior}, which yields $\mathcal{P}_n(\|\theta-\theta_0\|_\infty>\varepsilon\mid\mathcal D_n)\to0$ in $\Prob_{\theta_0}$-probability.
\end{proof}

\begin{lem}[Hellinger bounds]\label{lem:ch3:hellinger}
Let $h(\theta,\theta')$ be the Hellinger distance between the joint laws of $(\bY,\bX)$ under $\theta$ and $\theta'$, so that $h^2(\theta,\theta')=1-\E_{\bX}\{\rho_{\bX}(\theta,\theta')\}$ with $\rho_{\bX}$ the Hellinger affinity of $\Normal(\bPi\bX,\symbf{\Omega})$ and $\Normal(\bPi'\bX,\symbf{\Omega}')$. Under Assumptions~\ref{ass:ch3:design} and~\ref{ass:ch3:compact} there are constants $0<c_h\le C_h<\infty$ such that, for all $\theta,\theta'\in\Theta_M$,
$h(\theta,\theta_0)\ge c_h\min\{\|\bPi(\theta)-\bPi_0\|_F+\|\symbf{\Omega}(\theta)-\symbf{\Omega}_0\|_F,1\}$ and $h(\theta,\theta')\le C_h\|\theta-\theta'\|_\infty$.
\end{lem}

\begin{proof}
\textit{Lower bound.} For two Gaussians, $\rho_{\bX}=\rho_{\mathrm{cov}}\exp\{-q(\bX)\}$ with $\rho_{\mathrm{cov}}=\det(\symbf{\Omega})^{1/4}\det(\symbf{\Omega}_0)^{1/4}/\det(\bar{\symbf{\Omega}})^{1/2}\le1$, $q(\bX)=\tfrac18\bX^{\T}(\bPi-\bPi_0)^{\T}\bar{\symbf{\Omega}}^{-1}(\bPi-\bPi_0)\bX\ge0$ and $\bar{\symbf{\Omega}}:=\tfrac12(\symbf{\Omega}+\symbf{\Omega}_0)$, so $h^2\ge\max\{1-\rho_{\mathrm{cov}},1-\E e^{-q(\bX)}\}$. Writing $\symbf{\Omega}=\symbf{\Omega}_0+\mathbf{G}$ and expanding $\log\det$, $-\log\rho_{\mathrm{cov}}=\tfrac1{16}\tr\{(\symbf{\Omega}_0^{-1}\mathbf{G})^2\}+O(\|\mathbf{G}\|_F^3)\ge c\|\mathbf{G}\|_F^2$ for $\|\mathbf{G}\|_F\le\delta_1$, with $c,\delta_1$ depending only on $[\underline{c}_\Omega,\overline{c}_\Omega]$. For the mean part, $1-e^{-q}\ge q-q^2/2$ gives $1-\E e^{-q(\bX)}\ge\E q(\bX)-\tfrac12\E q(\bX)^2$, with $\E q(\bX)\ge(8\overline{c}_\Omega)^{-1}\lambda_{\min}(\bSig_{XX})\|\bPi-\bPi_0\|_F^2$ and $\E q(\bX)^2\le C\|\bPi-\bPi_0\|_F^4\E\|\bX\|_2^4$, so $1-\E e^{-q(\bX)}\ge c'\|\bPi-\bPi_0\|_F^2$ for $\|\bPi-\bPi_0\|_F\le\delta_2$. Hence $h^2(\theta,\theta_0)\ge c''(\|\bPi-\bPi_0\|_F+\|\symbf{\Omega}-\symbf{\Omega}_0\|_F)^2$ when both deviations are at most $\min(\delta_1,\delta_2)$. On the rest of $\Theta_M$, which is compact, $h^2(\theta,\theta_0)$ is continuous and positive ($h=0$ forces the reduced form of $\theta_0$), so it is bounded below by a positive constant. \textit{Upper bound.} Since $-2\log\rho_{\bX}\le\mathrm{KL}\{\Normal(\bPi'\bX,\symbf{\Omega}')\|\Normal(\bPi\bX,\symbf{\Omega})\}$ pointwise and $1-\rho\le-\log\rho$, averaging over $\bX$ gives $h^2(\theta,\theta')\le\tfrac12\mathrm{KL}(\theta'\|\theta)$ with the KL divergence of Lemma~\ref{lem:ch3:rfkl}. Its first term is the Bregman divergence of $-\log\det$ between $\symbf{\Omega}'$ and $\symbf{\Omega}$, whose Hessian is bounded on $[\underline{c}_\Omega,\overline{c}_\Omega]$, so it is at most $C\|\symbf{\Omega}-\symbf{\Omega}'\|_F^2$, and the second term is at most $C\|\bPi-\bPi'\|_F^2$. The reduced-form map is continuously differentiable on a neighborhood of the compact set $\Theta_M$ and is therefore Lipschitz on $\Theta_M$, which completes the proof.
\end{proof}

\begin{lem}[Local contraction at the rate $n^{-1/2}\log n$]\label{lem:ch3:localrate}
Let $\varepsilon_n:=n^{-1/2}\log n$. Under Assumptions~\ref{ass:ch3:design}--\ref{ass:ch3:priorpos}, \ref{ass:ch3:asympident} and~\ref{ass:ch3:localbilip}, there is a constant $K<\infty$ such that $\mathcal{P}_n(\|\theta-\theta_0\|_\infty>K\varepsilon_n\mid\mathcal D_n)\to 0$ in $\Prob_{\theta_0}$-probability.
\end{lem}

\begin{proof}
\textit{Tests.} By the lower bound of Lemma~\ref{lem:ch3:hellinger}, which bounds the Hellinger distance below by the distance of the reduced forms, together with Assumption~\ref{ass:ch3:localbilip} near $\theta_0$ and Lemma~\ref{lem:ch3:sep} away from it, $h(\theta,\theta_0)\ge c\min\{\|\theta-\theta_0\|_\infty,1\}$ on $\Theta'$. The alternative $\{\|\theta-\theta_0\|_\infty>K\varepsilon_n\}$ is therefore at Hellinger distance at least $cK\varepsilon_n$ from $\theta_0$. By the upper bound of Lemma~\ref{lem:ch3:hellinger} a $\|\cdot\|_\infty$-cover of $\Theta'$ at scale $\varepsilon_n/C_h$ is a Hellinger cover at scale $\varepsilon_n$, with $\log N(\varepsilon_n/C_h,\Theta',\|\cdot\|_\infty)=O(\log n)=o(n\varepsilon_n^2)$ because the model is finite-dimensional. The test construction of \citet[Theorem~7.1]{Ghosal2000}, which combines likelihood-ratio tests over such a cover (its covering factor $\exp\{O(\log n)\}$ is negligible for large $n$, since $\log n=o(\log^2 n)$), then gives tests $\varphi_n$ with $\E_{\theta_0}\varphi_n\le\exp(-c_1K^2n\varepsilon_n^2)$ and $\sup_{\|\theta-\theta_0\|_\infty>K\varepsilon_n}\Prob_\theta(1-\varphi_n)\le\exp(-c_1K^2n\varepsilon_n^2)$ for all large $K$, where $n\varepsilon_n^2=\log^2n$. The tests do not involve the prior.

\textit{Prior mass.} Each truly nonzero entry of $\bA$ and each free entry of $\bB$ has a slab density bounded below near its true value, so a ball of radius $c\varepsilon_n$ around it has prior mass of order $\varepsilon_n$. Each truly zero entry of $\bA$ has the spike $\Normal(0,\nu_{1,n}t)$, $t\in[\underline t,\overline t]$, so for $Z\sim\Normal(0,\nu_{1,n}t)$, $\Prob(|Z|\le c\varepsilon_n)\ge c'\min\{1,\varepsilon_n/\sqrt{\nu_{1,n}}\}\ge c'\varepsilon_n$ because $\nu_{1,n}<1$. Also, by the density condition of Assumption~\ref{ass:ch3:sigmaprior} the prior on $\bSigs$ gives a ball of radius $c\varepsilon_n$ mass at least a positive constant times $\varepsilon_n^{p(p+1)/2}$. With the constant factor of the indicator pattern, as in Lemma~\ref{lem:ch3:klmass},
\begin{equation}\label{eq:ch3:localpriormass}
\mathcal{P}_n\{\|\theta-\theta_0\|_\infty\le c_2\varepsilon_n\}\ge \exp(-C_2\log n).
\end{equation}

\textit{Denominator.} The log-likelihood ratio $\mathrm{LLR}_\theta:=\log\{p_{\theta_0}(\bY\mid\bX)/p_\theta(\bY\mid\bX)\}$ is a quadratic polynomial in $(\bY,\bX)$ whose coefficients are continuously differentiable in $\theta$ and vanish at $\theta_0$, so, by the fourth-moment bound, $\Var_{\theta_0}(\mathrm{LLR}_\theta)\le C\|\theta-\theta_0\|_\infty^2$ and $\mathrm{KL}(\theta_0\|\theta)\le C\|\theta-\theta_0\|_\infty^2$ near $\theta_0$. The ball in \eqref{eq:ch3:localpriormass} is thus a KL--variance neighborhood with both radii of order $\varepsilon_n^2$. Let $B_n$ denote this ball and let $Q_n=\mathcal P_n(\cdot\mid B_n)$ be the prior normalized on it. Uniformly over $\theta\in B_n$, Chebyshev's inequality gives probability at most $C/(n\varepsilon_n^2)$ that $\ell_n(\theta)-\ell_n(\theta_0)<-C'n\varepsilon_n^2$ for a sufficiently large constant $C'$. By Fubini's theorem, the expected $Q_n$-mass of these parameters is at most $C/(n\varepsilon_n^2)=o(1)$. With probability tending to one, at least half of $\mathcal P_n(B_n)$ therefore has likelihood ratio at least $\exp(-C'n\varepsilon_n^2)$. Applying \eqref{eq:ch3:localpriormass} gives
\begin{equation}\label{eq:ch3:localdenom}
Z_n:=\int_{\Theta'}\frac{p_\theta(\mathcal D_n)}{p_{\theta_0}(\mathcal D_n)}\,\mathcal{P}_n(d\theta)\ \ge\ \exp(-C_3\log^2 n-C_2\log n).
\end{equation}

\textit{Conclusion.} The tests give $(1-\varphi_n)\int_{\{\|\theta-\theta_0\|_\infty>K\varepsilon_n\}}\{p_\theta(\mathcal D_n)/p_{\theta_0}(\mathcal D_n)\}\,\mathcal{P}_n(d\theta)=O_{\Prob}\{\exp(-c_1K^2\log^2n)\}$, so $\mathcal{P}_n(\|\theta-\theta_0\|_\infty>K\varepsilon_n\mid\mathcal D_n)\le\varphi_n+O_{\Prob}[\exp\{-(c_1K^2-C_3)\log^2n+C_2\log n\}]$, which tends to zero once $c_1K^2>C_3$.
\end{proof}

Proposition~\ref{prop:ch3:thetagen} alone does not suffice for indicator recovery under an $n$-dependent variance ratio. The Bayes-factor argument of the next section needs the nuisance parameters in a shrinking neighborhood of their true values, which Lemma~\ref{lem:ch3:localrate} provides.

\subsection{Consistency of the Edge Indicators}
\label{sec:ch3:app-gammacons}

Fix an off-diagonal pair $(j,h)$ and write $\theta_{-(jh)}$ for $\theta$ with the coordinate $a_{jh}$ removed. Under the product prior the block $(a_{jh},\gamma_{jh},\rho_{jh},\tau_{jh})$ is independent of all other coordinates and hyperparameters, and the restriction to $\Theta'$ couples it to them only through the constraint $(a_{jh},\theta_{-(jh)})\in\Theta'$. Since the likelihood depends on the parameters only through $\theta$, the conditional posterior of $(a_{jh},\gamma_{jh})$ given $(\mathcal D_n,\theta_{-(jh)},\tau_{jh})$ does not depend on the other indicators, and $\rho_{jh}$ can be integrated out. In this section $\ell_n(a,\theta_{-(jh)})$ denotes the log-likelihood as a function of $a_{jh}=a$ with the nuisance block fixed, so that only the scalar score and curvature of Lemma~\ref{lem:ch3:onecoord} are needed. The conditional priors are $a_{jh}\sim\Normal(0,\tau_{jh})$ if $\gamma_{jh}=1$ and $\Normal(0,\nu_{1,n}\tau_{jh})$ if $\gamma_{jh}=0$, and the marginal likelihoods are taken over the admissible slice $\mathcal A_{jh}(\theta_{-(jh)}):=\{a\in\R:(a,\theta_{-(jh)})\in\Theta'\}$. Let $\varepsilon_n:=n^{-1/2}\log n$, let $K$ be the constant of Lemma~\ref{lem:ch3:localrate}, and let
\begin{equation*}
\mathcal N_n:=\{\theta_{-(jh)}:\|\theta_{-(jh)}-\theta_{0,-(jh)}\|_\infty\le K\varepsilon_n\}.
\end{equation*}
Since $\theta_0$ is interior to $\Theta_M$, there is $\delta_I>0$ such that, for all large $n$ and all $\theta_{-(jh)}\in\mathcal N_n$, the interval $I_0:=\{a:|a-a^0_{jh}|\le\delta_I\}$ lies in $\mathcal A_{jh}(\theta_{-(jh)})$. When the true graph contains an edge, let $\beta_{\min}:=\min\{|a^0_{jh}|:\gamma^0_{jh}=1\}$. This minimum is positive because $p$ is fixed.

\begin{lem}[Uniform separation away from the true coordinate]\label{lem:ch3:onecoordsep}
Under Assumptions~\ref{ass:ch3:design}, \ref{ass:ch3:compact} and~\ref{ass:ch3:asympident}, for any fixed $K$ in the definition of $\mathcal N_n$ and every $\delta>0$ there is $c_\delta>0$ such that, with $\Prob_{\theta_0}$-probability tending to one,
\begin{equation}\label{eq:ch3:onecoordsep}
\sup_{\theta_{-(jh)}\in\mathcal N_n}\ \sup_{a\in\mathcal A_{jh}(\theta_{-(jh)}):\ |a-a^0_{jh}|\ge \delta}\ \frac1n\big\{\ell_n(a,\theta_{-(jh)})-\ell_n(a^0_{jh},\theta_{-(jh)})\big\}\le -c_\delta .
\end{equation}
\end{lem}

\begin{proof}
By \eqref{eq:ch3:loglik}, $n^{-1}\ell_n(\theta)$ is linear in the sample moments $(\bS_{yy},\bS_{yx},\bS_{xx})$ with coefficients bounded uniformly on $\Theta_M$, so by the law of large numbers $\sup_{\theta\in\Theta_M}|n^{-1}\ell_n(\theta)-\bar\ell(\theta)|\to0$ in probability, where $\bar\ell$ is the same expression with the moments replaced by their expectations, and $\bar\ell(\theta)-\bar\ell(\theta_0)=-\mathrm{KL}(\theta_0\|\theta)$. By Lemma~\ref{lem:ch3:rfkl}, Assumption~\ref{ass:ch3:asympident} and compactness, $\zeta_\delta:=\inf\{\mathrm{KL}(\theta_0\|\theta):\theta\in\Theta',\|\theta-\theta_0\|_\infty\ge\delta\}>0$. If $|a-a^0_{jh}|\ge\delta$ then $\mathrm{KL}\{\theta_0\|(a,\theta_{-(jh)})\}\ge\zeta_\delta$, while $\sup_{\mathcal N_n}\mathrm{KL}\{\theta_0\|(a^0_{jh},\theta_{-(jh)})\}\le\zeta_\delta/3$ for large $n$ by uniform continuity, since $K\varepsilon_n\to0$. Hence the difference of the $\bar\ell$ values is at most $-\tfrac23\zeta_\delta$, and \eqref{eq:ch3:onecoordsep} holds with $c_\delta:=\zeta_\delta/3$.
\end{proof}

\begin{lem}[Localization and local quadratic expansion]\label{lem:ch3:localquad}
Under Assumptions~\ref{ass:ch3:design}, \ref{ass:ch3:compact} and~\ref{ass:ch3:asympident}, for any fixed $K$ in the definition of $\mathcal N_n$, let $\widehat a_n(\theta_{-(jh)}):=\argmax_{a\in\mathcal A_{jh}(\theta_{-(jh)})}\ell_n(a,\theta_{-(jh)})$. With probability tending to one, the following hold uniformly over $\theta_{-(jh)}\in\mathcal N_n$. (i) The maximizer $\widehat a_n$ is unique and $\sup_{\mathcal N_n}|\widehat a_n-a^0_{jh}|=O_{\Prob}(\varepsilon_n)$. (ii) The curvature $J_n:=-\partial^2\ell_n(a,\theta_{-(jh)})/\partial a^2|_{a=\widehat a_n}$ satisfies $c_Jn\le J_n\le C_Jn$ for constants $0<c_J<C_J<\infty$, and $-\ell_n''(a,\theta_{-(jh)})\ge c_Jn$ for every admissible $a$. (iii) Uniformly for $|a-\widehat a_n|\le\varepsilon_n$, $\ell_n(a,\theta_{-(jh)})=\ell_n(\widehat a_n,\theta_{-(jh)})-\tfrac12J_n(a-\widehat a_n)^2+o_{\Prob}(1)$.
\end{lem}

\begin{proof}
By \eqref{eq:ch3:scoreA}, $\ell_n''(a,\theta_{-(jh)})=-n\{(\mathbf{M}(a)^{-1})_{hj}\}^2-\{(\bSigs)^{-1}\}_{jj}\sum_iY_{ih}^2$, where $\mathbf{M}(a)$ is $\bI_p-\bA$ with $a_{jh}=a$. The first term is nonpositive, $\{(\bSigs)^{-1}\}_{jj}\ge\overline{c}_\Sigma^{-1}$ on $\Theta_M$, and $n^{-1}\sum_iY_{ih}^2\to\E_{\theta_0}(Y_h^2)>0$, so on an event of probability tending to one $\ell_n''\le-c_Jn$ with $c_J:=\E_{\theta_0}(Y_h^2)/(2\overline{c}_\Sigma)$ for every admissible $(a,\theta_{-(jh)})$. The upper bound $C_Jn$ follows from the bounds on $\mathbf{M}(a)^{-1}$ and $\bSigs$ and $n^{-1}\sum_iY_{ih}^2=O_{\Prob}(1)$. This proves (ii) and shows that $\ell_n(\cdot,\theta_{-(jh)})$ is strictly concave on every interval of the slice.

For (i), the score at $(a^0_{jh},\theta_{0,-(jh)})$ is $O_{\Prob}(\sqrt n)$ by Lemma~\ref{lem:ch3:lan}. For $\theta_{-(jh)}\in\mathcal N_n$, with $\mathbf{P}:=(\bSigs)^{-1}$ and $\mathbf{P}_0:=(\bSig_0^*)^{-1}$, \eqref{eq:ch3:scoreA} gives
\begin{multline*}
\ell_n'(a^0_{jh},\theta_{-(jh)})-\ell_n'(a^0_{jh},\theta_{0,-(jh)})
=-n\{(\mathbf{M}^{-1})_{hj}-(\mathbf{M}_0^{-1})_{hj}\}\\
+\sum_{i=1}^nY_{ih}\mathbf{e}_j^{\T}\big[(\mathbf{P}-\mathbf{P}_0)\bE_i^*+\mathbf{P}\{(\mathbf{M}-\mathbf{M}_0)\bY_i-(\bB-\bB_0)\bX_i\}\big].
\end{multline*}
On $\mathcal N_n$, $\|\mathbf{P}-\mathbf{P}_0\|+\|\mathbf{M}-\mathbf{M}_0\|+\|\bB-\bB_0\|+|(\mathbf{M}^{-1})_{hj}-(\mathbf{M}_0^{-1})_{hj}|\le C\varepsilon_n$ by smoothness on $\Theta_M$, so the right side is bounded by $Cn\varepsilon_n+C\varepsilon_n\sum_i|Y_{ih}|(\|\bE_i^*\|+\|\bY_i\|+\|\bX_i\|)=O_{\Prob}(n\varepsilon_n)$ by the law of large numbers and the fourth-moment bound. Hence $\sup_{\mathcal N_n}|\ell_n'(a^0_{jh},\theta_{-(jh)})|=O_{\Prob}(n\varepsilon_n)$. On $I_0$, $\ell_n''\le-c_Jn$, so by the mean value theorem $\ell_n'(a,\theta_{-(jh)})$ has the sign of $a^0_{jh}-a$ whenever $|a-a^0_{jh}|>|\ell_n'(a^0_{jh},\theta_{-(jh)})|/(c_Jn)$: the maximizer over $I_0$ is unique, lies within $O_{\Prob}(\varepsilon_n)$ of $a^0_{jh}$ uniformly over $\mathcal N_n$, and is interior to $I_0$ with probability tending to one, so $\ell_n'(\widehat a_n,\theta_{-(jh)})=0$. Outside $I_0$, Lemma~\ref{lem:ch3:onecoordsep} with $\delta=\delta_I$ gives $\ell_n(a,\theta_{-(jh)})\le\ell_n(a^0_{jh},\theta_{-(jh)})-c_{\delta_I}n$, below the maximum over $I_0$. Hence the maximizer over $I_0$ is the global maximizer.

For (iii), the quadratic part of $\ell_n$ is exactly quadratic in $a$ and the log-determinant part has third derivative $-2n\{(\mathbf{M}(a)^{-1})_{hj}\}^3$, bounded by $Cn$ on $\Theta_M$. Taylor's formula around $\widehat a_n$, where the first derivative vanishes, leaves a remainder of at most $Cn\varepsilon_n^3=Cn^{-1/2}\log^3n=o(1)$.
\end{proof}

For $g\in\{0,1\}$ let $v_1:=\tau_{jh}$, $v_0:=\nu_{1,n}\tau_{jh}$ and
\begin{equation*}
m_{g,n}(\theta_{-(jh)},\tau_{jh}):=\int_{\mathcal A_{jh}(\theta_{-(jh)})}\exp\{\ell_n(a,\theta_{-(jh)})\}\,\Normal(a\mid 0,v_g)\,da,
\end{equation*}
where $\Normal(a\mid0,v)$ is the $\Normal(0,v)$ density. Completing the square, for $J>0$, $v>0$ and $\widehat a\in\R$,
\begin{equation}\label{eq:ch3:completedsquare}
\int_{\R}\exp\{-\tfrac12J(a-\widehat a)^2\}\Normal(a\mid 0,v)\,da=\frac{1}{\sqrt{1+Jv}}\exp\Big(-\frac{\widehat a^2}{2(v+J^{-1})}\Big),
\end{equation}
and the integrand is proportional to the $\Normal(\mu,s^2)$ density with $s^2=(J+v^{-1})^{-1}$ and $|\mu-\widehat a|=|\widehat a|/(1+Jv)$.

\begin{lem}[Component marginal likelihoods]\label{lem:ch3:marginals}
Under Assumptions~\ref{ass:ch3:design}--\ref{ass:ch3:nurate}, \ref{ass:ch3:asympident} and~\ref{ass:ch3:localbilip}, uniformly over $\theta_{-(jh)}\in\mathcal N_n$ and $\tau_{jh}\in[\underline t,\overline t]$, with $J_n$ and $\widehat a_n$ as in Lemma~\ref{lem:ch3:localquad}:
\begin{enumerate}
\item For $g=1$, and also for $g=0$ when $\gamma^0_{jh}=0$,
\begin{equation}\label{eq:ch3:marginalclosed}
m_{g,n}=\exp\{\ell_n(\widehat a_n,\theta_{-(jh)})\}\frac{1}{\sqrt{1+J_n v_g}}\exp\Big(-\frac{\widehat a_n^2}{2(v_g+J_n^{-1})}\Big)\{1+o_{\Prob}(1)\}.
\end{equation}
\item Whatever the value of $\gamma^0_{jh}$, for some $c>0$ and with probability tending to one,
\begin{equation}\label{eq:ch3:marginalupper}
m_{0,n}\le\exp\{\ell_n(\widehat a_n,\theta_{-(jh)})\}\Big\{\frac{1}{\sqrt{1+c_Jnv_0}}\exp\Big(-\frac{\widehat a_n^2}{2\{v_0+(c_Jn)^{-1}\}}\Big)+e^{-cn}\Big\}.
\end{equation}
\end{enumerate}
\end{lem}

\begin{proof}
Work on the event, of probability tending to one, on which Lemmas~\ref{lem:ch3:onecoordsep} (with $\delta=\delta_I/4$) and~\ref{lem:ch3:localquad} hold and $\sup_{\mathcal N_n}|\widehat a_n-a^0_{jh}|\le\delta_I/4$, so that $\{|a-\widehat a_n|\le\delta_I/2\}\subseteq I_0$ and every admissible $a$ with $|a-\widehat a_n|>\delta_I/2$ has $|a-a^0_{jh}|>\delta_I/4$. Write $\ell_n(a)$ for $\ell_n(a,\theta_{-(jh)})$.

\textit{Part 2.} For $|a-\widehat a_n|\le\delta_I/2$, concavity with $\ell_n''\le-c_Jn$ and $\ell_n'(\widehat a_n)=0$ give $\ell_n(a)\le\ell_n(\widehat a_n)-\tfrac12c_Jn(a-\widehat a_n)^2$, so this region contributes at most the first term of \eqref{eq:ch3:marginalupper} by \eqref{eq:ch3:completedsquare} with $J=c_Jn$. On the rest of the slice, Lemma~\ref{lem:ch3:onecoordsep} gives $\ell_n(a)\le\ell_n(a^0_{jh})-cn\le\ell_n(\widehat a_n)-cn$ with $c:=c_{\delta_I/4}$, and the prior density integrates to at most one.

\textit{Part 1.} Split the slice into $\mathcal I_n:=\{|a-\widehat a_n|\le\varepsilon_n\}$, the near region $\{\varepsilon_n<|a-\widehat a_n|\le\delta_I/2\}$ and the far region. On $\mathcal I_n$, part (iii) of Lemma~\ref{lem:ch3:localquad} shows that the contribution equals $\exp\{\ell_n(\widehat a_n)\}\{1+o_{\Prob}(1)\}\int_{\mathcal I_n}\exp\{-\tfrac12J_n(a-\widehat a_n)^2\}\Normal(a\mid0,v_g)\,da$. The integrand is proportional to a $\Normal(\mu_g,s_g^2)$ density with $s_g^2\le J_n^{-1}\le(c_Jn)^{-1}$ and $|\mu_g-\widehat a_n|=|\widehat a_n|/(1+J_nv_g)$. For $g=1$, $|\widehat a_n|\le C_A$ and $J_nv_1\ge c_J\underline t\,n$, so $|\mu_1-\widehat a_n|=O(n^{-1})=o(\varepsilon_n)$. For $g=0$ with $\gamma^0_{jh}=0$, $|\widehat a_n|=O_{\Prob}(\varepsilon_n)$ and $J_nv_0\ge c_J\underline t\,n\nu_{1,n}\to\infty$ by Assumption~\ref{ass:ch3:nurate}, so again $|\mu_0-\widehat a_n|=o_{\Prob}(\varepsilon_n)$. In both cases $(\varepsilon_n-|\mu_g-\widehat a_n|)/s_g\ge\tfrac12c_J^{1/2}\log n$ for large $n$, so the $\Normal(\mu_g,s_g^2)$ mass outside $\mathcal I_n$ is at most $2\exp(-c_J\log^2n/8)=o(1)$, and restricting \eqref{eq:ch3:completedsquare} to $\mathcal I_n$ changes it by a relative $o(1)$ term: the local contribution equals the right side of \eqref{eq:ch3:marginalclosed}. The leading term $L_g:=\exp\{\ell_n(\widehat a_n)\}(1+J_nv_g)^{-1/2}\exp\{-\widehat a_n^2/(2(v_g+J_n^{-1}))\}$ satisfies $L_g\ge\exp\{\ell_n(\widehat a_n)\}n^{-1}$ with probability tending to one, because $J_nv_g\le C_J\overline t\,n$ and the exponent is at most $C_A^2/(2\underline t)$ for $g=1$ and at most $\widehat a_n^2/(2v_0)=O_{\Prob}\{\log^2n/(n\nu_{1,n})\}=o_{\Prob}(\log n)$ for $g=0$ with $\gamma^0_{jh}=0$ (Assumption~\ref{ass:ch3:nurate}). The near region contributes at most $\exp\{\ell_n(\widehat a_n)-\tfrac12c_Jn\varepsilon_n^2\}=\exp\{\ell_n(\widehat a_n)-\tfrac12c_J\log^2n\}=o(L_g)$, as in Part~2, and the far region at most $\exp\{\ell_n(\widehat a_n)-cn\}=o(L_g)$.
\end{proof}

Under the Beta--Bernoulli prior the prior odds of $\gamma_{jh}$ are $a_\rho/b_\rho$, so after integrating out $\rho_{jh}$,
\begin{equation}\label{eq:ch3:postodds}
\frac{\mathcal{P}_n(\gamma_{jh}=1\mid \mathcal D_n,\theta_{-(jh)},\tau_{jh})}{\mathcal{P}_n(\gamma_{jh}=0\mid \mathcal D_n,\theta_{-(jh)},\tau_{jh})}=\frac{a_\rho}{b_\rho}\cdot\frac{m_{1,n}}{m_{0,n}}.
\end{equation}

\begin{lem}[Bayes factor of a causal edge]\label{lem:ch3:bfA}
Under Assumptions~\ref{ass:ch3:design}--\ref{ass:ch3:nurate}, \ref{ass:ch3:asympident} and~\ref{ass:ch3:localbilip}, in $\Prob_{\theta_0}$-probability,
\begin{equation*}
\inf_{\theta_{-(jh)}\in\mathcal N_n,\,\tau_{jh}\in[\underline t,\overline t]}\frac{m_{1,n}}{m_{0,n}}\to\infty\ \text{ if }\gamma^0_{jh}=1,\qquad
\sup_{\theta_{-(jh)}\in\mathcal N_n,\,\tau_{jh}\in[\underline t,\overline t]}\frac{m_{1,n}}{m_{0,n}}\to0\ \text{ if }\gamma^0_{jh}=0,
\end{equation*}
and, if $\gamma^0_{jh}=0$, uniformly over the same set,
\begin{equation}\label{eq:ch3:BFlog}
\log\frac{m_{1,n}}{m_{0,n}}=-\frac12\log\frac{1+J_n\tau_{jh}}{1+J_n\nu_{1,n}\tau_{jh}}-\frac{\widehat a_n^2}{2}\Big(\frac{1}{\tau_{jh}+J_n^{-1}}-\frac{1}{\nu_{1,n}\tau_{jh}+J_n^{-1}}\Big)+o_{\Prob}(1).
\end{equation}
\end{lem}

\begin{proof}
If $\gamma^0_{jh}=1$, then $|a^0_{jh}|\ge\beta_{\min}$ and Lemma~\ref{lem:ch3:localquad} gives $\inf_{\mathcal N_n}|\widehat a_n|\ge\beta_{\min}/2$ with probability tending to one. Part~1 of Lemma~\ref{lem:ch3:marginals} with $g=1$, $J_n\le C_Jn$ and $|\widehat a_n|\le C_A$ gives $m_{1,n}\ge\tfrac12\exp\{\ell_n(\widehat a_n)\}(1+C_J\overline t\,n)^{-1/2}\exp\{-C_A^2/(2\underline t)\}$, and part~2 with $v_0\in[\nu_{1,n}\underline t,\nu_{1,n}\overline t]$ and $\nu_{1,n}\overline t\ge(c_Jn)^{-1}$ for large $n$ gives $m_{0,n}\le\exp\{\ell_n(\widehat a_n)\}[(1+c_J\underline t\,n\nu_{1,n})^{-1/2}\exp\{-\beta_{\min}^2/(16\overline t\nu_{1,n})\}+e^{-cn}]$. Since $m_{1,n}/(x+y)\ge\tfrac12\min(m_{1,n}/x,m_{1,n}/y)$ and $(1+c_J\underline t\,n\nu_{1,n})/(1+C_J\overline t\,n)\ge c'\nu_{1,n}$ for large $n$,
\begin{equation*}
\frac{m_{1,n}}{m_{0,n}}\ge c''\min\Big\{\nu_{1,n}^{1/2}\exp\Big(\frac{\beta_{\min}^2}{16\overline t\,\nu_{1,n}}\Big),\ n^{-1/2}e^{cn}\Big\}\to\infty,
\end{equation*}
uniformly, because $\tfrac12\log\nu_{1,n}+\beta_{\min}^2/(16\overline t\nu_{1,n})\to\infty$ as $\nu_{1,n}\to0$.

If $\gamma^0_{jh}=0$, part~1 of Lemma~\ref{lem:ch3:marginals} applies to both $g$, and the ratio of the two versions of \eqref{eq:ch3:marginalclosed} gives \eqref{eq:ch3:BFlog}. Since $a^0_{jh}=0$, Lemma~\ref{lem:ch3:localquad} gives $\sup_{\mathcal N_n}\widehat a_n^2=O_{\Prob}(n^{-1}\log^2n)$, so the second term of \eqref{eq:ch3:BFlog} is nonnegative and at most $\widehat a_n^2/(2\nu_{1,n}\tau_{jh})=O_{\Prob}\{\log^2n/(n\nu_{1,n})\}$. Also, $J_n\nu_{1,n}\tau_{jh}\to\infty$ uniformly, so the first term equals $-\tfrac12\log(1/\nu_{1,n})+o(1)$. By Assumption~\ref{ass:ch3:nurate}, $\log^2n/(n\nu_{1,n})=o\{\log(1/\nu_{1,n})\}$ and $\log(1/\nu_{1,n})\to\infty$, so $\log(m_{1,n}/m_{0,n})\to-\infty$ uniformly.
\end{proof}

\begin{theorem}[Posterior consistency of a single edge indicator]\label{thm:ch3:gammacons}
Under Assumptions~\ref{ass:ch3:design}--\ref{ass:ch3:nurate}, \ref{ass:ch3:asympident} and~\ref{ass:ch3:localbilip}, for each off-diagonal $(j,h)$, $\mathcal{P}_n(\gamma_{jh}=\gamma^0_{jh}\mid\mathcal D_n)\to1$ in $\Prob_{\theta_0}$-probability.
\end{theorem}

\begin{proof}
Let $G_n:=\{\theta_{-(jh)}\in\mathcal N_n,\ \tau_{jh}\in[\underline t,\overline t]\}$. By Lemma~\ref{lem:ch3:localrate} and the truncation of the slab scales, $\mathcal{P}_n(G_n^c\mid\mathcal D_n)\to0$ in probability. By \eqref{eq:ch3:postodds} and Lemma~\ref{lem:ch3:bfA}, the conditional probability $\mathcal{P}_n(\gamma_{jh}=1\mid\mathcal D_n,\theta_{-(jh)},\tau_{jh})=(a_\rho/b_\rho)(m_{1,n}/m_{0,n})/\{1+(a_\rho/b_\rho)(m_{1,n}/m_{0,n})\}$ converges to $\gamma^0_{jh}$ in probability, uniformly over $G_n$. Integrating this conditional probability over the posterior of $(\theta_{-(jh)},\tau_{jh})$ gives the marginal probability $\mathcal{P}_n(\gamma_{jh}=1\mid\mathcal D_n)$. The contribution of $G_n^c$ is at most $\mathcal{P}_n(G_n^c\mid\mathcal D_n)=o_{\Prob}(1)$ and that of $G_n$ has the stated limit.
\end{proof}

\begin{proof}[Proof of Proposition~\ref{prop:ch3:indgen}]
Since $p$ is fixed, the union bound and Theorem~\ref{thm:ch3:gammacons} give $\mathcal{P}_n(\exists\,j\neq h:\gamma_{jh}\neq\gamma^0_{jh}\mid\mathcal D_n)\le\sum_{j\neq h}\mathcal{P}_n(\gamma_{jh}\neq\gamma^0_{jh}\mid\mathcal D_n)\to0$ in $\Prob_{\theta_0}$-probability.
\end{proof}

\subsection{The Special Case of \texorpdfstring{\MRRGM{}}{IV-RGM}}
\label{sec:ch3:app-rgmcons}

In \MRRGM{} the parameter set is $\Theta_M^{\symbf{\Delta}}$ and the prior is that of Section~\ref{sec:ch3:app-asymp-priors}, with the slab prior $b_{jh}\mid\eta_{jh}\sim\Normal(0,\eta_{jh})$ (truncated half-Cauchy prior on $\sqrt{\eta_{jh}}$) on each free entry of $\bB$.

\begin{proof}[Proof of Theorems~\ref{thm:ch3:thetacons} and~\ref{thm:ch3:indicators}]
The lemmas of Sections~\ref{sec:ch3:app-asymp-likelihood}--\ref{sec:ch3:app-gammacons} are stated on $\Theta'$. We check their hypotheses and the prior-mass factors for $\Theta'=\Theta_M^{\symbf{\Delta}}$ and this prior. \textit{Parameter set.} $\Theta_M^{\symbf{\Delta}}$ is closed in $\Theta_M$, hence compact. Since $\bB_0$ satisfies the exclusions and $\theta_0$ is interior to $\Theta_M$, every small ball around $\theta_0$ in the free coordinates lies in $\Theta_M^{\symbf{\Delta}}$. Assumptions~\ref{ass:ch3:design} and~\ref{ass:ch3:compact} continue to hold, and Assumptions~\ref{ass:ch3:asympident} and~\ref{ass:ch3:localbilip} hold by Lemma~\ref{lem:ch3:bilip}. \textit{Continuous parameter.} Lemmas~\ref{lem:ch3:sep} and~\ref{lem:ch3:tests} use only compactness, continuity and Assumption~\ref{ass:ch3:asympident}. In Lemma~\ref{lem:ch3:klmass}, only the pattern of $\bA$ contributes the constant $c_\gamma$, and each free entry of $\bB$ has the slab prior. The bound \eqref{eq:ch3:Vbound} holds because the supremum is over a smaller set. Lemma~\ref{lem:ch3:changingprior} then gives Proposition~\ref{prop:ch3:thetagen}. \textit{Local rate.} In Lemma~\ref{lem:ch3:localrate} the tests use Lemma~\ref{lem:ch3:sep} and Assumption~\ref{ass:ch3:localbilip} on $\Theta_M^{\symbf{\Delta}}$, whose log covering number at scale $\varepsilon_n$ is again $O(\log n)$. In \eqref{eq:ch3:localpriormass} each free entry of $\bB$ contributes a factor of order $\varepsilon_n$ because its slab density is bounded below near $b^0_{jh}$. The factors for $\bA$ and $\bSigs$ are as before, and \eqref{eq:ch3:localdenom} is unchanged. \textit{Edge indicators.} The arguments of Section~\ref{sec:ch3:app-gammacons} concern a single coordinate $a_{jh}$, whose admissible slice is the same in $\Theta_M$ and $\Theta_M^{\symbf{\Delta}}$ because the exclusions do not involve $\bA$. They use Lemma~\ref{lem:ch3:localrate}, compactness, Assumption~\ref{ass:ch3:asympident} (in Lemma~\ref{lem:ch3:onecoordsep}) and the scalar expansion in $a_{jh}$, all of which hold on $\Theta_M^{\symbf{\Delta}}$ with the nuisance block consisting of the remaining free coordinates. Hence Theorem~\ref{thm:ch3:gammacons} holds for every $j\neq h$, and the union bound gives Proposition~\ref{prop:ch3:indgen}.
\end{proof}

\subsection{Posterior Consistency of \texorpdfstring{\MRRGMp{}}{IV-RGM+}}
\label{sec:ch3:app-pluscons}

\paragraph{Setting.} A \emph{map} is a set $S\subseteq\{1,\ldots,p\}\times\{1,\ldots,k\}$ of instrument--response pairs, with rows $S_j:=\{h:(j,h)\in S\}$. The true map is $S_0:=\{(j,h):b^0_{jh}\neq0\}$, with rows $S_{0j}$. In \MRRGMp{}, the indicators $\phi_{jh}$ define the map $S=\{(j,h):\phi_{jh}=1\}$. After integrating out $\psi_{jh}$, the indicators are independent with $\Pr(\phi_{jh}=1)=a_\psi/(a_\psi+b_\psi)$, so every map has a positive prior probability $\pi(S)$. Given $S$, the entries $b_{jh}$ with $(j,h)\in S$ have the slab prior $\Normal(0,\eta_{jh})$ and the other entries are zero. The priors of $\bA$ and $\bSigs$ are as in Section~\ref{sec:ch3:app-asymp-priors}. The prior is restricted to the set $\Theta_M^+$ of Assumption~\ref{ass:ch3:compactplus}.

The set $\Theta_M^{+}$ is compact, because the largest absolute cycle gain is a continuous function of $\bA$. For a map $S$, let $\Theta^S:=\{\theta\in\Theta_M^{+}:b_{jh}=0\text{ for }(j,h)\notin S\}$, and let $\mathcal{P}_n^{S}$ be the product of the priors of the entries of $\bA$, of the entries of $\bB$ in $S$ and of $\bSigs$, as a measure on $\Theta^S$ that is not normalized. The prior of \MRRGMp{} gives a measurable set $E$ the mass $\sum_S\pi(S)\mathcal{P}_n^{S}(E\cap\Theta^S)$, divided by its total mass. Write $g(\theta):=(\bPi(\theta),\symbf{\Omega}(\theta))$ for the reduced form, so that $g(\theta_0)=\vartheta_0=(\bPi_0,\symbf{\Omega}_0)$, and let
\begin{equation*}
m_S:=\int_{\Theta^S}\exp\{\ell_n(\theta)-\ell_n(\theta_0)\}\,\mathcal{P}_n^{S}(\mathrm{d}\theta).
\end{equation*}
The posterior probability of the map $S$ is then $\pi(S)m_S/\sum_{S'}\pi(S')m_{S'}$. The dimension of the free coordinates of $\Theta^{S}$ is $p(p-1)+|S|+p(p+1)/2$, and we write $q_0:=p(p-1)+|S_0|+p(p+1)/2$.

\begin{example}[Own and shared instruments]\label{ex:ch3:overid}
In the design of Example~\ref{ex:ch3:blockdiag}, suppose that every response has $m_i\ge2$ own instruments and that the number $u_i$ of shared instruments that act on response $i$ is smaller than $m_{(1)}=\min_lm_l$. Then Assumption~\ref{ass:ch3:overid} holds. Indeed, let $I(T):=\{l:G_l\subseteq T\}$. If $w_l\neq0$, then $\mathbf{w}^{\T}\bB_0$ is nonzero on $G_l$, so every $\mathbf{w}\in\mathcal{V}(T)$ is supported in $I(T)$ and $d(T)\le|I(T)|$. Let $i$ enter $\mathcal{V}(T)$, so that $i\in I(T)$, and let $v_i\le u_i$ be the number of shared instruments of response $i$ that are not in $T$. Each of them gives a linear constraint $\sum_lw_lb^0_{lh}=0$ on $\mathcal{V}(T)$ with $b^0_{ih}\neq0$. If $v_i=0$, then $T$ contains $S_{0i}$ and the sets $G_l$, $l\in I(T)$, so $|T|-d(T)\ge|S_{0i}|-1+\sum_{l\in I(T),l\neq i}(m_l-1)$, with equality to $|S_{0i}|-1$ only if $I(T)=\{i\}$ and $T=S_{0i}$. If $v_i\ge1$, the constraints lower the dimension by at least one, so $d(T)\le|I(T)|-1$, and they require $w_l\neq0$ for some $l\neq i$, so $|I(T)|\ge2$. Then $|T|-d(T)\ge m_i+(u_i-v_i)+\sum_{l\in I(T),l\neq i}m_l-|I(T)|+1\ge|S_{0i}|-1+(m_{(1)}-v_i)>|S_{0i}|-1$. The designs of Section~\ref{sec:sim} have $m_i\in\{3,10\}$ and $u_i\le2$. The condition $m_i\ge2$ cannot be dropped: with one own instrument per response and no shared instruments, $T=G_i\cup G_l$ gives $|T|-d(T)=0=|S_{0i}|-1$ with $T\neq S_{0i}$. More generally, the assumption fails whenever some response $i$ has a single instrument. For $l\neq i$ and $T=S_{0l}\cup S_{0i}$, both responses enter $\mathcal{V}(T)$, $d(T)\ge2$ and $|T|-d(T)\le|S_{0l}|-1$. If $T\neq S_{0l}$, this contradicts the assumption for response $l$. If $T=S_{0l}$, then $d(S_{0l})\ge2$, which contradicts it as well.
\end{example}

\begin{lem}[Dimension of the models that contain the truth]\label{lem:ch3:plusrank}
With $\mathcal V(T)$ and $d(T)$ as defined in Section~\ref{subsec:asymptotics}, let $S$ be a map and $\theta^*\in\Theta^S$ with $g(\theta^*)=\vartheta_0$. Let $J_S(\theta^*)$ be the Jacobian of the map from the off-diagonal entries of $\bA$ and the entries of $\bB$ in $S$ to $\bPi=(\bI_p-\bA)^{-1}\bB$, at $\theta^*$. Then
\begin{equation}\label{eq:ch3:plusrank}
\rank J_S(\theta^*)=p^2+|S|-\sum_{j=1}^pd(S_j).
\end{equation}
Under Assumptions~\ref{ass:ch3:compactplus} and~\ref{ass:ch3:overid}, $\rank J_S(\theta^*)\ge p(p-1)+|S_0|+1$ unless $S=S_0$ and $\theta^*=\theta_0$, and $\theta_0$ is the only point of $\Theta^{S_0}$ with reduced form $\vartheta_0$.
\end{lem}

\begin{proof}
Write $\theta^*=(\bA^*,\bB^*,\bSig^{**})$, $\mathbf{M}^*:=\bI_p-\bA^*$ and $\symbf{\Xi}:=\mathbf{M}^*\mathbf{M}_0^{-1}$, so that $\bB^*=\symbf{\Xi}\bB_0$ by \eqref{eq:ch3:xiform}. Since $\bPi(\theta^*)=\bPi_0$, a direction $(\mathrm{d}\bA,\mathrm{d}\bB)$ with $\diag(\mathrm{d}\bA)=\bzero$ and $\mathrm{d}\bB$ supported in $S$ has $\mathrm{d}\bPi=(\mathbf{M}^*)^{-1}\{(\mathrm{d}\bA)\bPi_0+\mathrm{d}\bB\}$. It lies in the kernel of $J_S(\theta^*)$ if and only if $\mathrm{d}\bB=-(\mathrm{d}\bA)\bPi_0$. For row $j$, put $\mathbf{w}^{\T}:=(\mathrm{d}\bA)_{j,:}\mathbf{M}_0^{-1}$, which is a one-to-one linear function of row $j$ of $\mathrm{d}\bA$. Row $j$ of $(\mathrm{d}\bA)\bPi_0$ is $\mathbf{w}^{\T}\bB_0$. The kernel therefore consists of the directions whose rows satisfy $\mathbf{w}\in\mathcal{V}(S_j)$ and $(\mathbf{w}^{\T}\mathbf{M}_0)_j=0$, the latter being the zero diagonal of $\mathrm{d}\bA$. Row $j$ of $\symbf{\Xi}$ lies in $\mathcal{V}(S_j)$, because row $j$ of $\bB^*$ is supported in $S_j$, and it satisfies $(\symbf{\xi}_j^{\T}\mathbf{M}_0)_j=(\mathbf{M}^*)_{jj}=1$. The linear constraint is thus not identically zero on $\mathcal{V}(S_j)$, and the kernel has dimension $\sum_j\{d(S_j)-1\}$. Subtracting it from the number $p(p-1)+|S|$ of columns gives \eqref{eq:ch3:plusrank}.

Since $\symbf{\Xi}$ is invertible, the Leibniz formula for its determinant gives a permutation $\sigma$ with $\xi_{j\sigma(j)}\neq0$ for every $j$. Response $\sigma(j)$ then enters $\mathcal{V}(S_j)$, and Assumption~\ref{ass:ch3:overid} gives $|S_j|-d(S_j)\ge|S_{0\sigma(j)}|-1$. Summing over $j$ and using \eqref{eq:ch3:plusrank},
\begin{equation*}
\rank J_S(\theta^*)\ge p^2+\sum_j\{|S_{0\sigma(j)}|-1\}=p(p-1)+|S_0|.
\end{equation*}
Equality requires $S_j=S_{0\sigma(j)}$ for every $j$. Then $\symbf{\xi}_j\in\mathcal{V}(S_{0\sigma(j)})$, which is spanned by $\mathbf{e}_{\sigma(j)}$ because $d(S_{0\sigma(j)})=1$. Every row of $\symbf{\Xi}$ is then pure in the sense of the proof of Theorem~\ref{thm:ch3:ident}, and Step~2 of that proof, with the bound on the cycle gains on $\Theta_M^{+}$, gives $\symbf{\Xi}=\bI_p$. Hence $\theta^*=\theta_0$ and $S=S_0$. For $S=S_0$ the rank cannot exceed the number $p(p-1)+|S_0|$ of columns of $J_{S_0}$, so every point of $\Theta^{S_0}$ with reduced form $\vartheta_0$ attains the equality and is therefore $\theta_0$.
\end{proof}

\begin{lem}[Likelihood bounds in the reduced form]\label{lem:ch3:plusquad}
Under Assumptions~\ref{ass:ch3:design} and~\ref{ass:ch3:compact}, there are constants $0<c_1\le C_1<\infty$ and a random variable $D_n=O_{\Prob}(1)$ such that, with probability tending to one, for every $\theta\in\Theta_M$,
\begin{equation*}
-C_1\{n\|g(\theta)-\vartheta_0\|^2+D_n^2\}\ \le\ \ell_n(\theta)-\ell_n(\theta_0)\ \le\ C_1D_n^2-c_1n\|g(\theta)-\widehat\vartheta_n\|^2,
\end{equation*}
where $\widehat\vartheta_n=(\widehat\bPi_n,\widehat{\symbf{\Omega}}_n)$ is the least-squares estimator of the reduced form and $\|\widehat\vartheta_n-\vartheta_0\|=D_n/\sqrt n$. In particular, if $\|g(\theta)-\vartheta_0\|\ge r\ge2D_n/\sqrt n$, then $\ell_n(\theta)-\ell_n(\theta_0)\le C_1D_n^2-c_1nr^2/4$.
\end{lem}

\begin{proof}
With $\widehat\bPi_n=\bS_{yx}\bS_{xx}^{-1}$ and $\widehat{\symbf{\Omega}}_n=\bS_{yy}-\widehat\bPi_n\bS_{xx}\widehat\bPi_n^{\T}$, the log-likelihood \eqref{eq:ch3:loglik} depends on $\theta$ through $(\bPi,\symbf{\Omega})=g(\theta)$ only, and
\begin{equation*}
\ell_n(\theta)-\ell_n(\widehat\vartheta_n)=-\frac n2\Big[\tr(\symbf{\Omega}^{-1}\widehat{\symbf{\Omega}}_n)-\log|\symbf{\Omega}^{-1}\widehat{\symbf{\Omega}}_n|-p+\tr\{\symbf{\Omega}^{-1}(\bPi-\widehat\bPi_n)\bS_{xx}(\bPi-\widehat\bPi_n)^{\T}\}\Big],
\end{equation*}
where $\ell_n(\widehat\vartheta_n)$ denotes the value at $(\widehat\bPi_n,\widehat{\symbf{\Omega}}_n)$. The first three terms equal $\sum_i(\lambda_i-\log\lambda_i-1)$ for the eigenvalues $\lambda_i$ of $\symbf{\Omega}^{-1/2}\widehat{\symbf{\Omega}}_n\symbf{\Omega}^{-1/2}$. On $\Theta_M$ the eigenvalues of $\symbf{\Omega}$ lie in a fixed compact subset of $(0,\infty)$, and with probability tending to one so do those of $\widehat{\symbf{\Omega}}_n$ and $\bS_{xx}$. On a compact subset of $(0,\infty)$, $\lambda-\log\lambda-1$ lies between two positive multiples of $(\lambda-1)^2$. Hence the bracket lies between two positive multiples of $\|g(\theta)-\widehat\vartheta_n\|^2$, that is,
$-C_1'n\|g(\theta)-\widehat\vartheta_n\|^2\le\ell_n(\theta)-\ell_n(\widehat\vartheta_n)\le-c_1n\|g(\theta)-\widehat\vartheta_n\|^2$. By the central limit theorem, $D_n=\sqrt n\|\widehat\vartheta_n-\vartheta_0\|=O_{\Prob}(1)$. Applying the lower bound at $\theta_0$ gives $\ell_n(\widehat\vartheta_n)-\ell_n(\theta_0)\le C_1'D_n^2$, and the upper bound gives $\ell_n(\theta_0)\le\ell_n(\widehat\vartheta_n)$. The two displayed inequalities follow, with $\|g(\theta)-\widehat\vartheta_n\|^2\le2\|g(\theta)-\vartheta_0\|^2+2D_n^2/n$ for the first. For the last statement, $\|g(\theta)-\widehat\vartheta_n\|\ge r-D_n/\sqrt n\ge r/2$.
\end{proof}

\begin{lem}[Volume of the parameters with a nearly true reduced form]\label{lem:ch3:plusball}
Let $G$ be a continuously differentiable map from an open subset of $\R^D$ to $\R^m$ with $G(\theta^*)=\vartheta_0$ and $\rank\,\mathrm{d}G(\theta^*)\ge q$. Then there are a neighborhood $\mathcal{U}$ of $\theta^*$ and a constant $C$ such that the Lebesgue measure of $\{\theta\in\mathcal{U}:\|G(\theta)-\vartheta_0\|\le r\}$ is at most $Cr^q$ for all $0<r\le1$.
\end{lem}

\begin{proof}
Choose $q$ coordinates $G_I$ of $G$ whose differential at $\theta^*$ has rank $q$. By the local submersion theorem there are a bounded neighborhood $\mathcal{U}$ of $\theta^*$ and a diffeomorphism $\varphi$ of $\mathcal{U}$ onto a bounded open set such that $G_I(\varphi^{-1}(z))=G_I(\theta^*)+(z_1,\ldots,z_q)$. The set in question is contained in $\varphi^{-1}(\{z:|z_l|\le r,\ l\le q\})$, whose measure is at most $(2r)^q$ times a constant that bounds the Jacobian determinant of $\varphi^{-1}$ and the volume of the remaining coordinates.
\end{proof}

\begin{proof}[Proof of Theorem~\ref{thm:ch3:pluscons}]
\textit{Part (i).} There are finitely many maps and $\pi(S)>0$ for each, so it suffices to show that $m_S/m_{S_0}\to0$ in probability for every $S\neq S_0$.

\textit{Lower bound for $S_0$.} Let $\mathcal{B}_n$ be the set of $\theta\in\Theta^{S_0}$ with $\|\theta-\theta_0\|_\infty\le n^{-1/2}$. For large $n$ it is a full ball in the free coordinates, because $\theta_0$ is interior to $\Theta_M$ and its cycle gains are smaller than $c_g$. On it, $\|g(\theta)-\vartheta_0\|\le C_gn^{-1/2}$ for a constant $C_g$, because $g$ has bounded derivatives on $\Theta_M$, and Lemma~\ref{lem:ch3:plusquad} gives $\ell_n(\theta)-\ell_n(\theta_0)\ge-C_1(C_g^2+D_n^2)$. Each coordinate has prior mass at least $cn^{-1/2}$ on its interval of half-width $n^{-1/2}$ around the true value, for a constant $c>0$ that does not depend on $\nu_{1,n}$. For an entry of $\bA$ this holds because the slab component has probability $a_\rho/(a_\rho+b_\rho)$ and a density bounded below on $[-C_A,C_A]$, the slab scales being truncated. For an entry of $\bB$ in $S_0$ it holds for the same reason. The prior of $\bSigs$ has a density bounded below near $\bSig_0^*$ (Assumption~\ref{ass:ch3:sigmaprior}), which gives the factor for its $p(p+1)/2$ coordinates. Hence
\begin{equation}\label{eq:ch3:pluslower}
m_{S_0}\ \ge\ \exp\{-C_1(C_g^2+D_n^2)\}\,c^{q_0}\,n^{-q_0/2}.
\end{equation}

\textit{Maps that cannot reproduce $\vartheta_0$.} Let $S$ be a map with $\vartheta_0\notin g(\Theta^S)$. The set $g(\Theta^S)$ is compact, so its distance $\delta_S$ to $\vartheta_0$ is positive. By Lemma~\ref{lem:ch3:plusquad}, with probability tending to one, $\ell_n(\theta)-\ell_n(\theta_0)\le C_1D_n^2-c_1n\delta_S^2/4$ on $\Theta^S$, so $m_S\le\exp(C_1D_n^2-c_1n\delta_S^2/4)$ and $m_S/m_{S_0}\to0$ by \eqref{eq:ch3:pluslower}.

\textit{Other maps.} Let $S\neq S_0$ be a map for which $F_S:=\{\theta\in\Theta^S:g(\theta)=\vartheta_0\}$ is not empty. In the free coordinates of $\Theta^S$, the differential of $g$ at a point $\theta^*\in F_S$ has rank $\rank J_S(\theta^*)+p(p+1)/2$, because $\bPi$ does not depend on $\bSigs$ and $\mathrm{d}\bSigs\mapsto(\mathbf{M}^*)^{-1}(\mathrm{d}\bSigs)(\mathbf{M}^*)^{-\T}$ is one-to-one. By Lemma~\ref{lem:ch3:plusrank} this rank is at least $q_0+1$. The map $g$ is smooth on an open set that contains $\Theta_M$. By Lemma~\ref{lem:ch3:plusball} and the compactness of $F_S$, there are finitely many open sets $\mathcal{U}_1,\ldots,\mathcal{U}_L$ that cover $F_S$ and a constant $C$ such that the Lebesgue measure of $\{\theta\in\mathcal{U}_l\cap\Theta^S:\|g(\theta)-\vartheta_0\|\le r\}$ is at most $Cr^{q_0+1}$. On the compact set $\Theta^S\setminus\bigcup_l\mathcal{U}_l$ the continuous function $\|g(\theta)-\vartheta_0\|$ has no zero, so it is bounded below by some $\delta>0$. Let $r_n:=K'\sqrt{\log n/n}$ with $c_1K'^2/4\ge q_0/2+1$. For large $n$, $r_n<\delta$ and, with probability tending to one, $r_n\ge2D_n/\sqrt n$. Split $\Theta^S$ into the set where $\|g(\theta)-\vartheta_0\|\le r_n$, which is contained in $\bigcup_l\mathcal{U}_l$, and its complement. On the complement, Lemma~\ref{lem:ch3:plusquad} gives $\ell_n(\theta)-\ell_n(\theta_0)\le C_1D_n^2-(q_0/2+1)\log n$. On the first set, $\ell_n(\theta)-\ell_n(\theta_0)\le C_1D_n^2$, and the prior measure is at most its Lebesgue measure times a bound on the prior density. The density of an entry of $\bA$ is a mixture of normal densities with variances of at least $\nu_{1,n}\underline t$, so it is at most $C\nu_{1,n}^{-1/2}$. The densities of the entries of $\bB$ and of $\bSigs$ are bounded. Hence
\begin{equation*}
m_S\ \le\ e^{C_1D_n^2}\Big\{C\,\nu_{1,n}^{-p(p-1)/2}\,(K'^2\log n/n)^{(q_0+1)/2}+n^{-q_0/2-1}\Big\}.
\end{equation*}
With \eqref{eq:ch3:pluslower},
\begin{equation*}
\frac{m_S}{m_{S_0}}\ \le\ C'e^{2C_1D_n^2}\Big\{\nu_{1,n}^{-p(p-1)/2}(\log n)^{(q_0+1)/2}\,n^{-1/2}+n^{-1}\Big\},
\end{equation*}
which tends to zero in probability because $\nu_{1,n}\ge n^{-\alpha}$ with $\alpha p(p-1)<1$.

\textit{Parts (ii) and (iii).} Given the map $S_0$, the posterior of \MRRGMp{} is the posterior of \MRRGM{} with the map $\symbf{\Delta}$ that equals one on $S_0$ and zero elsewhere, on the parameter set $\Theta^{S_0}$. The matrix $\bB_0$ satisfies the exclusions of this map, and the rank condition of Theorem~\ref{thm:ch3:identrgm} holds. Indeed, $d(S_{0j})=1$ means that no nonzero $\mathbf{w}$ with $w_j=0$ has $\mathbf{w}^{\T}\bB_0$ supported in $S_{0j}$, so the rows of $\bB_{0,-j,H_j}$ are linearly independent, and \eqref{eq:ch3:pirank} gives the rank of $\bPi_{0,-j,H_j}$. The set $\Theta^{S_0}$ is a compact subset of $\Theta_M^{\symbf{\Delta}}$ that contains a neighborhood of $\theta_0$, and the two identification conditions hold on it by Lemma~\ref{lem:ch3:bilip}. The proofs of Theorems~\ref{thm:ch3:thetacons} and~\ref{thm:ch3:indicators} in Section~\ref{sec:ch3:app-rgmcons} then apply on $\Theta^{S_0}$. The only change is that the admissible values of a single coordinate $a_{jh}$ are now those in $\Theta^{S_0}$. They contain a neighborhood of $a^0_{jh}$, because the cycle gains of $\theta_0$ are smaller than $c_g$, and the lemmas of Section~\ref{sec:ch3:app-gammacons} use nothing else about them besides compactness and identifiability. For any event $E$,
\begin{equation*}
\mathcal{P}_n(E\mid\mathcal{D}_n)\le\mathcal{P}_n(S\neq S_0\mid\mathcal{D}_n)+\mathcal{P}_n(E\mid S=S_0,\mathcal{D}_n),
\end{equation*}
and part~(i) gives (ii) and (iii).
\end{proof}

The rate $\nu_{1,n}\ge n^{-\alpha}$ is compatible with Assumption~\ref{ass:ch3:nurate}: both hold for $\nu_{1,n}=1/\log n$ and for $\nu_{1,n}=n^{-\alpha}$ with $\alpha<1/\{p(p-1)\}$. The restriction arises because the proof uses a simple bound on the prior density of $\bA$, which is larger than necessary.

\FloatBarrier
\vskip 0.2in
\bibliography{Bibliography-MM-MC}

\clearpage
\includepdf[pages=-]{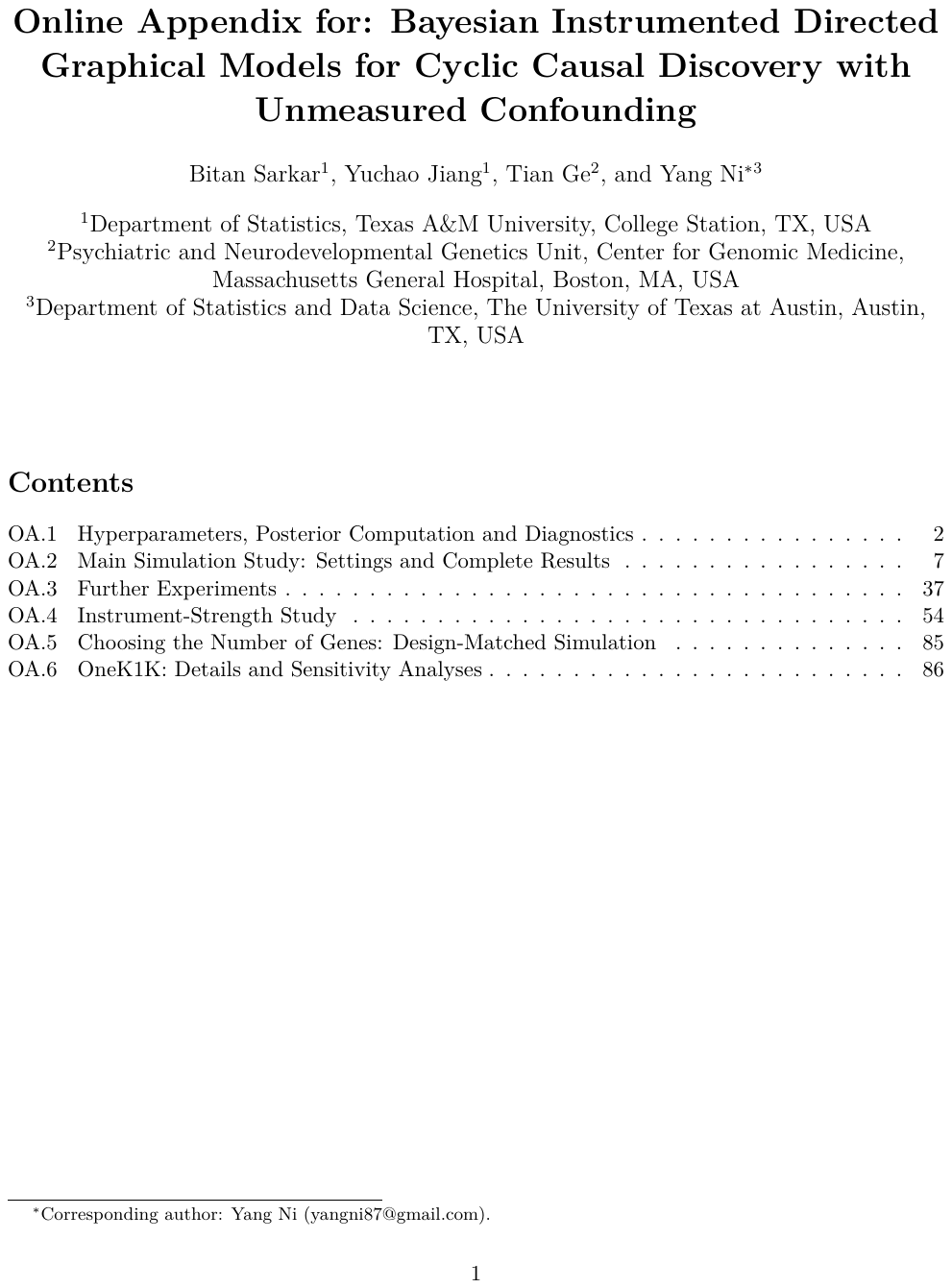}

\end{document}